\documentclass[aps,nofootinbib,prd,superscriptaddress, tightenlines, notitlepage, twocolumn, floatfix]{revtex4-2}

\usepackage{amsmath}
\usepackage{mathtools}
\usepackage{graphicx}
\usepackage[normalem]{ulem}
\usepackage[com pat=1.1.0]{tikz-feynman}
\usepackage{tikz}
\usepackage{aas_macros}
\usetikzlibrary{external}
\usepackage{comment}
\usetikzlibrary{arrows}
\usepackage[colorlinks=true, allcolors=blue]{hyperref}

\begin{document}

\title{Neutrino flavor instabilities in a binary neutron star merger remnant: \\
Roles of a long-lived hypermassive neutron star}

\author{Hiroki Nagakura}
\email{hiroki.nagakura@nao.ac.jp}
\affiliation{Division of Science, National Astronomical Observatory of Japan, 2-21-1 Osawa, Mitaka, Tokyo 181-8588, Japan}

\author{Kohsuke Sumiyoshi}
\affiliation{National Institute of Technology, Numazu College, Ooka 3600, Numazu, Shizuoka 410-8501, Japan}

\author{Sho Fujibayashi}
\affiliation{Frontier Research Institute of Interdisciplinary Sciences, Tohoku University, Sendai 980-8578}
\affiliation{Astronomical Institute, Graduate School of Science, Tohoku University, Sendai 980-8578}
\affiliation{Max-Planck-Institut f\"ur Gravitationsphysik (Albert-Einstein-Institut), Am M\"uhlenberg 1, D-14476 Potsdam-Golm, Germany}

\author{Yuichiro Sekiguchi}
\affiliation{Toho University, Funabashi, Chiba, 274-8510, Japan}
\affiliation{Center for Gravitational Physics and Quantum Information, Yukawa Institute for Theoretical Physics, Kyoto University, Kyoto, 606-8502, Japan}

\author{Masaru Shibata}
\affiliation{Max-Planck-Institut f\"ur Gravitationsphysik (Albert-Einstein-Institut), Am M\"uhlenberg 1, D-14476 Potsdam-Golm, Germany}
\affiliation{Center for Gravitational Physics and Quantum Information, Yukawa Institute for Theoretical Physics, Kyoto University, Kyoto, 606-8502, Japan}

\begin{abstract}
Understanding the post-merger evolution of binary neutron star merger (BNSM) requires accurate modeling of neutrino transport and microphysics including neutrino flavor conversions. Many previous studies have suggested that fast flavor instability (FFI) and collisional flavor instability (CFI) pervade inner regions of BNSM remnant, and they could impact on fluid dynamics and r-process nucleosynthesis. In this work, we re-examine prospects of occurrences of FFI and CFI using Boltzmann neutrino transport, assuming a frozen fluid background obtained from a numerical relativity simulation of BNSM. We pay special attention to a case involving a long-lived ($>1\,$s) hypermassive neutron star (HMNS). Apart from confirming the claim that these flavor instabilities can occur in BNSM remnants, some new insights are revealed. We identify multiple mechanisms responsible for generating electron neutrino lepton number (ELN) angular crossings, corresponding to a key indicator of FFI onset, which differ notably from those in black hole (BH) accretion disk systems. We argue that the appearance of positive chemical potential of electron-type neutrinos plays important roles on generating ELN angular crossings. For CFI, their growth rates are generally lower than FFI, but they can persistently occur in most of the accretion disk up to $\sim 1\,$s. We also find that neglecting contributions of heavy-leptonic neutrinos results in overestimating growth rate and area of unstable regions of CFI. Our result suggests that FFI (CFI) tends to occur transiently (persistently) and locally (widespread in the disk), and FFI is more sensitive to the central compact object (HMNS or BH) than CFI, though more self-consistent simulations with incorporating effects of flavor conversions are needed to confirm these claims.
\end{abstract}
\maketitle

\section{Introduction}\label{sec:intro}
Binary neutron star merger (BNSM) is one of the most intense collisions in our universe. For many years, it has been considered a promising gravitational wave (GW) source in the frequency range of $\sim$kHz \cite{1993PhRvL..70.2984C}, triggering short gamma-ray bursts (GRB) \cite{1991AcA....41..257P} and kilonova \cite{1998ApJ...507L..59L}, and one of the most efficient cosmic factories of r-process elements \cite{1974ApJ...192L.145L}. The first multi-messenger observation involving gravitational waves and electromagnetic radiations, GW~170817/GRB~170817A/AT~2017gfo, provided a convincing evidence linking BNSM to these phenomena, confirming many aspects of our current understanding of BNSM theory \cite{2017PhRvL.119p1101A,2017ApJ...848L..12A}. However, considerable uncertainties still exist in the theory, particularly regarding underlying physical processes (see recent reviews, e.g., \cite{2016nure.book.....S,2017RPPh...80i6901B,2019ARNPS..69...41S,2020ARNPS..70...95R,2023LRCA....9....1F}). As GW detector sensitivity continues to be improved \cite{2021arXiv210909882E,2025arXiv250312263A} and the electromagnetic detector network expands \cite{2024arXiv241102342L}, we expect many more BNSM observations in the coming decades, while theoretical uncertainties may constitute a major obstacle to extract detailed physical information from observed data. Improving our knowledge of input physics and precise modeling of the interplay among physical processes of BNSMs are indispensable for the further progress of multi-messenger astronomy with BNSMs.

One of lingering uncertainties in the BNSM theory is neutrino kinetics (see, e.g., \cite{2023LRCA....9....1F} for a recent review of numerical techniques of neutrino transport in BNSM simulations). Neutrinos play a central role in carrying away energy and leptons from the hot dense matter. These transport processes play an important role on the thermal evolution of BNSM remnants and affect the r-process nucleosynthesis in the ejecta. This exhibits the importance of accurate modeling of neutrino radiation field to link electromagnetic signals to the dynamics of BNSM. Modeling the neutrino radiation field requires solving kinetic equations that account for both macroscopic transport and microphysical processes. Despite significant progress in recent decades, self-consistent neutrino transport simulations incorporating all necessary input physics from first-principles remain intractable due to technical challenges (e.g., handling macro- and microphysical processes with vastly different characteristic scales) and computational costs. It is, hence, essential to combine different approaches, each employing specific approximations for different physical processes, in order to gain insights into roles of neutrinos in BNSM systems. In the present study, we pay special attention to neutrino flavor instabilities, one of the quantum kinetic features of neutrinos. This issue has been studied considerably as a key piece of missing physics in theories of high energy astrophysical systems including BNSM and core-collapse supernova (CCSN); see recent reviews \cite{2021ARNPS..71..165T,2022Univ....8...94C,2022arXiv220703561R,2024RvMP...96b5004V,2024PrPNP.13704107F,2025arXiv250305959J}.

The neutrino flavor instabilities discussed in this paper arise from neutrino self-interactions. In environments with high neutrino densities, neutrinos can undergo coherent forward scattering off one another \cite{1992PhLB..287..128P}. These neutrino self-interactions alter the neutrino dispersion relation through refractive effects, in which the mechanism is analogous to the neutrino oscillation in medium. However, neutrinos are not always in flavor eigenstates during flavor conversions, implying that the self-interaction can induce flavor off-diagonal refractive effects. As flavor conversions enhance flavor coherence, they can, in turn, amplify these off-diagonal refractive effects. Flavor instabilities emerge from such a positive feedback cycle between them, while the stability hinges on details of neutrino distributions and collision term, as we shall discuss below.

The fast flavor instability (FFI) \cite{2005PhRvD..72d5003S} and collisional flavor instability (CFI) \cite{2023PhRvL.130s1001J} are currently matter of intense investigations in BNSM systems; see, e.g., \cite{2017PhRvD..95j3007W,2017PhRvD..96l3015W,2020PhRvD.102j3015G,2021JCAP...01..017P,2021PhRvL.126y1101L,2022PhRvD.106h3005R,2022PhRvD.105h3024J,2022PhRvD.106j3003F,2023PhRvD.108j3014N,2024PhRvD.110j3019R,2024ApJ...974..110M,2024ApJ...963...11G,2024PhRvD.109h3031Z,2024PhRvD.109d3046F,2025PhRvD.111b3015K,2025arXiv250311758Q,2025arXiv250323727L} for FFI and \cite{2023PhRvD.108h3002X} for CFI. The growth rate of FFI can be solely characterized by neutrino self-interaction potential ($\mu$), and the resultant timescale of flavor conversions could be several orders of magnitude faster than any other relevant physical processes in BNSM. On the other hand, the growth rate of CFI can be roughly estimated by $\Gamma$ for non-resonance or $\sqrt{\Gamma \mu}$ for resonance, where $\Gamma$ denotes collision frequency (see Sec.~\ref{subsec:flavorinstaana} for more formal definitions). This suggests that CFI could be especially active in optically thick regions, which potentially have an impact on BNSM dynamics.

Assuming neutrinos are nearly in flavor eigenstates, the instability criteria for both FFI and CFI can be given analytically; see \cite{2019ApJ...886..139N,2020PhRvR...2a2046M} for FFI and \cite{2023PhRvD.107l3011L,2024PhRvD.109b3012A} for CFI. The former (FFI) is linked to a so-called zero-angular ELN-XLN crossings, where ELN and XLN denote electron- and heavy-leptonic neutrino lepton number, respectively \cite{2022PhRvD.105j1301M,2023PhRvD.107j3022Z}. It is worthy to note that the authors in \cite{2024JHEP...08..225F} offered a more physical intuition on the instability criterion in a manner analogous to plasma waves in Vlasov equation. CFI, on the other hand, arises from differences in decoherence rate between neutrinos and antineutrinos due to their unequal matter-collision frequencies. As suggested by \cite{2023PhRvD.108h3002X,2023PhRvD.107l3011L}, CFI is associated with neutrino energy crossings weighted by reaction rate accounts, which aligns with a claim by \cite{2022PhRvL.128h1102D} that all flavor instabilities (including slow modes; see, e.g., \cite{2010ARNPS..60..569D} and references therein) are fundamentally associated with crossings in momentum space of neutrinos.

There is broad agreement in the community that FFI occurs in BNSM remnants. The authors in \cite{2017PhRvD..95j3007W} offer a simple geometric argument of how ELN angular crossings\footnote{In general, heavy-leptonic neutrinos and their antipartners have different distributions, implying that XLN contributions need to be taken into account for FFI. On the other hand, XLN is expected to be much less than ELN in BNSM remnant systems. For this reason, all heavy-leptonic neutrinos and their antipartners have been collectively treated even in the modern BNSM simulations (but see \cite{2023A&A...672A.124L,2024arXiv241119178H,2024arXiv240904420G,2025PhRvD.111d3013P}).} can form. Assuming that there are optically thick regions for both $\nu_e$ and $\bar{\nu}_e$ in accretion disks (where $\nu_e$ and $\bar{\nu}_e$ denote electron-type neutrinos and their antipartners, respectively), they suggest that differences in the shapes of these neutrino spheres (or energy spheres) can lead to ELN angular crossings at any spatial positions above the disk. Further investigation in \cite{2022PhRvD.106h3005R}, which studies ELN angular crossings based on multi-angle neutrino transport on top of a fluid distribution obtained by realistic BNSM simulations in the early phase ($\sim 5\,$ms after merger), confirm the overall trend. Similarly, the authors in \cite{2022PhRvD.105h3024J,2024ApJ...974..110M,2025PhRvD.111b3015K} analyze ELN angular crossings based on simulations for black hole (BH) accretion disk system as a model of BNSM remnants. These studies consistently show the presence of ELN crossings, but they also find that underlying mechanisms are more complex than the simple geometric argument \cite{2017PhRvD..95j3007W}. The complexity mirrors findings in CCSN, where multiple mechanisms can give rise to ELN crossings (see, e.g., \cite{2019ApJ...886..139N,2021PhRvD.103f3033A,2021PhRvD.104h3025N}). Regarding CFI, on the other hand, the authors in \cite{2023PhRvL.130s1001J} show that both non-resonance and resonance CFI can occur in BNSM remnant based on a simulation data of BNSM. However, there are no systematic survey regarding occurrences of CFI. We also note that the study in \cite{2023PhRvL.130s1001J} ignore contributions of heavy-leptonic neutrinos, which potentially leads to false-positive in identifying CFI \cite{2023PhRvD.108l3024L,2024PhRvD.109b3012A}.

In this paper, we present the first systematic study for occurrences of both FFI and CFI based on a long-term simulation of BNSM remnant. We pay special attention to a case with a long-lived HMNS ($> 1$ second), which could emerge in coalescenses of relatively low mass two neutron stars ($\lesssim 1.4 M_{\odot}$). Such systems are expected to eject larger amounts of neutrinos and baryons from the system compared to short-lived HMNS or prompt formation of BH \cite{2018ApJ...860...64F,2018MNRAS.481.3670R,2020ApJ...901..122F,2021ApJ...913..100K}, indicating that flavor instabilities could give large impacts on nucleosynthesis in ejecta. For FFI, we provide a concise interpretation of complex features in generating ELN angular crossings. In addition to this, we propose a necessary (though not sufficient) condition of occurrences for FFI, which will be useful to identify FFI in simulations with approximate (non-multi-angle) neutrino transport. We also discuss some new insights on CFI; for instances their persistency and ubiquity in the accretion disk.

This paper is organized as follows. In Sec.~\ref{sec:methodsandmodels}, we summarize our BNSM remnant model (Sec.~\ref{subsec:BNSMremmodel}), computational setup and method for Boltzmann neutrino transport simulations (Sec.~\ref{subsec:Boltzmann}), and stability analysis of FFI and CFI (Sec.~\ref{subsec:flavorinstaana}). All results in the present study are encapsulated in Sec.~\ref{sec:results}. In the section, we first briefly describe the global and time-dependent properties of FFI and CFI (Sec.~\ref{subsec:overall}), and then delve into the detailed structures for FFI and CFI in Sec.~\ref{subsec:mechaELNgene}~and~\ref{subsec:CFIanalysis}, respectively. Finally, we conclude and summarize our results in Sec.~\ref{sec:summary}. Throughout this paper, we use the unit with $c = \hbar = 1$, where $c$ and $\hbar$ are the light speed and the reduced Planck constant, respectively.

\section{Methods and models}\label{sec:methodsandmodels}

\subsection{Remnant model}\label{subsec:BNSMremmodel}

\begin{figure*}[ht]
\begin{minipage}{1.0\textwidth}
\centering
\includegraphics[width=0.32\linewidth]{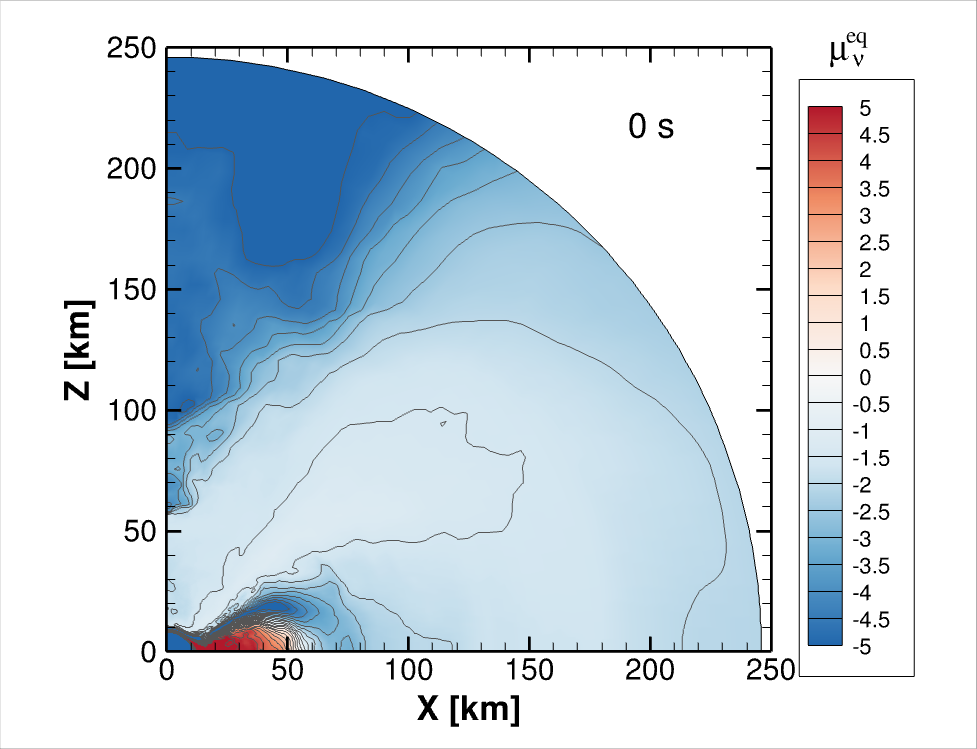}
\includegraphics[width=0.32\linewidth]{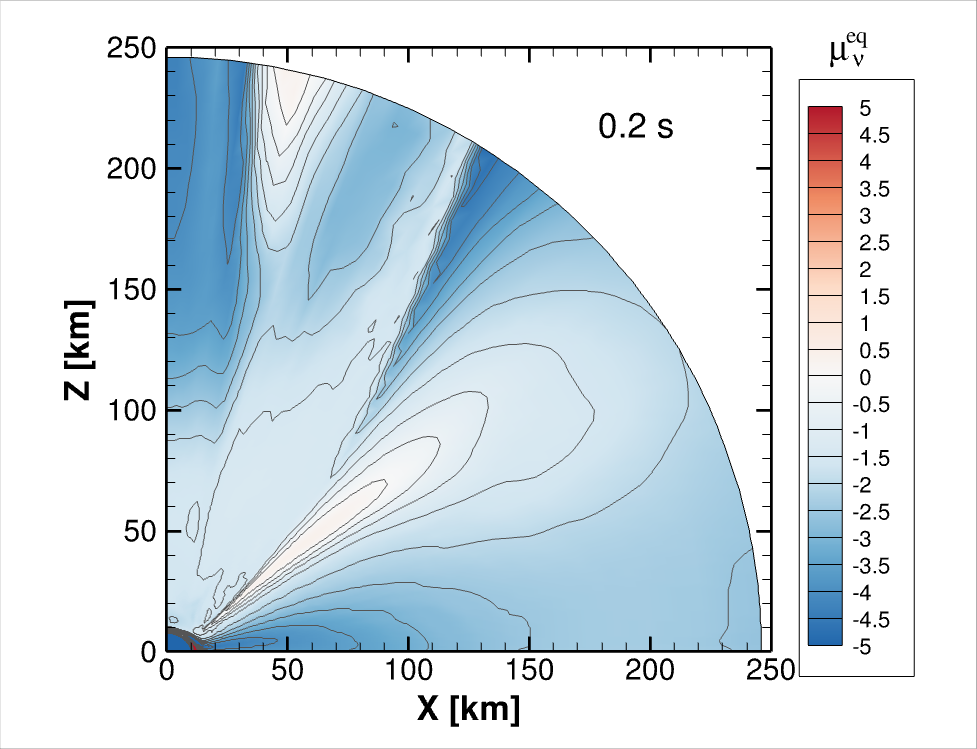}
\includegraphics[width=0.32\linewidth]{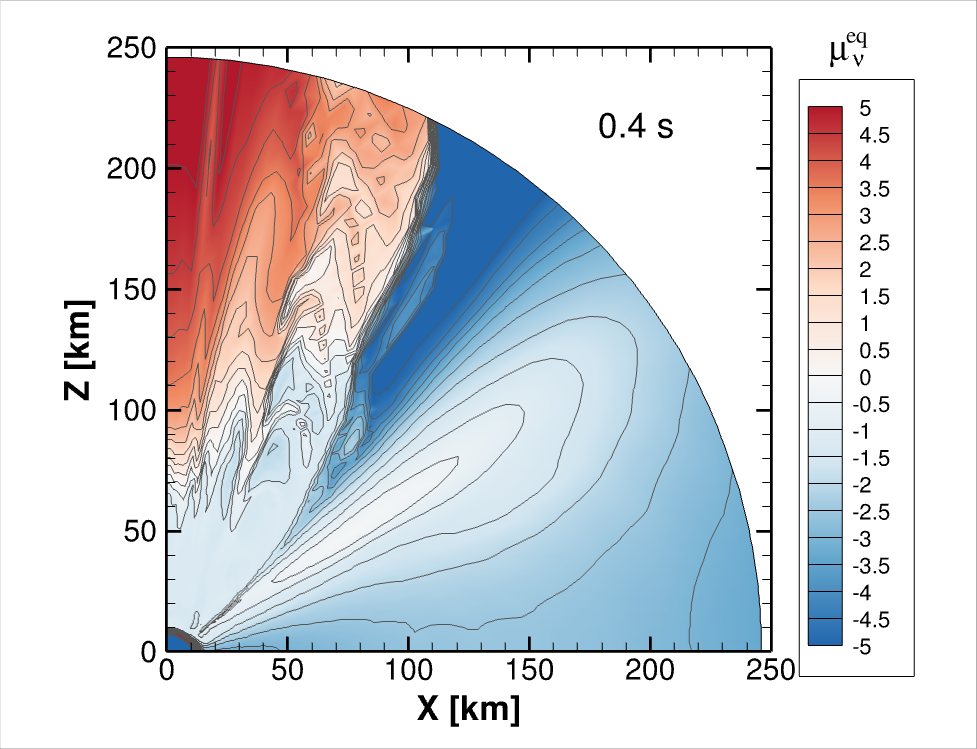}
\includegraphics[width=0.32\linewidth]{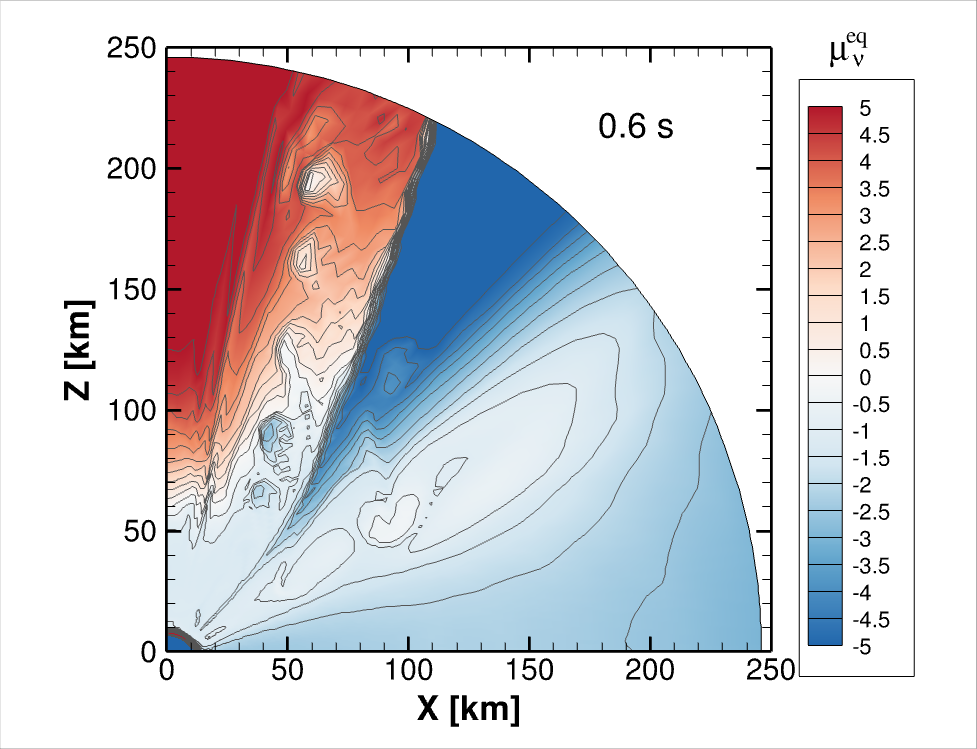}
\includegraphics[width=0.32\linewidth]{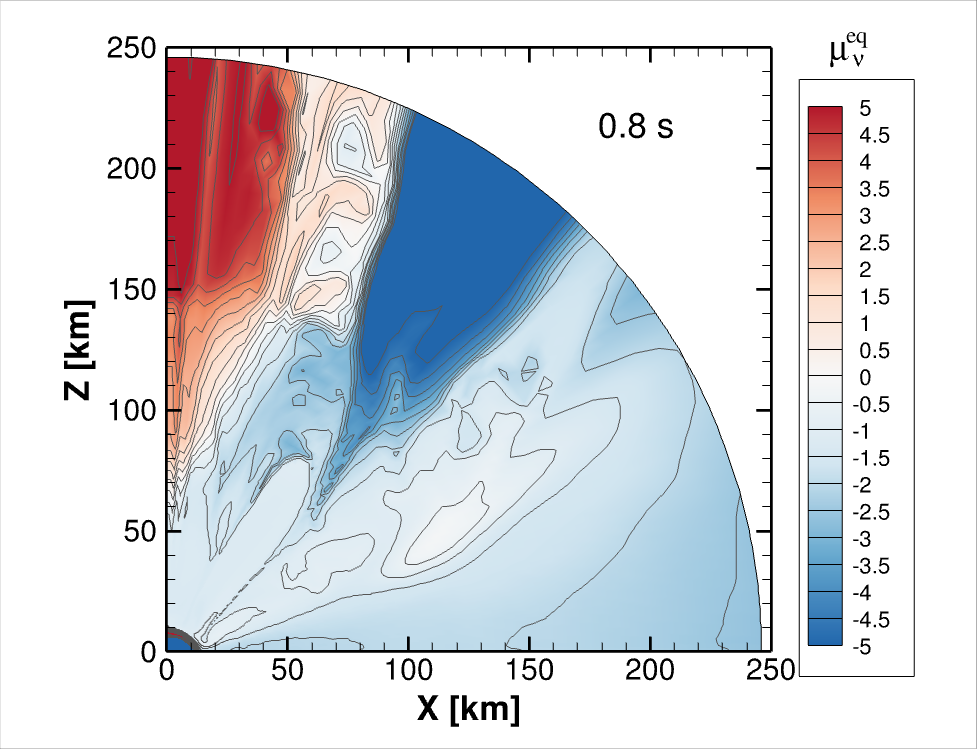}
\includegraphics[width=0.32\linewidth]{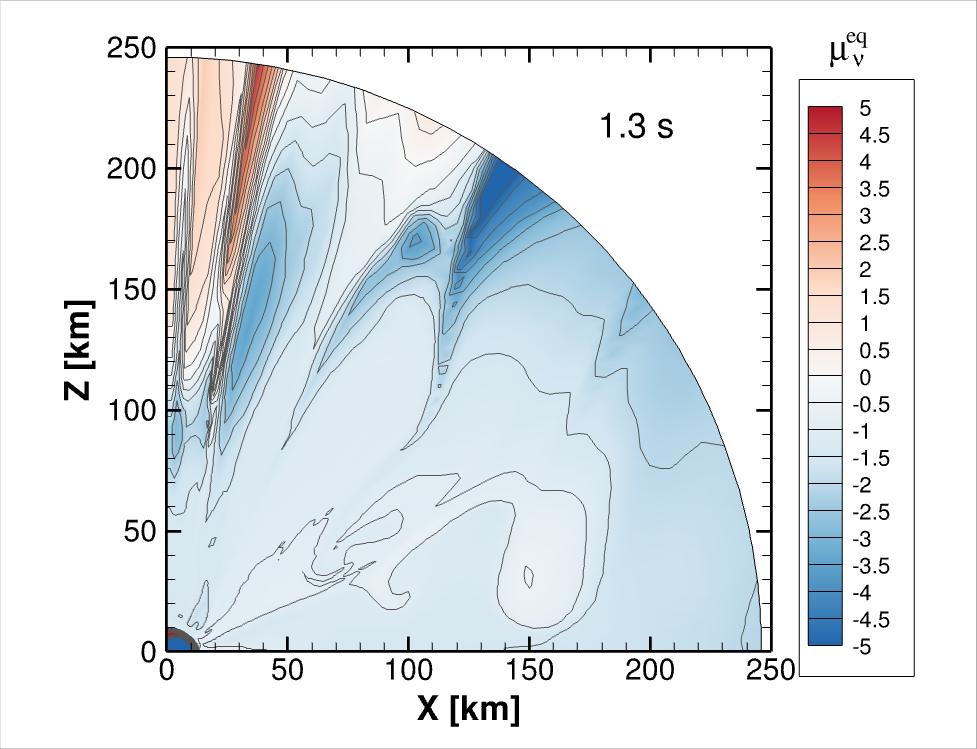}
\end{minipage}
\caption{Profiles of $\nu_e$ chemical potential under beta equilibrium ($\mu_{\nu}^{\rm eq}$) in the unit of $[\mathrm{MeV}]$. This can be computed from matter profiles as $\mu_{e} + \mu_{p} - \mu_{n}$. The time displayed in each panel is measured from the timing when we switch to viscous-radiation hydrodynamic simulation. The post-merger time can be obtained by adding $\sim 50\,$ms to them.}
\label{fig:nuechemipote}
\end{figure*}

\begin{figure*}[ht]
\begin{minipage}{1.0\textwidth}
\centering
\includegraphics[width=0.32\linewidth]{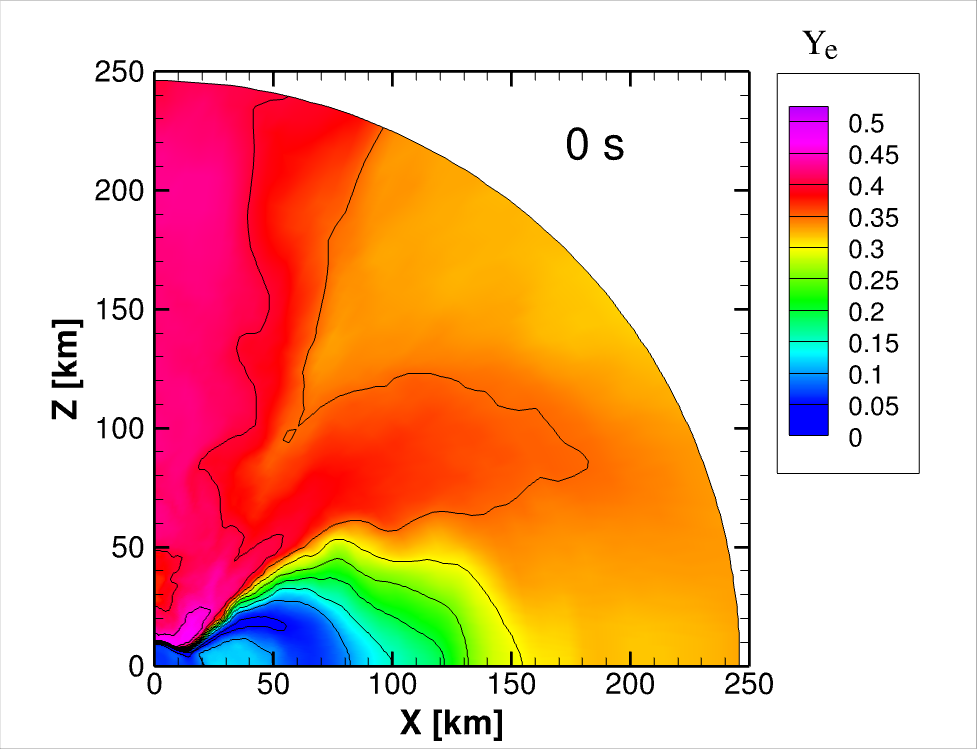}
\includegraphics[width=0.32\linewidth]{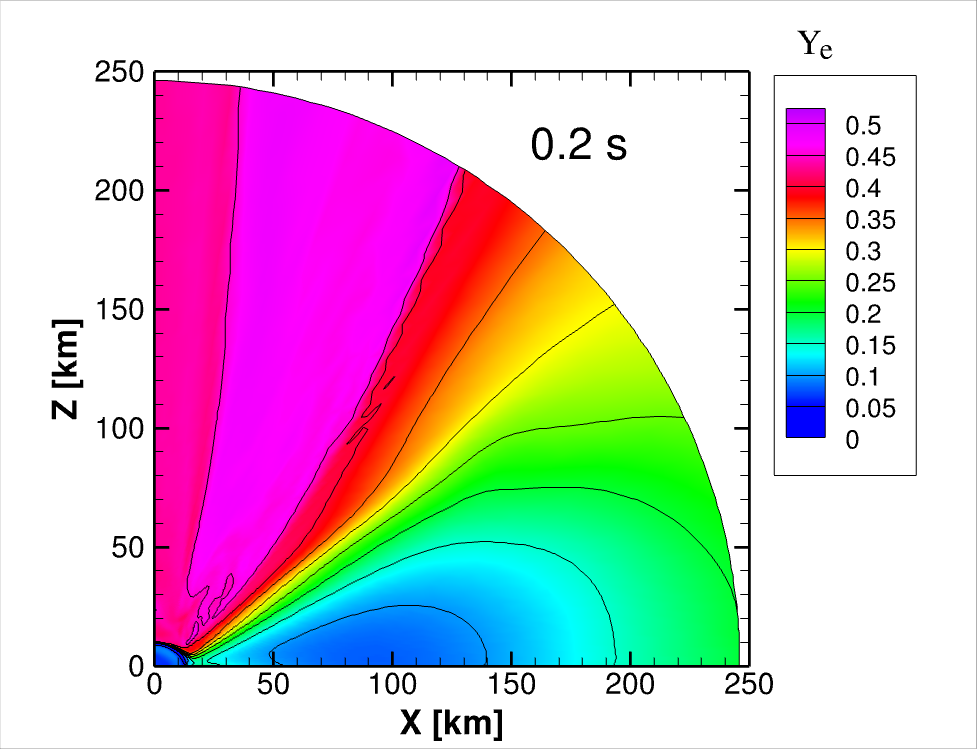}
\includegraphics[width=0.32\linewidth]{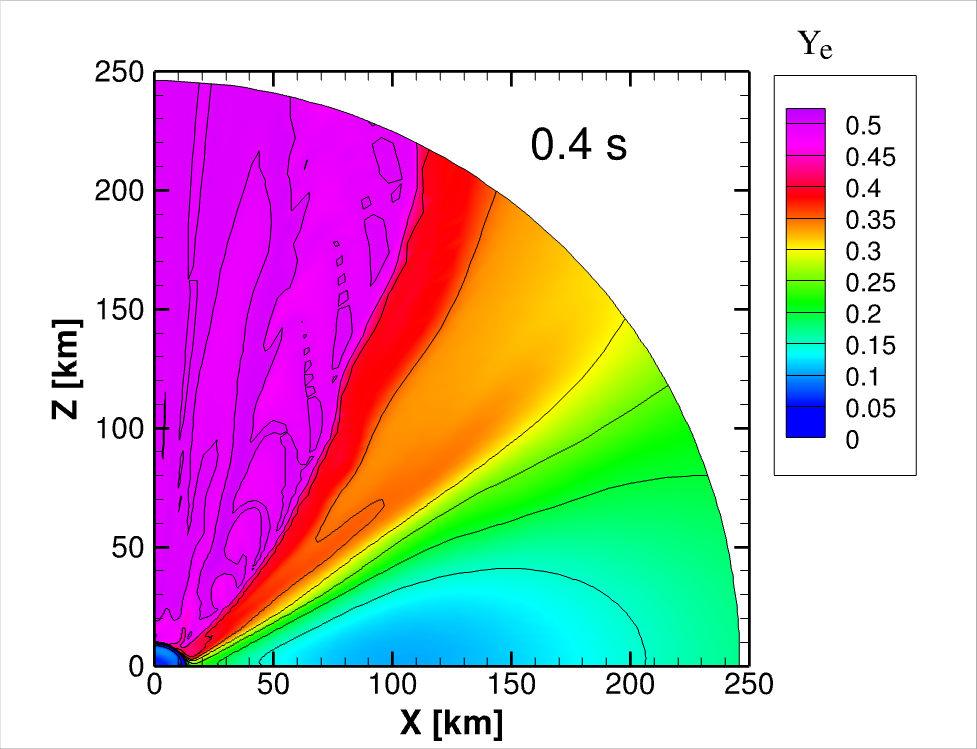}
\includegraphics[width=0.32\linewidth]{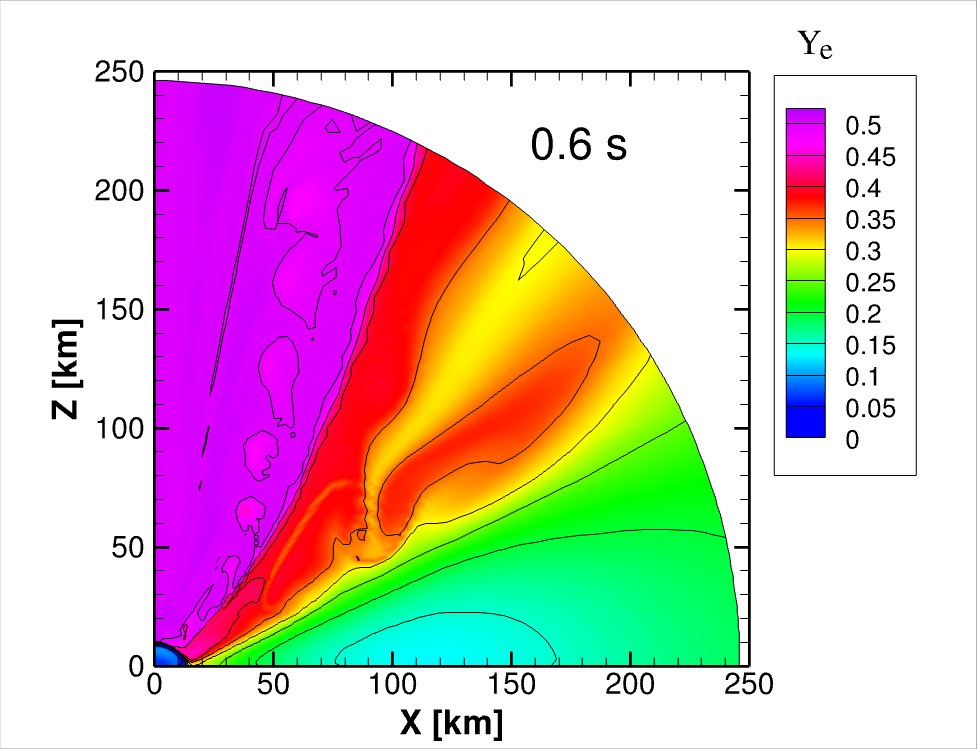}
\includegraphics[width=0.32\linewidth]{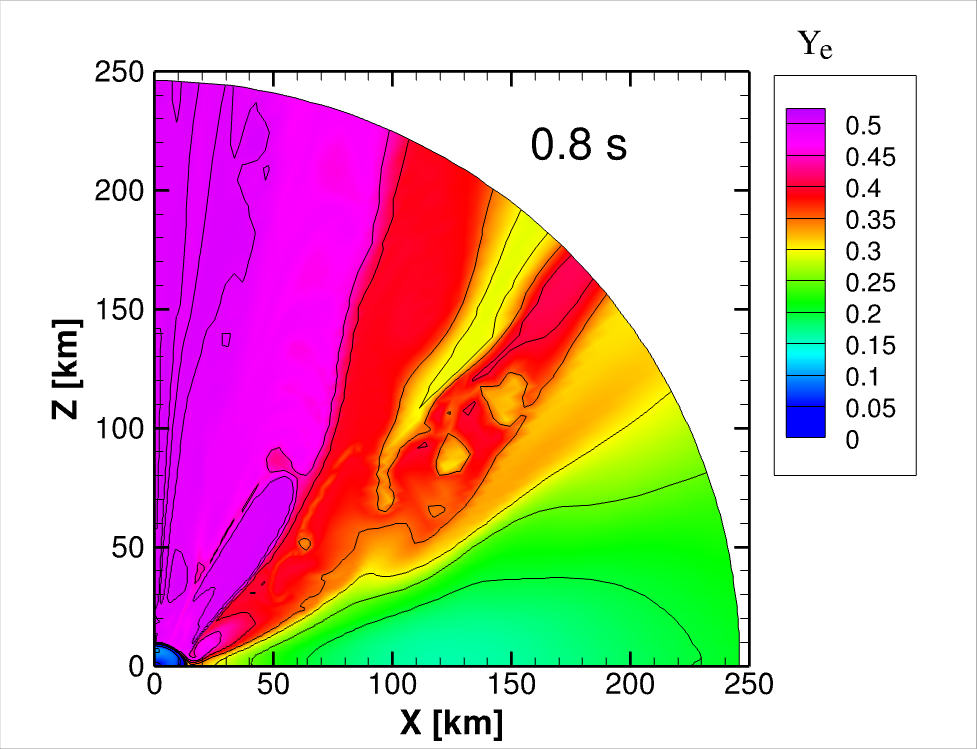}
\includegraphics[width=0.32\linewidth]{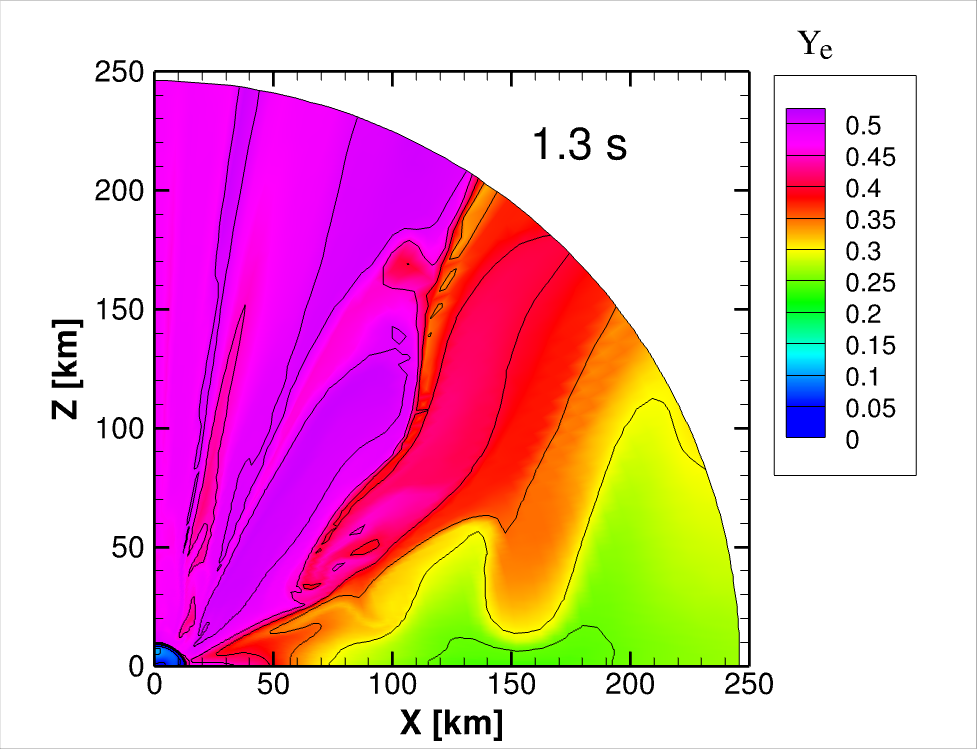}
\end{minipage}
\caption{Same as Fig.~\ref{fig:nuechemipote} but for electron fraction ($Y_e$).}
\label{fig:Ye}
\end{figure*}

In this study, we employ a long-term post-merger BNSM model developed through a hybrid approach of numerical relativity simulations. Assuming equal-mass ($1.35 M_{\odot}$) binary neutron stars, the late-inspiral, merger, and early post-merger dynamics (up to $\sim 50\,$ms after merger) were modeled by three dimensional numerical relativity simulations \cite{2015PhRvD..91f4059S}. To efficiently capture the longer-term evolution beyond this point (lasting over 1 second), we switched them to axisymmetric viscous-radiation hydrodynamic simulations \cite{2020ApJ...901..122F}. In this study, we employ a model with DD2 equation-of-state (EOS) \cite{2014ApJS..214...22B} corresponding to one of the stiff EOSs among those used in our simulations (DD2-135M model in \cite{2020ApJ...901..122F}). The HMNS in this model does not enter runaway gravitational collapse, at least not until the end of our simulation ($\sim 1.3\,$s).

In the simulation, neutrino transport is computed based on an energy-integrated formalism as described in \cite{2010PThPh.124..331S,2017ApJ...846..114F}. To compute neutrino number density and neutrino-matter interactions (or collision term) within gray neutrino transport schemes, the average energy of neutrinos is estimated by using a prescription in \cite{2017ApJ...846..114F}. We incorporate a set of weak interactions in BNSM simulations: electron and positron capture of both nucleons and heavy nuclei, electron–positron pair, nucleon–nucleon bremsstrahlung, and plasmon decay; see \cite{1985ApJ...293....1F,1986ApJ...309..653C,2006NuPhA.777..356B,1996A&A...311..532R} for details of each weak process.

We note that the present study does not utilize neutrino data directly from the numerical relativity simulation. Instead, we run Boltzmann neutrino transport simulations on top of fixed fluid backgrounds at selected time snapshots ($0, 0.2, 0.4, 0.6, 0.8,$ and $1.3\,$s) taken from the viscous-radiation hydrodynamic simulation. These simulations are run for $\sim 3\,$ms, which is roughly three times longer than the light crossing time of the computational domain ($r \sim 250\,$km). We confirm that the neutrino radiation field reaches nearly a quasi-steady state. It should be cautioned that our analysis may overlook ELN angular crossings caused by short-time ($\lesssim 1\,$ms) temporal variations in the vicinity of HMNS. Although addressing this issue is beyond the scope of this paper, we plan to extend the present work for more self-consistent treatments with time-dependent Boltzmann neutrino transport (see, e.g., \cite{2025PhRvD.111b3015K}); the results will be reported in a separate paper.

One might think that this post-processing approach is in some sense redundant, but it is essential for accurately capturing angular crossings, which require multi-angle treatment in neutrino transport. In Sec.~\ref{subsec:Boltzmann}, we describe some essential information on our numerical scheme for solving Boltzmann equation. It is worth noting that, by avoiding any uncertainties in neutrino angular distributions, we can thoroughly investigate mechanisms for generating ELN angular crossings. These detailed inspections would be difficult for other surrogate prescriptions of crossing search \cite{2018PhRvD..98j3001D,2020JCAP...05..027A,2021PhRvD.104f3014N,2021PhRvD.103l3012J,2022PhRvD.106h3005R}.

In Appendix~\ref{appendix:matterdist}, we provide time-dependent matter profiles obtained from the numerical relativity simulation, but we refer readers to \cite{2018ApJ...860...64F} for details of the model. Here, we highlight two important physical quantities, that are closely related to FFI and CFI. One of them is the chemical potential of $\nu_e$ ($\mu_{\nu}^{\rm eq}$) that can be obtained with beta equilibrium condition ($\mu_{\nu}^{\rm eq} \equiv \mu_{e} + \mu_{p} - \mu_{n}$ where $\mu_{e}$, $\mu_{p}$, and $\mu_{n}$ denote the chemical potential of electron, proton, and neutron, respectively), which is displayed in Fig.~\ref{fig:nuechemipote}. As we shall discuss in Sec.~\ref{subsec:mechaELNgene}, $\nu_e$ emission in regions with $\mu_{\nu}^{\rm eq}>0$ plays pivotal roles on generating ELN angular crossings. The other quantity is electron-fraction ($Y_e$), displayed in Fig.~\ref{fig:Ye}. As we shall see in Sec.~\ref{subsec:CFIanalysis}, it offers a useful reference for analysis of CFI, as it reflects a disparity in charged-current reactions between $\nu_e$ and $\bar{\nu}_e$.

\subsection{Boltzmann neutrino transport}\label{subsec:Boltzmann}
The computational setup and methodology for our Boltzmann neutrino transport simulations follow those described in \cite{2021ApJ...907...92S}. Here, we briefly summarize the key elements and refer interested readers to our previous papers \cite{2012ApJS..199...17S,2021ApJ...907...92S} for a comprehensive explanation. We solve Boltzmann equation for three species of neutrinos: $\nu_e$, $\bar{\nu}_e$, and $\nu_x$, where $\mu$ and $\tau$ neutrinos, and their antipartners are collectively treated as $\nu_x$ in our simulations. In this framework, XLN is always zero, indicating that ELN angular crossings provide a necessary and sufficient condition for occurrences of FFI. On the other hand, $\nu_x$ can have an influence on triggering CFI, even if XLN is zero. We will describe them explicitly in Sec.~\ref{subsec:flavorinstaana}.

We employ a discrete ordinate (${\rm S_n}$) method to solve Boltzmann equation in flat spacetimes\footnote{Neglecting general relativistic effects, in particular light-bending effects, causes a systematic error in locating regions occurring ELN angular crossings, because neutrino angular distributions in momentum space hinges on global neutrino transport. As we shall show below, however, ELN angular crossings in optically thick regions, corresponding to strong curvature regimes nearby HMNS, are determined by local properties of matter rather than global ones, whereas general relativistic effects are less important in optically thin region. This suggests that the flat spacetime condition would be reasonable to study qualitative trends in occurrences of ELN angular crossings.} under multi-angle, multi-energy, and multi-species treatments \cite{2012ApJS..199...17S}. The code was originally designed for the study of CCSN, but it has recently been extended to study neutrino radiation field in BNSM remnants \cite{2021ApJ...907...92S}. Thermodynamical quantities and nuclear abundances, which are necessary information on neutrino matter interactions, are determined with DD2 EOS to maintain consistency with numerical relativity simulations. In the collision term, the majority of weak reaction rates are computed based on the formulae in \cite{1985ApJS...58..771B}, while an extended rate for nucleon–nucleon bremsstrahlung \cite{2005ApJ...629..922S} is also adopted. These are the same as those used in our previous study in BNSM remnant \cite{2021ApJ...907...92S}.


It is important to note that the Boltzmann scheme is different from a neutrino transport module implemented in a CCSN neutrino-radiation hydro code \cite{2014ApJS..214...16N,2017ApJS..229...42N,2019ApJ...878..160N,2023ApJ...944...60A}. More specifically, this scheme does not implement the two-energy grid technique \cite{2014ApJS..214...16N}, updated neutrino-matter interactions \cite{2019ApJS..240...38N}, and general relativistic effects \cite{2021ApJ...909..210A}. Although these improvements have not yet been incorporated, the scheme remains suited to capture qualitative trends of neutrino radiation fields in BNSM remnants. We note that the code capability for simulating neutrino transport in BNSM remnant systems has been well tested in \cite{2021ApJ...907...92S}.


Assuming axial- and equatorial symmetries in space, we discretize the spatial domain from $0$ to $250\,$km in radius ($r$) on a spherical polar grid, using $512$ grid points. For the zenith angle ($\theta$), we cover the domain of $0 \le \theta \le \pi/2$ by $96$ grid points. We note that the resolution is higher than our previous study \cite{2021ApJ...907...92S}. This is because steep density gradients appear near the surface of HMNS in the late phase, which is mainly due to neutrino cooling. These gradients cause rapid changes in neutrino opacities, especially in regions where neutrinos transition from optically thick to thin medium. To address this, we increase the number of grid points by following a technique similar to that used in our CCSN simulations \cite{2018ApJ...854..136N}. The details of the grid structure and our resolution study are found in Appendix~\ref{appendix:resodepe}. In neutrino momentum space, the neutrino energy ($\epsilon_\nu$) is discritized with 14 grid points in the range of $0\,$MeV $< \epsilon_\nu < 300\,$MeV, and the full solid angle is covered by two angles ($\theta_{\nu}$ for zenith angle and $\phi_{\nu}$ for azimuthal one\footnote{The azimuthal angle in neutrino momentum space is measured from $\boldsymbol{e}_{\theta}$, which corresponds to the spatial coordinate basis of $\theta$; see Fig. 1 in \cite{2012ApJS..199...17S}.}) with 24 ($0 < \theta_{\nu} < \pi $), and 12 ($0 < \phi_{\nu} < 2 \pi $) grid points. For more detail of grid structure in neutrino momentum space, we refer readers to \cite{2012ApJS..199...17S}.


\subsection{Approximate assessments of FFI and CFI}\label{subsec:flavorinstaana}

\begin{figure*}[ht]
\begin{minipage}{1.0\textwidth}
\centering
\includegraphics[width=0.32\linewidth]{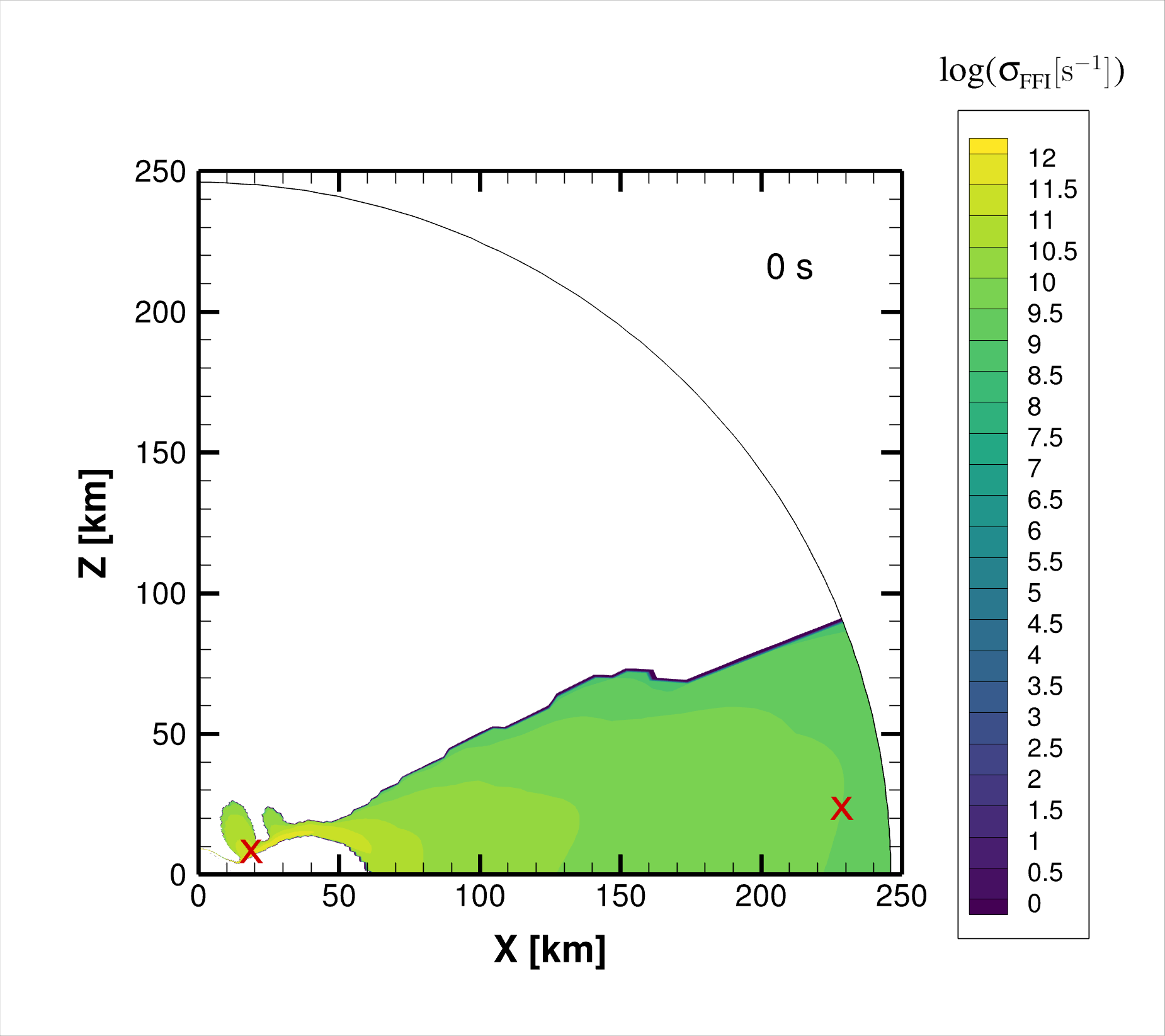}
\includegraphics[width=0.32\linewidth]{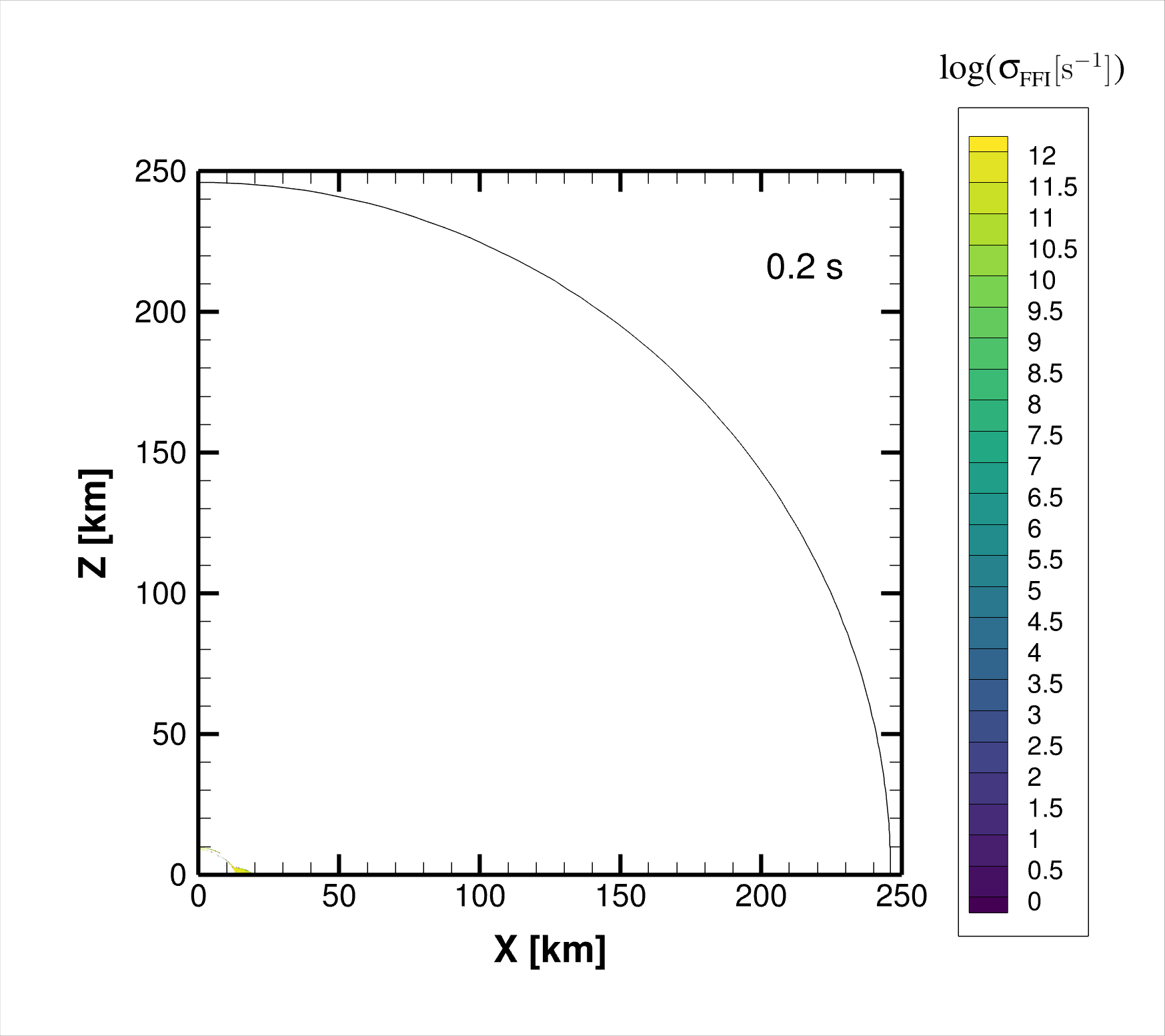}
\includegraphics[width=0.32\linewidth]{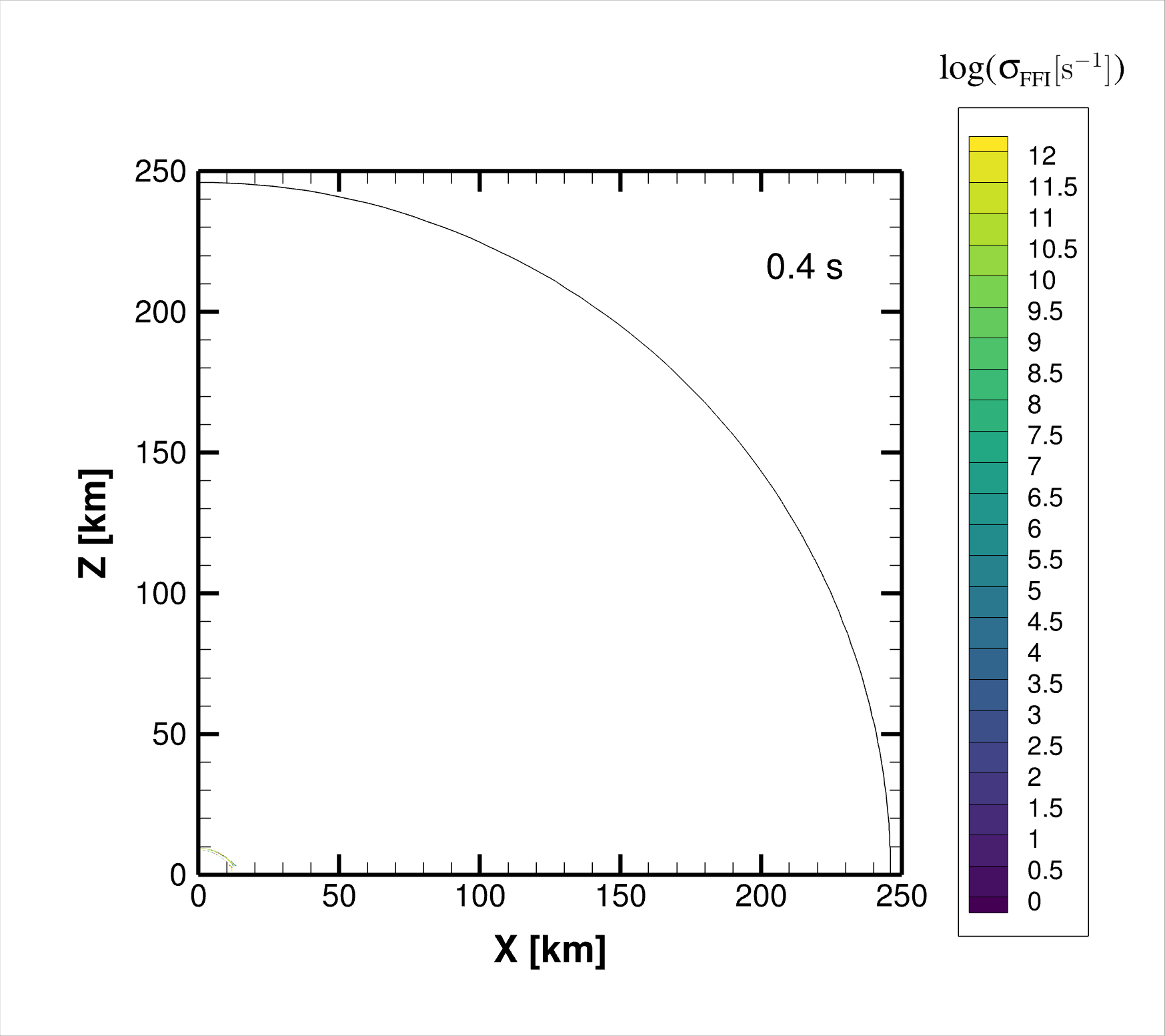}
\includegraphics[width=0.32\linewidth]{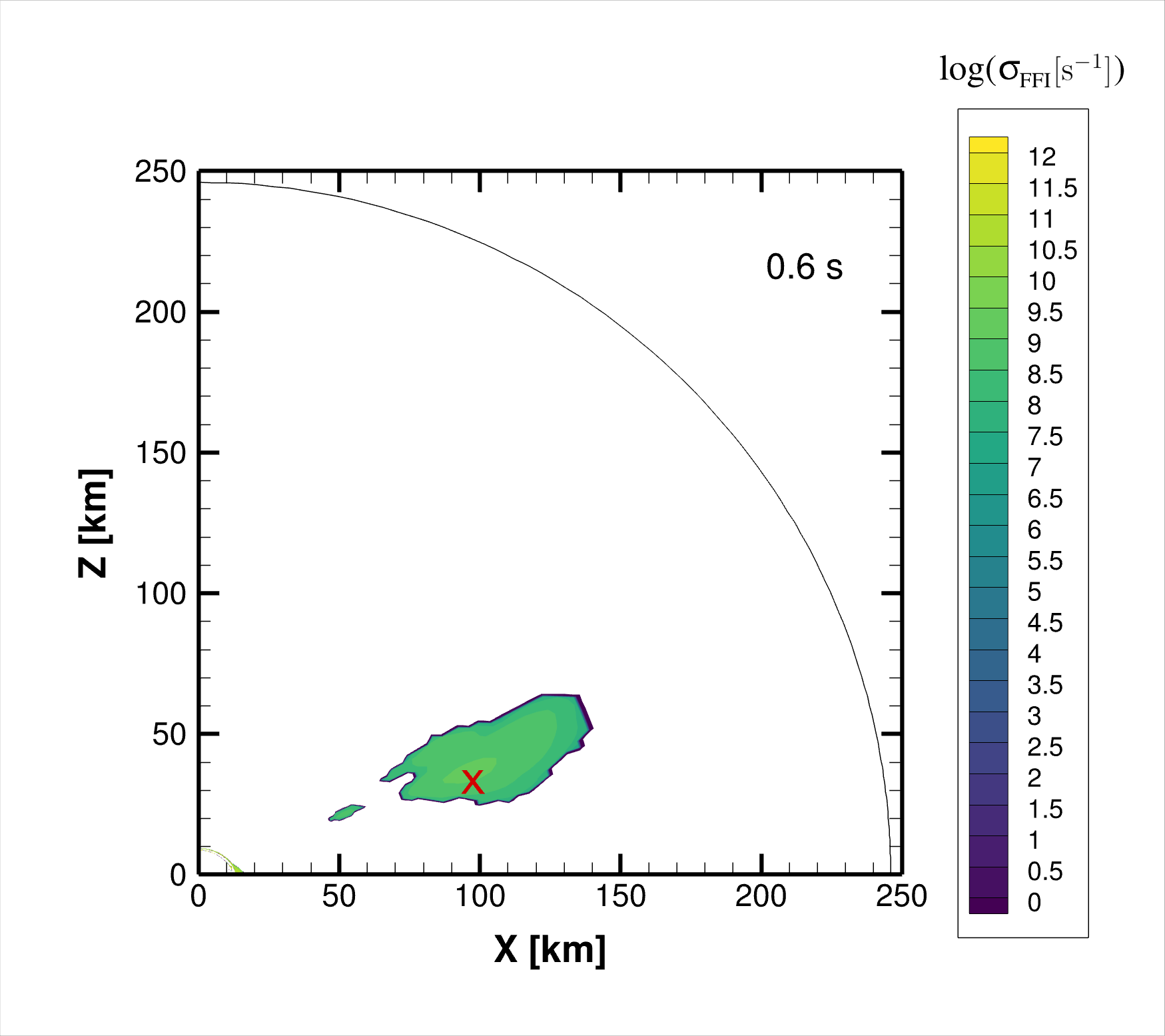}
\includegraphics[width=0.32\linewidth]{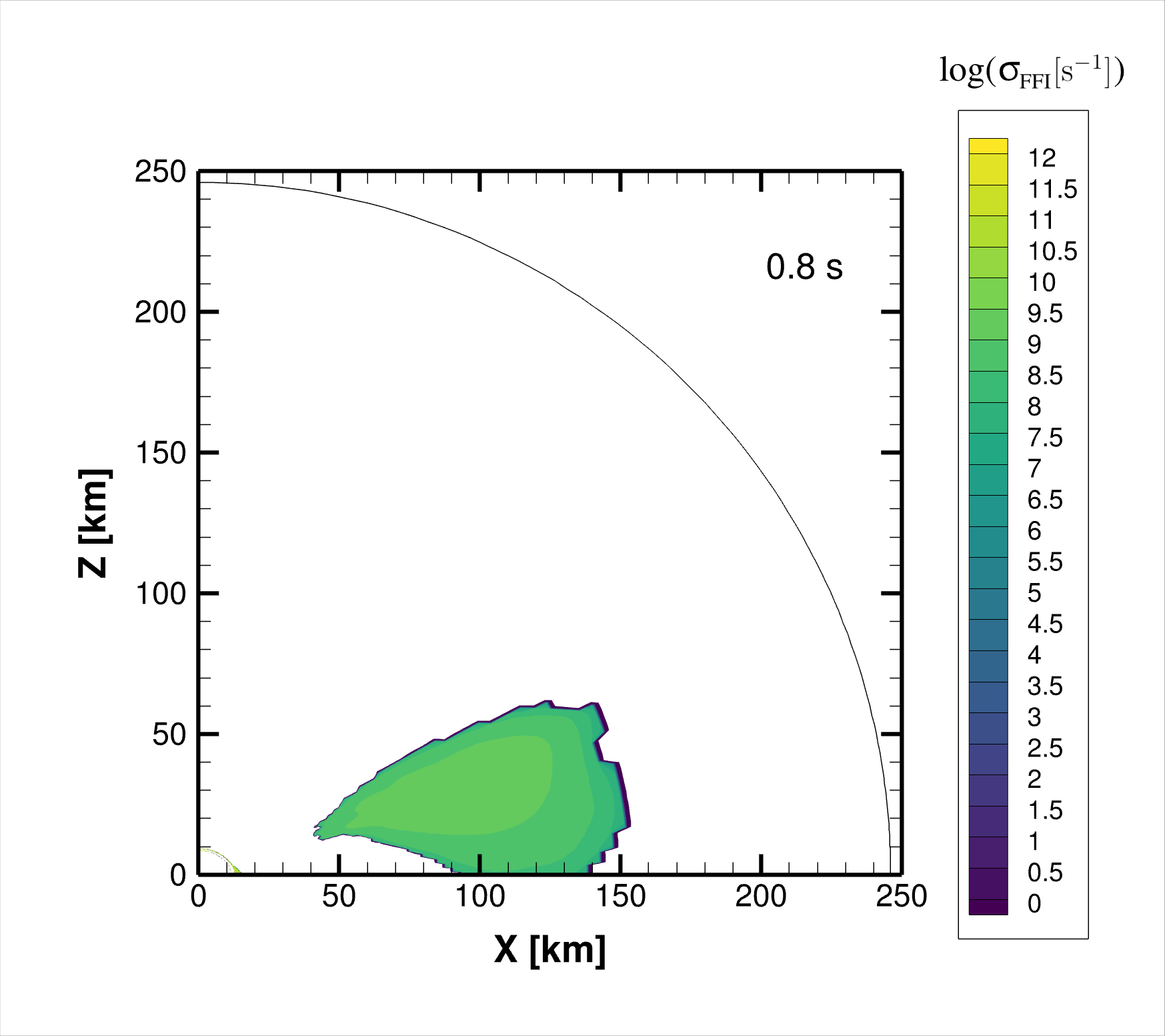}
\includegraphics[width=0.32\linewidth]{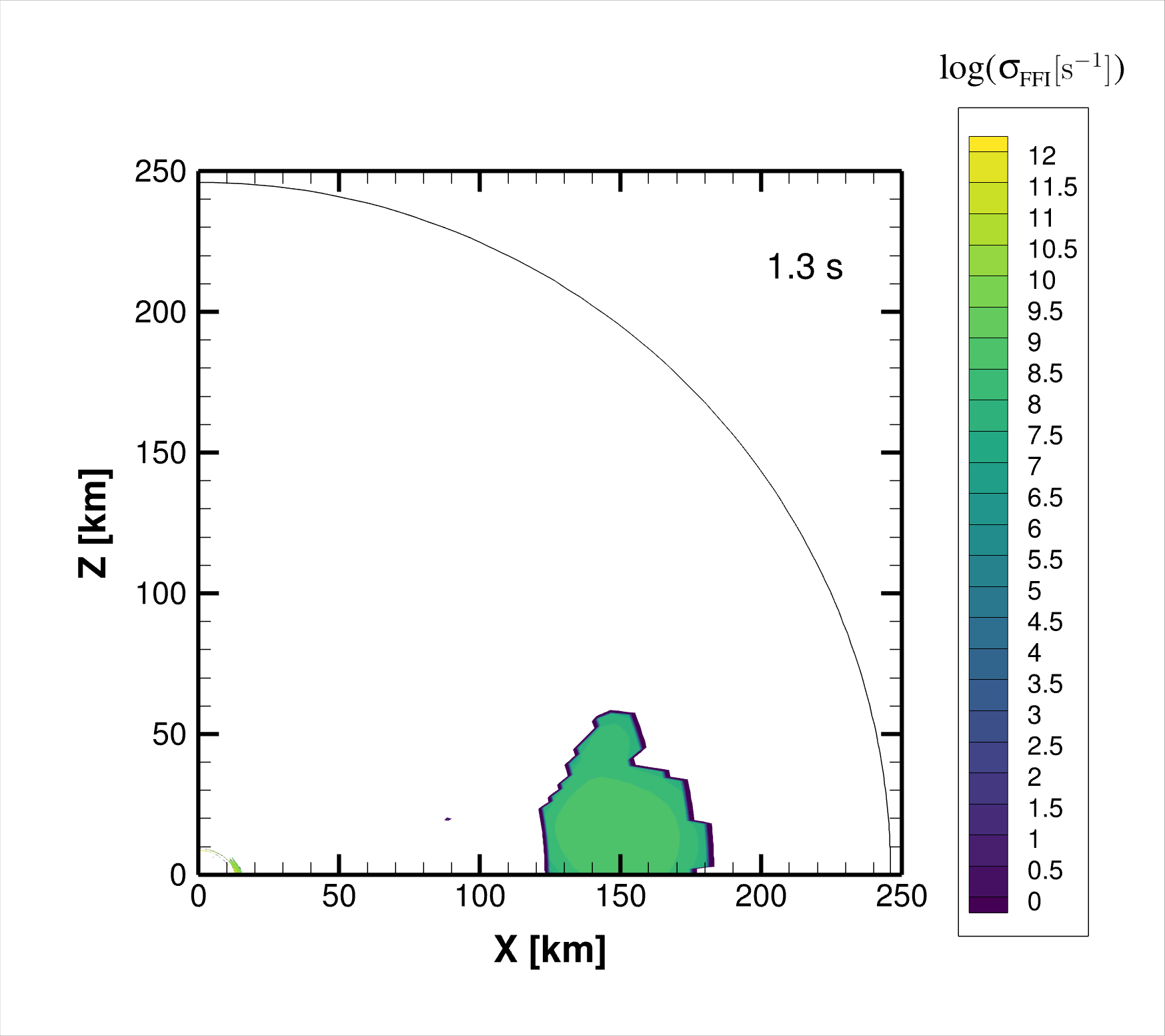}
\end{minipage}
\caption{Color maps of growth rates of FFI in the unit of $[\mathrm{s}^{-1}]$. Each panel represents a different time snapshot. Red cross marks in panels of $t=0\,$s and $0.6\,$s represent spatial positions where ELN angular distributions in neutrino momentum space are analyzed to validate our interpretation of the mechanisms for crossing generation; see the text for more details.}
\label{fig:FFIgrowthdistri}
\end{figure*}

\begin{figure*}[ht]
\begin{minipage}{1.0\textwidth}
\centering
\includegraphics[width=0.32\linewidth]{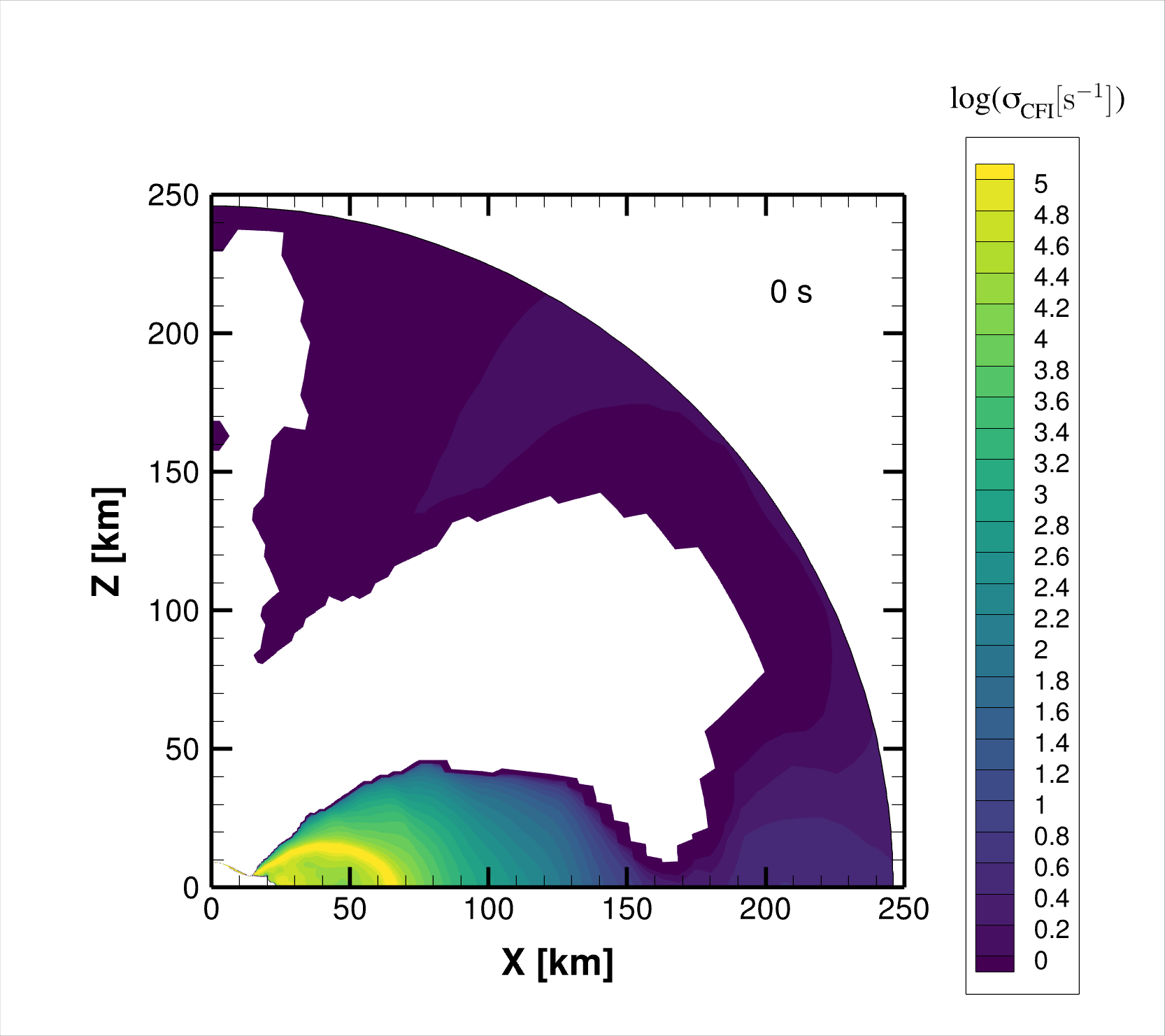}
\includegraphics[width=0.32\linewidth]{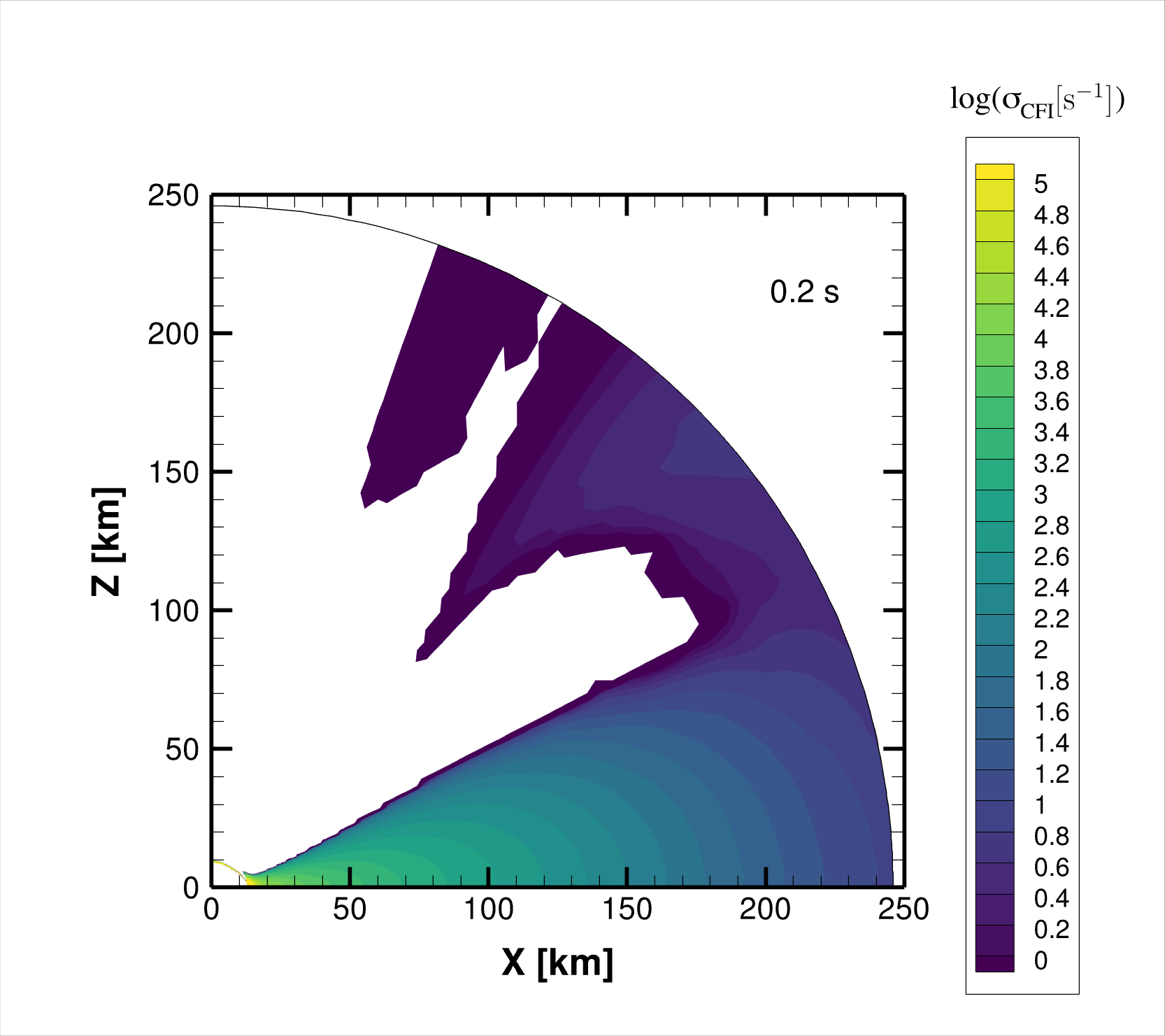}
\includegraphics[width=0.32\linewidth]{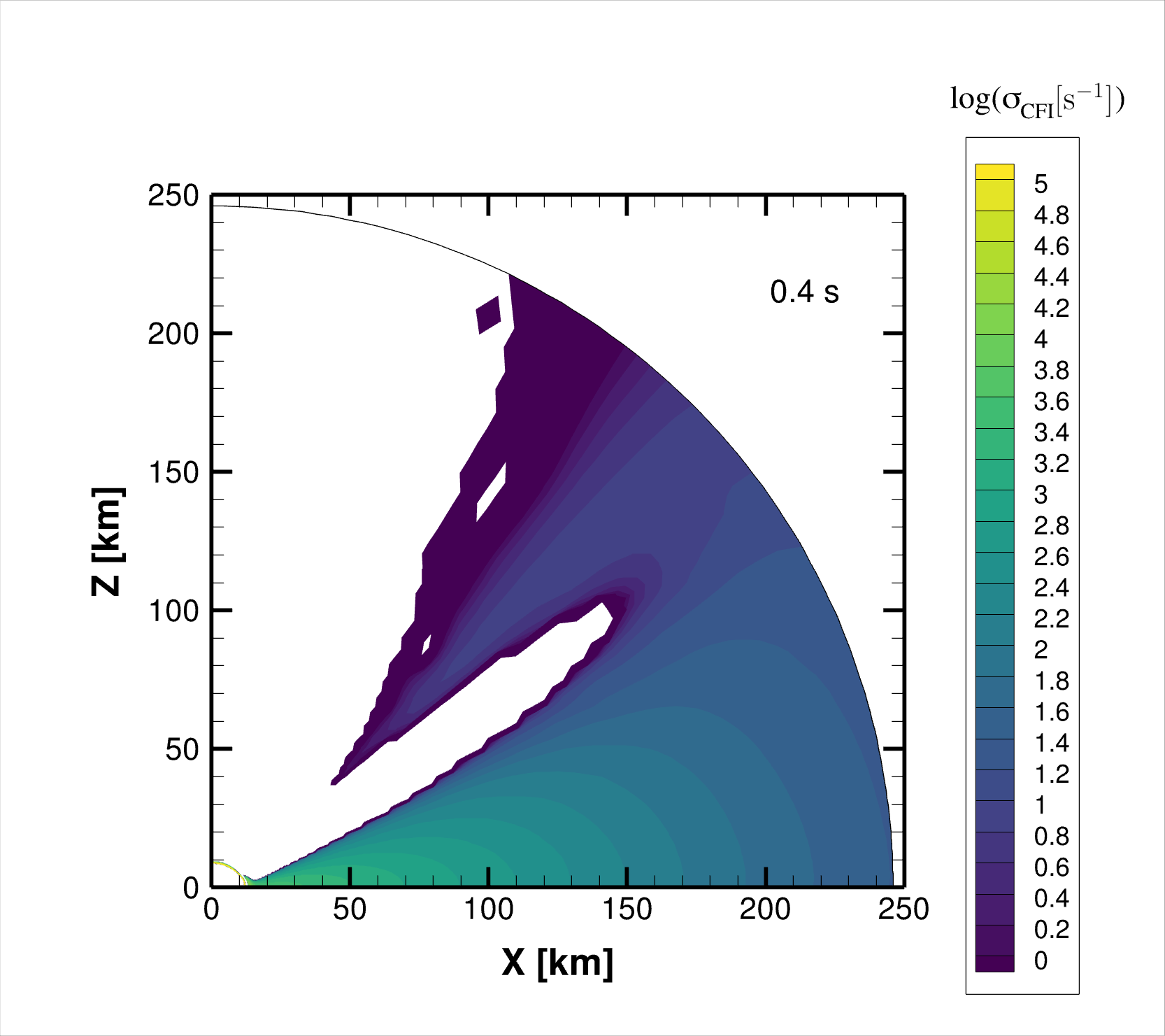}
\includegraphics[width=0.32\linewidth]{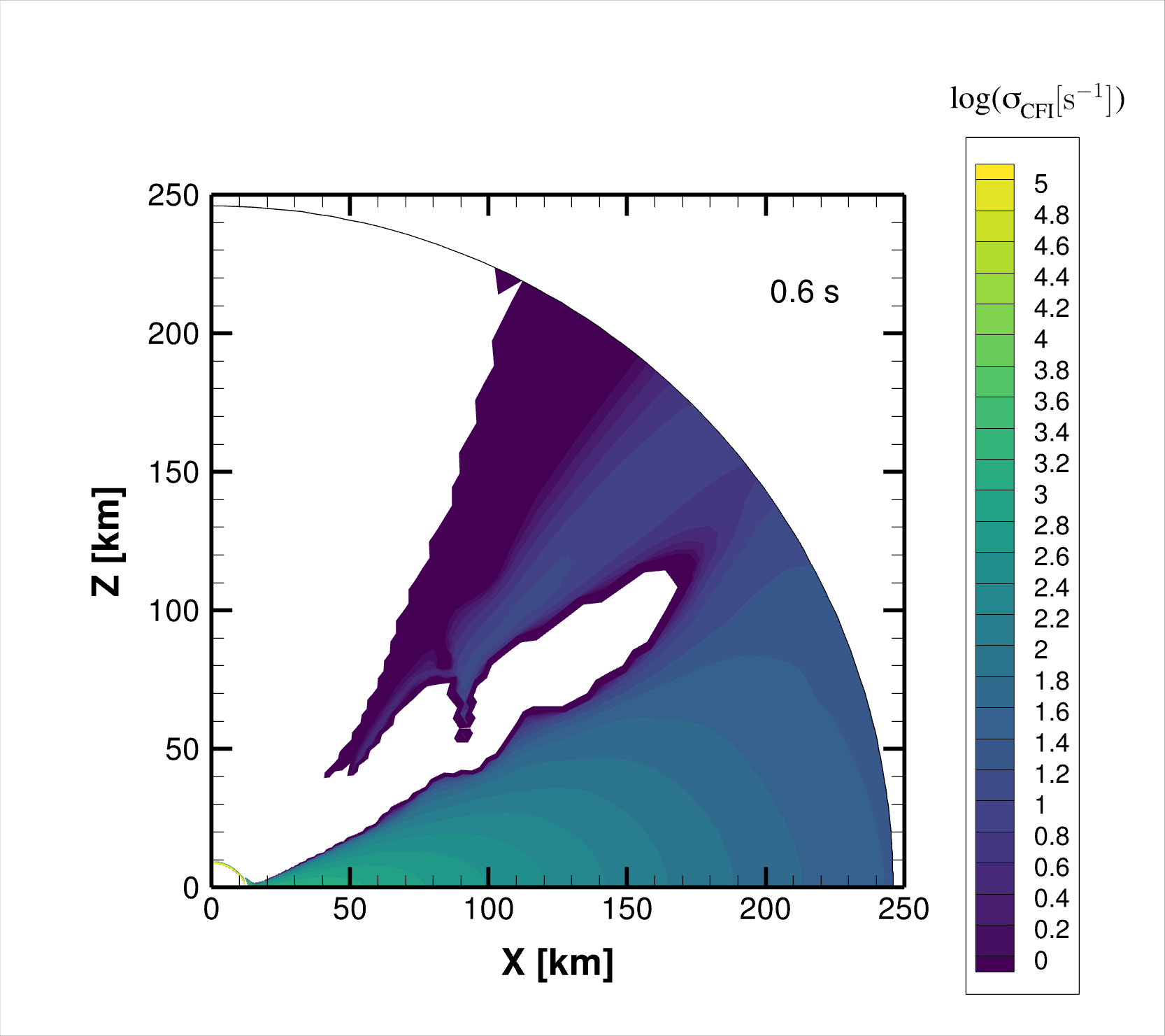}
\includegraphics[width=0.32\linewidth]{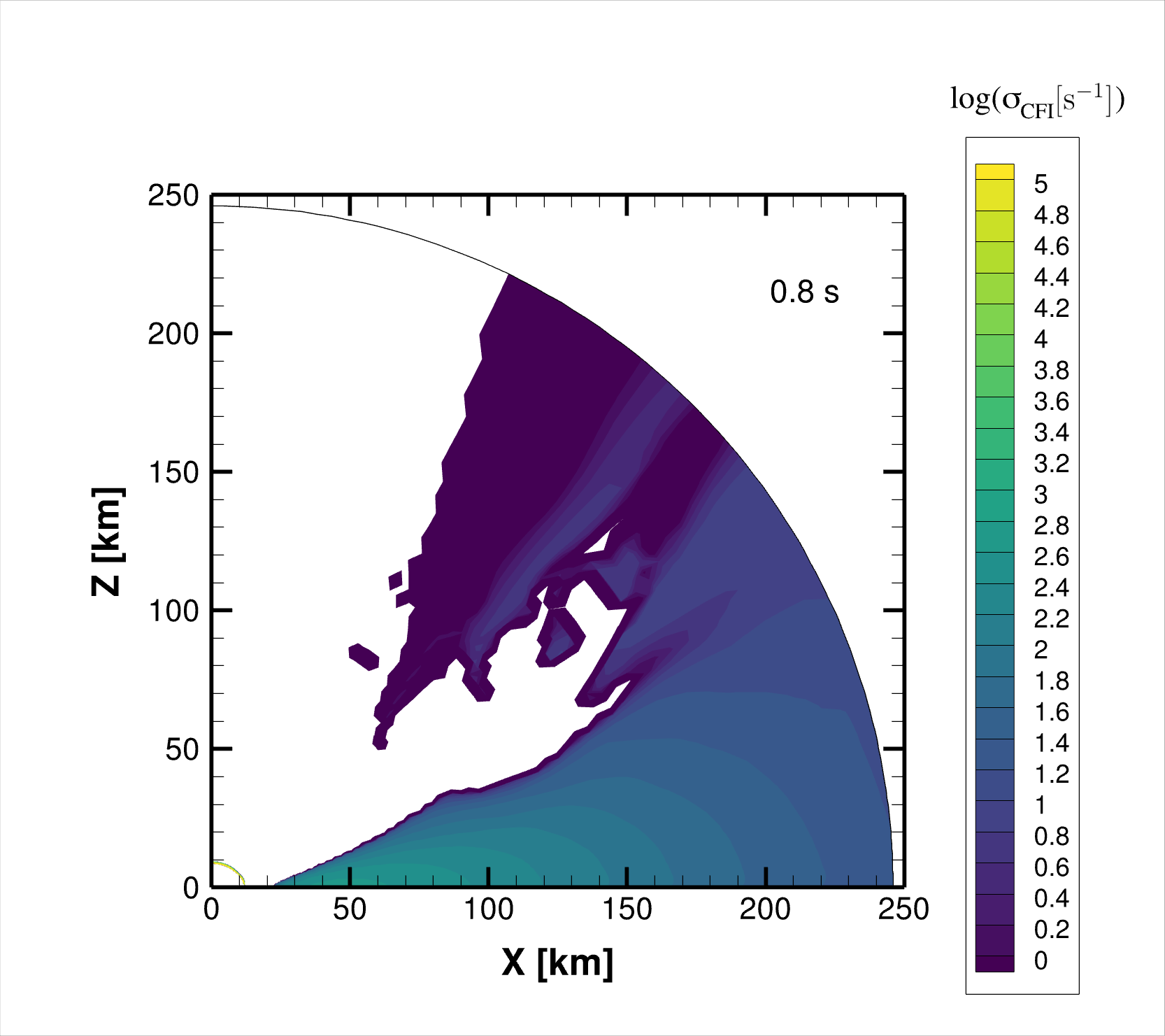}
\includegraphics[width=0.32\linewidth]{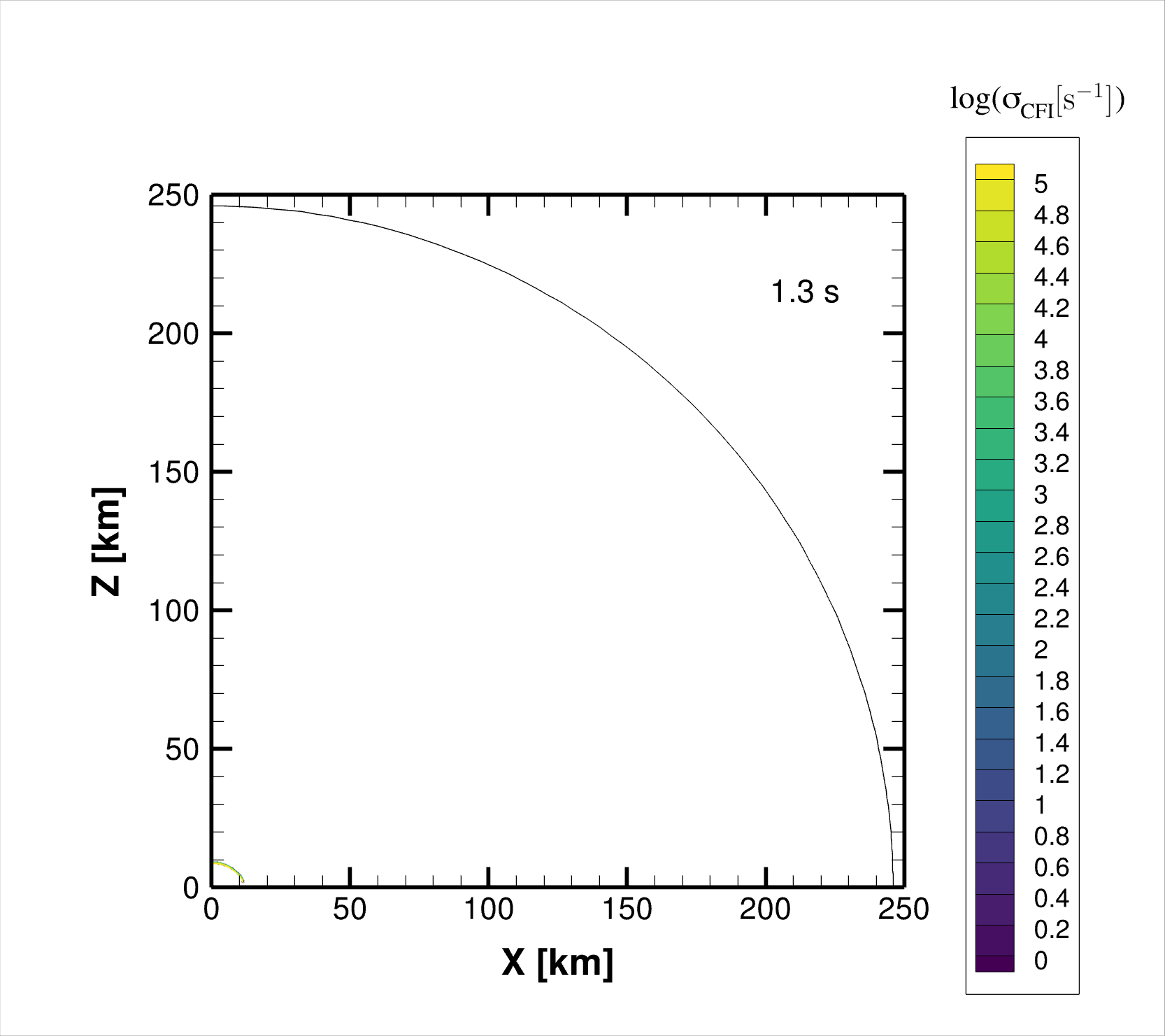}
\end{minipage}
\caption{Same as Fig.~\ref{fig:FFIgrowthdistri} but for CFI. We note that the color range is narrower than that in Fig.~\ref{fig:FFIgrowthdistri}.}
\label{fig:CFIgrowthdistri}
\end{figure*}

Neutrino flavor instabilities can be studied through a stability analysis of quantum kinetic neutrino transport. Assuming neutrinos are initially close to flavor eigenstates, one can identify stable and unstable modes, corresponding to decaying and growing flavor coherence (or off-diagonal elements of density matrix of neutrinos), respectively, by solving a dispersion relation derived from a linearized quantum kinetic equation (QKE); see, e.g., \cite{2017PhRvL.118b1101I,2020JCAP...01..005C,2023PhRvD.107h3034L,2023PhRvD.107l3011L}. FFI corresponds to an instability mode with no vacuum (or neutrino mass) contribution, and their eigen value (or growth rate) and eigen vector can be determined from energy-integrated neutrino angular distributions. In contrast, CFI is driven by the interplay between the neutrino self-interactions and off-diagonal elements of collision term, and the dispersion relation can be found in \cite{2023PhRvD.107l3011L,2025arXiv250209260Z}. Notably, both flavor instabilities can occur regardless of the neutrino mass hierarchy.

Although it is straightforward to solve the neutrino dispersion relation, we do not adopt the approach in the present study for practical reasons. As noted in previous works (see, e.g., \cite{2018PhRvD..97b3024M}), spurious modes of flavor instabilities are involved when we solve the dispersion relation numerically. This implies that careful inspections are necessary to distinguish physical modes from artifacts, which place a significant burden on our systematic survey of flavor instabilities. In addition to this, identifying spurious modes typically demands very high resolution analyses in neutrino momentum space, exhibiting that this approach is not suited for the present study. Importantly, the primary purpose of this study is not to precisely quantify growth rates and eigen vectors of flavor instabilities, but rather to gain insights into the overall features of occurrences of flavor instabilities and their underlying physical mechanisms.

For these purposes, we employ approximate methods for identifying FFI and CFI, following the procedure used in \cite{2024PhRvD.109b3012A}. The growth rate of FFI is approximately estimated by an empirical formula in \cite{2019ApJ...886..139N,2020PhRvR...2a2046M}. This can be given as,
\begin{equation}
\sigma_\mathrm{FFI} = \sqrt{-\left(\int_{\Delta G_v>0}\frac{d\Omega_p}{4\pi}\Delta G_v \right)\left(\int_{\Delta G_v<0}\frac{d\Omega_p}{4\pi}\Delta G_v\right)},
\label{eq:FFIgrowthrate}
\end{equation}
where $\Omega_p$ denotes the solid angle in neutrino momentum space. In the expression, $\Delta G_v$ is defined as 
\begin{equation}
\Delta G_v \equiv \sqrt{2}G_F \int \frac{\epsilon_\nu^2 d\epsilon_\nu}{2\pi^2}\left(f_{\nu_e}-f_{\bar{\nu}_e}\right),
\label{eq:def_G_v}
\end{equation}
where $G_F$ denotes Fermi constant and $f_{i}$ represents a distribution function of $"i"$-species neutrinos. 

It is important to note that the absence of ELN angular crossings leads to $\sigma_\mathrm{FFI}\le0$ according to the formal stability analysis \cite{2022PhRvD.105j1301M}. However, the empirical formula yields $\sigma_\mathrm{FFI}=0$ in such cases, as can be seen from Eq.~(\ref{eq:FFIgrowthrate}); the lack of crossings makes either the first parenthesis or second one inside the square root in the right hand side of Eq.~(\ref{eq:FFIgrowthrate}) be zero. We, hence, keep in mind that Eq.~(\ref{eq:FFIgrowthrate}) serves only for an approximate estimate for the growth rate of FFI but not for decaying rates of stable modes.

Following \cite{2024PhRvD.109b3012A}, we approximately determine the growth rate of CFI as,
\begin{equation}
\sigma_\mathrm{CFI} = \mathrm{max}\left(\mathrm{Im}(\omega_\pm^\mathrm{pres}),\mathrm{Im}(\omega_\pm^\mathrm{break})\right),
\label{eq:CFIgrowthrate}
\end{equation}
where
\begin{equation}
\text{max\,(Im\,}\omega_\pm^\mathrm{pres})=\begin{cases}
    -\gamma+\frac{|G\alpha|}{|A|},& \quad (A^2\ge |G\alpha|),\\
    -\gamma+\sqrt{|G\alpha|},& \quad (A^2< |G\alpha|),
\end{cases}
\label{eq:CFIgrowthrate_isopre}
\end{equation}
for the isotropy-preserving branch and
\begin{equation}
\text{max\,(Im\,}\omega_\pm^\mathrm{break})=\begin{cases}
    -\gamma+\frac{|G\alpha|}{|A|},&\quad (A^2\ge |G\alpha|),\\
    -\gamma+\frac{\sqrt{|G\alpha|}}{\sqrt{3}},& \quad (A^2< |G\alpha|),
\end{cases}
\label{eq:CFIgrowthrate_isobreak}
\end{equation}
for the isotropy-breaking branch. In the expression, $G$, $A$, $\gamma$, and $\alpha$ are defined as
\begin{equation}
G\equiv\frac{\mathfrak{g}+\bar{\mathfrak{g}}}{2},\; 
A\equiv\frac{\mathfrak{g}-\bar{\mathfrak{g}}}{2},\;
\gamma\equiv\frac{\Gamma+\bar{\Gamma}}{2},\;
\alpha\equiv\frac{\Gamma-\bar{\Gamma}}{2},
\label{eq_G_A}
\end{equation}
with
\begin{equation}
\mathfrak{g}\equiv \sqrt{2}G_F (n_{\nu_e}-n_{\nu_x}),\;
\bar{\mathfrak{g}}\equiv \sqrt{2}G_F (n_{\bar{\nu}_e}-n_{\bar{\nu}_x}),
\label{eq:gdef}
\end{equation}
and
\begin{equation}
\Gamma\equiv\frac{\Gamma_{\nu_e}+\Gamma_{\nu_x}}{2},\;
\bar\Gamma\equiv\frac{\Gamma_{\bar{\nu}_e}+\Gamma_{\bar{\nu}_x}}{2}.
\label{eq:gammadef}
\end{equation}
We note that $n_{\nu_i}$ and $\Gamma_i$ represent the number density of neutrinos and the energy-averaged collision rates, respectively. They can be computed as,
\begin{equation}
\label{eq:numberdensityofneutrinos}
n_i= \frac{1}{8 \pi^3} \int \epsilon_\nu^2 d\epsilon_\nu d\Omega_p f_i(\epsilon_\nu,\Omega_p),
\end{equation}
\begin{equation}
\label{eq:aveReacrate}
\Gamma_i \equiv \frac{1}{8 \pi^3 n_i} \int \epsilon_\nu^2 d\epsilon_\nu d\Omega_p \Gamma(\epsilon_\nu,\Omega_p)f_i(\epsilon_\nu,\Omega_p),
\end{equation}
where $\Gamma$ denotes the effective absorption rates\footnote{We note that there is a typo in Eq.~(20) of \cite{2024PhRvD.109b3012A}.} ($\Gamma \equiv j + \kappa$ with $j$ and $\kappa$ being emissivity and absorption coefficient, respectively). We note that, when $\sigma_\mathrm{CFI}$ is less than zero, all modes are decaying (i.e., stable). In this analysis, we take into account charged-current reactions for $\nu_e$ and $\bar{\nu}_e$ and pair processes for all species of neutrinos. On the other hand, we neglect neutral-current processes (scatterings with nucleons and heavy nuclei). This is because the neutrinos are nearly isotropic in optically thick regions where CFI is expected to play the most important role. For this reason, although ignoring contributions of scatterings would introduce a systematic error in searching for CFI, the qualitative trends we shall discuss in the following sections are expected to be robust. For details of these approximate formula, we refer readers to \cite{2023PhRvD.107l3011L}.

\section{Results}\label{sec:results}

\subsection{Global properties}\label{subsec:overall}

We begin by summarizing overall features of FFI and CFI based on our stability analysis. Figures~\ref{fig:FFIgrowthdistri}~and~\ref{fig:CFIgrowthdistri} show spatial color maps of growth rates of FFI and CFI, respectively. We confirm that both instabilities can arise in the BNSM remnant with the growth rate of FFI being much higher than that of CFI. It is worthy to note that there are no (or extremely narrow) regions where the growth rate of FFI becomes less than $\sim 10^7~\mathrm{s}^{-1}$. This is because the characteristic growth rate of FFI is governed by the ELN number density and the depth of crossing\footnote{Although there are no formal definitions of the depth of crossing, it can be estimated as,
\begin{equation}
1-\frac{|\left(\int_{\Delta G_v>0}\frac{d\Omega_p}{4\pi}\Delta G_v \right) - \left(\int_{\Delta G_v<0}\frac{d\Omega_p}{4\pi}\Delta |G_v| \right)|}{(\left(\int_{\Delta G_v>0}\frac{d\Omega_p}{4\pi}\Delta G_v \right) + \left(\int_{\Delta G_v<0}\frac{d\Omega_p}{4\pi}\Delta |G_v| \right))}.
\end{equation}
}. Assuming that the depth is higher than the subpercent level, the corresponding frequency should be above $10^7~\mathrm{s}^{-1}$ across relevant spatial regions and times. In contrast, the growth rate of CFI can be very low, which is due to low collisional frequencies (see also Eqs.~(\ref{eq:CFIgrowthrate_isopre})~and~(\ref{eq:CFIgrowthrate_isobreak})). These overall trends are consistent with previous studies (see, e.g., \cite{2023PhRvD.108h3002X,2024ApJ...974..110M}).

As illustrated in these figures, however, the spatial distributions of these instabilities are highly complex and exhibit significant time dependence. For instances, FFI is absent in the outer region of HMNS ($\gtrsim 20\,$km) during the time interval $0.2\,$s $\lesssim t \lesssim 0.4\,$s; no ELN angular crossings emerge around the pole (z-axis) at any of the time snapshots analyzed in this study; CFI becomes always unstable at the surface of HMNS, while the inner region is always stable; CFI arises in the disk region persistently but they are suppressed at $t \gtrsim 1.3\,$s. These results highlight the need for detailed analyses. In the following, we delve into them for FFI and CFI in Sec.~\ref{subsec:mechaELNgene}~and~\ref{subsec:CFIanalysis}, respectively. To facilitate readers' understanding, we simplify our argument by illustrating the key points graphically for complex mechanisms.

\subsection{Mechanisms of ELN angular crossing generations}\label{subsec:mechaELNgene}

\subsubsection{ELN crossings in optically thick regions}\label{subsubsec:alphadrivenmecha}

\begin{figure}
\begin{minipage}{0.45\textwidth}
\centering
\includegraphics[width=1.0\linewidth]{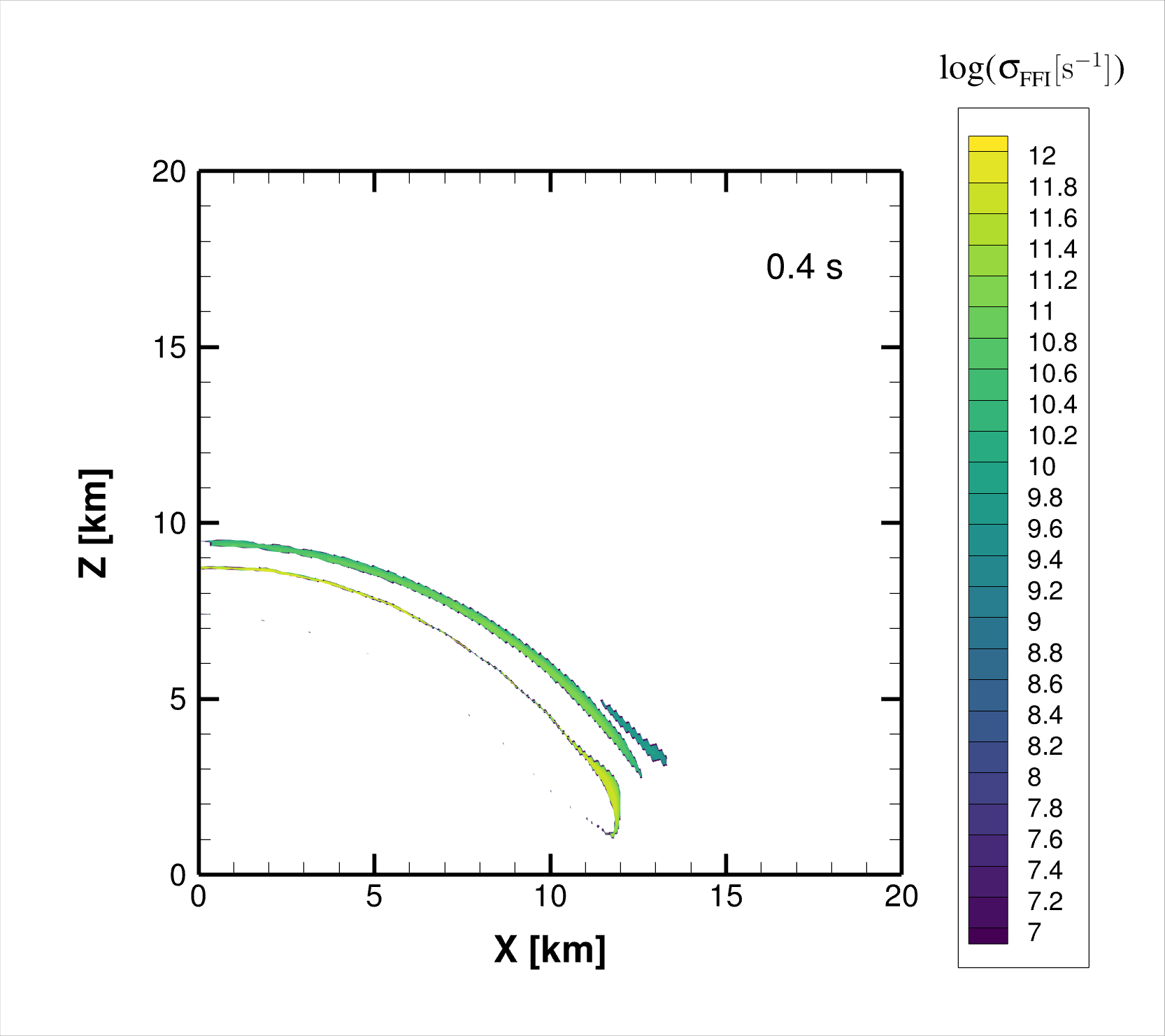}
\end{minipage}
\caption{Same as Fig.~\ref{fig:FFIgrowthdistri} (a color map of growth rates of FFI) but we zoom in the central region for the time snapshot of $t=0.4\,$s.}
\label{fig:nu-FFI-0.4s-c1}
\end{figure}

We find that ELN angular crossings persistently occur in a thin layer near the surface of HMNS; see Fig.~\ref{fig:nu-FFI-0.4s-c1} as an example. In these optically thick regions, neutrinos are trapped and reach in thermal equilibrium with surrounding medium, suggesting that neutrinos are isotropically distributed. Strictly speaking, however, they have a slight anisotropy in neutrino momentum space, and their angular distributions vary depending on neutrino species due to different opacities. This indicates that ELN angular crossings should arise in regions where $n_{\nu_e}$ and $n_{\bar{\nu}_e}$ are nearly equal. We note that the similar type of ELN angular crossings have already been reported in FFI analyses on CCSN systems. They appear in convective envelope of proto-neutron star \cite{2020PhRvD.101f3001G,2020PhRvD.101b3018D}.

Before moving on to discussions of other mechanisms, two important points should be noted. First, the spatial regions where these ELN angular crossings occur are extremely narrow \cite{2020PhRvD.101f3001G}, making them easy to overlook in Boltzmann simulation data. Indeed, Fig.~\ref{fig:nu-FFI-0.4s-c1} shows some spot-like structures of crossing regions in space. However, these are numerical artifacts due to limited spatial resolutions. The number of spots would increase with resolutions, and the crossing region is expected to converge into a loop-like morphology in the limit of infinite resolution. 

Second, one might guess that the impact of FFI originated from these ELN angular crossings is insignificant due to the small size of the spatial region, the shallow depth of the crossings in neutrino momentum space, and nearly equal number densities of all species of neutrinos. However, it is premature to make the judgement. While the depth of crossings does affect the growth rate of FFI (see also Eq.~(\ref{eq:FFIgrowthrate})), recent studies have shown that even shallow crossings can lead to strong fast flavor conversions (see Sec.~III.C in \cite{2023PhRvD.107j3022Z}). It is also important to note that, despite similar number densities, neutrino fluxes vary by species due to species-dependent opacities. As shown in previous studies \cite{2023PhRvL.130u1401N,2024PhRvD.109l3008X,2025arXiv250311758Q,2025arXiv250323727L}, FFC in optically thick regions enhances flavor-integrated fluxes, which facilitate neutrino diffusion and induce a secular cooling effect on fluid dynamics. In addition to this, effects of flavor conversions cannot be confined locally but rather spread by neutrino advections. In fact recent quantum kinetic simulations demonstrate that FFI in small spatial domain can substantially alter global neutrino radiation field \cite{2023PhRvL.130u1401N,2023PhRvD.108j3014N,2024PhRvD.109l3008X}. Although more self-consistent studies by neutrino-radiation hydrodynamics incorporating effects of flavor conversions are certainly needed to reach firm conclusion, the FFI might break chemical- and thermal equilibrium with matter, thereby affecting fluid dynamics, chemical compositions, and neutrino emission in BNSM remnant.

\subsubsection{ELN crossings induced by $\nu_e$ diffusion}\label{subsubsec:nuedif}

\begin{figure}
\begin{minipage}{0.45\textwidth}
\centering
\includegraphics[width=1.0\linewidth]{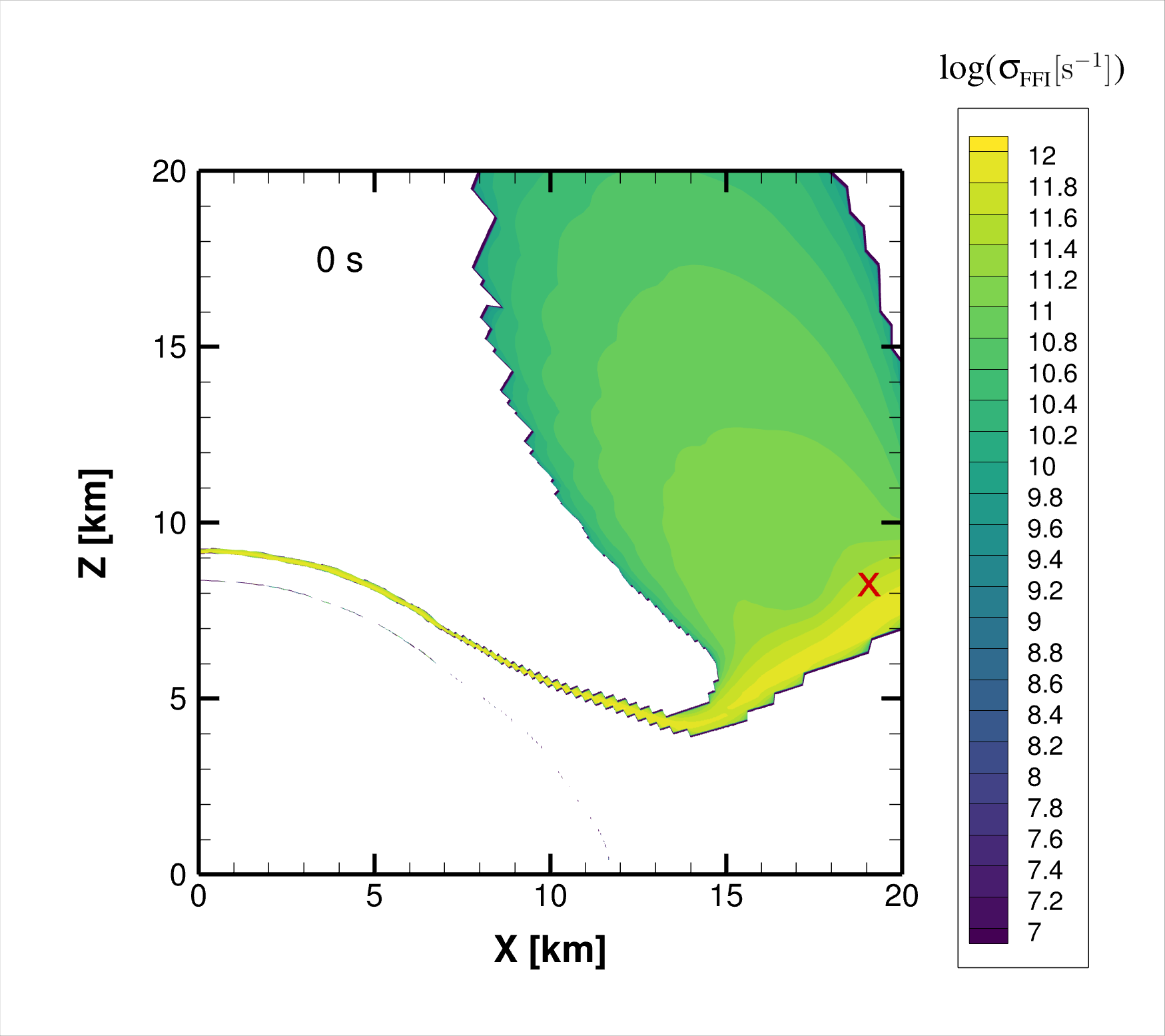}
\end{minipage}
\caption{Same as Fig.~\ref{fig:nu-FFI-0.4s-c1} but for $t=0\,$s ($\sim 50\,$ms after merger) time snapshot. The red cross mark corresponds to the spatial position where we display the ELN angular distribution in Fig.~\ref{fig:DifELNAngdist}; see the text for more details.}
\label{fig:nu-FFI-0.0s-c1.mark}
\end{figure}

\begin{figure}
\begin{minipage}{0.5\textwidth}
    \includegraphics[width=\linewidth]{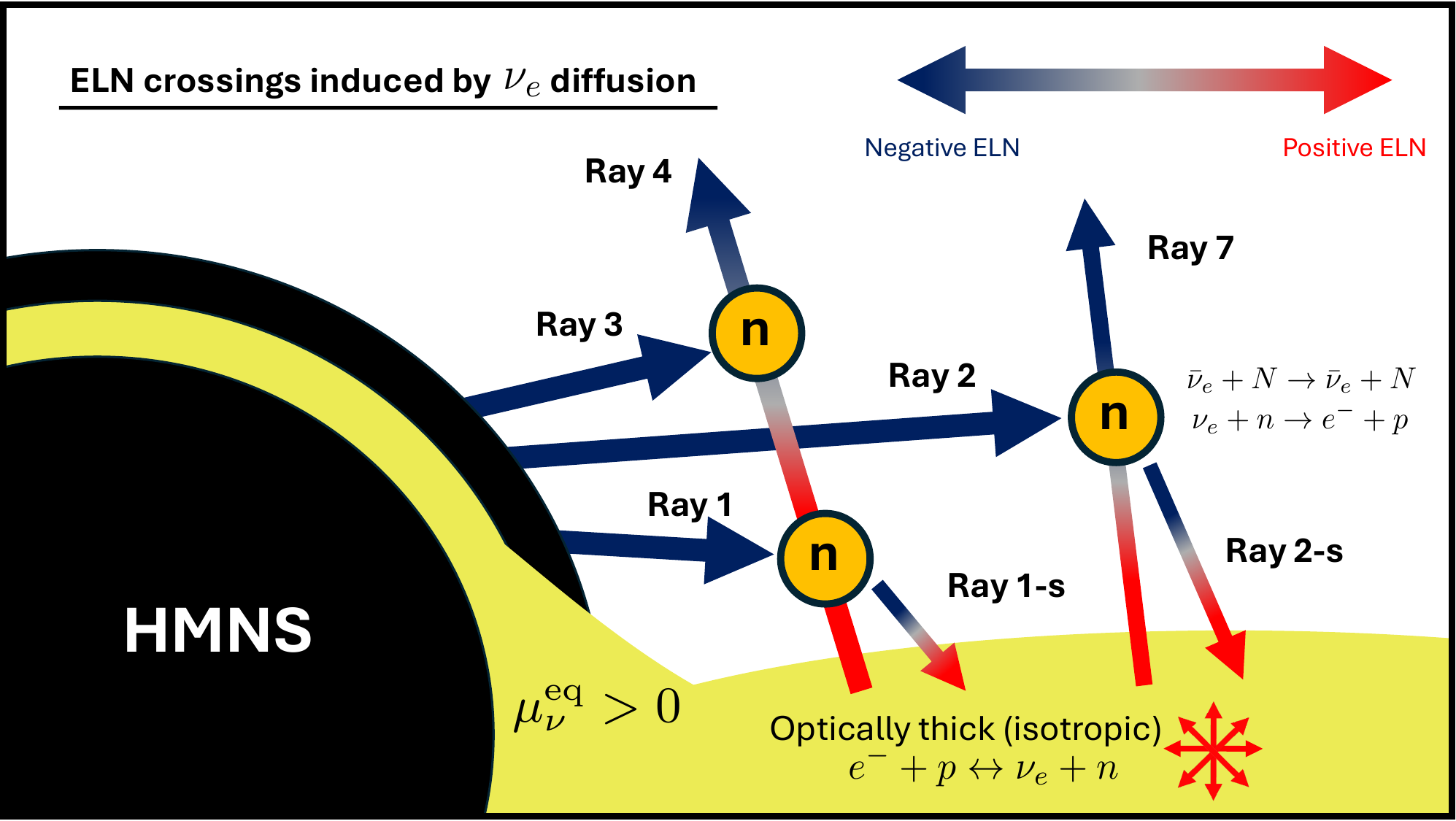}
\end{minipage}
    \caption{Schematic picture for ELN crossings induced by $\nu_e$ diffusion. The red and blue arrows represent positive ($\nu_e > \bar{\nu}_e$) and negative ($\nu_e < \bar{\nu}_e$) ELN, respectively. For convenience in referring to them in the text, the arrows are individually numbered. We note that the letter "-s" is appended to some arrows to denote scattered neutrinos. The width of each arrow represents the intensity of neutrinos. For this reason, the rays for scattered neutrinos are drawn with thinner lines than those for emitted neutrinos from HMNS. The black filled region represents the high matter density with $\mu_{\nu}^{\rm eq}<0$, corresponds to the major part of HMNS. The yellow one portrays an optically thick region for neutrinos with $\mu_{\nu}^{\rm eq}>0$ (both $\nu_e$ and $\bar{\nu}_e$ nearly achieve beta equilibrium). The orange filled circles denote neutrons. In this figure, we omit to show protons, since more neutrons than protons exist in the region. It should also be noted that the neutrino trajectories drawn in this schematic figure are simplified so as for readers to capture the essential trend. We also note that low energy neutrinos ($\lesssim 10\,$MeV) play an important roles on generating ELN angular crossings. See text for more details.}
    \label{fig:CrossingbyNuedif}
\end{figure}

Here, we propose a new mechanism to generate ELN angular crossings in the vicinity of HMNS. This type of crossings are evident in the time snapshot of $t=0\,$s (see Fig.~\ref{fig:nu-FFI-0.0s-c1.mark}). Of particular interest in this figure is an ELN crossing region extending vertically from the edge of disk surface ($X \sim 20\,$km, $Z \sim 8\,$km). Below, we explain the physical process with a schematic picture (Fig.~\ref{fig:CrossingbyNuedif}).

In general, the surface of HMNS emits more $\bar{\nu}_e$ than $\nu_e$, as illustrated by blue arrows labeled Ray1, 2, and 3 in Fig.~\ref{fig:CrossingbyNuedif}. On the other hand, a large number of $\nu_e$ diffuse out from an optically thick disk near the equatorial plane, where $\mu_{\nu}^{\rm eq}$ is positive (indicated by yellow-shaded regions in the figure). This neutrino emission pattern gives rise to ELN angular crossings above the disk. It should be mentioned that neutrons are more abundant than protons above the disk, implying that $\nu_e$ are absorbed by neutrons through a charged current reaction. In addition to this, $\bar{\nu}_e$ emitted from HMNS experience more frequent scatterings by nucleons than $\nu_e$, since both the average energy and number flux of $\bar{\nu}_e$ are higher than those of $\nu_e$. These two effects attenuate the positive ELN along rays coming from the disk (see Ray 4 and 7 in Fig.~\ref{fig:CrossingbyNuedif}), which eventually erase ELN angular crossings near the polar region.

An important remark should be made here. Although the schematic illustration in Fig.~\ref{fig:CrossingbyNuedif} and the accompanying argument are based on a simple ray-tracing description, this is primarily for conceptual clarity. In reality, the situation is complex. Neutrinos experience multiple scatterings as they propagate through the accretion disk, indicating that the neutrinos can not escape freely from the disk but rather diffuse outward. In addition to this, the above discussion omits the importance of energy dependent feature. More specifically, low energy neutrinos can escape from the deeper and dense regions of accretion disk with $\mu_{\nu}^{\rm eq}>0$, whereas high-energy neutrinos decouple at less dense region where $\mu_{\nu}^{\rm eq}<0$. In our model, we confirm that $\nu_e$ dominates over $\bar{\nu}_e$ for $\lesssim 10$MeV, generating positive ELN angular regions. This consideration suggests that the appearance of positive ELN angular crossing depends on the competition between low- and high energy neutrinos. If the excess of $\nu_e$ in the low energy is insufficient to overcome the negative ELN in high energy region, crossings are not generated even if the optically thick disk with $\mu_{\nu}^{\rm eq}>0$ is present.

\begin{figure}
\begin{minipage}{0.45\textwidth}
\centering
\includegraphics[width=1.0\linewidth]{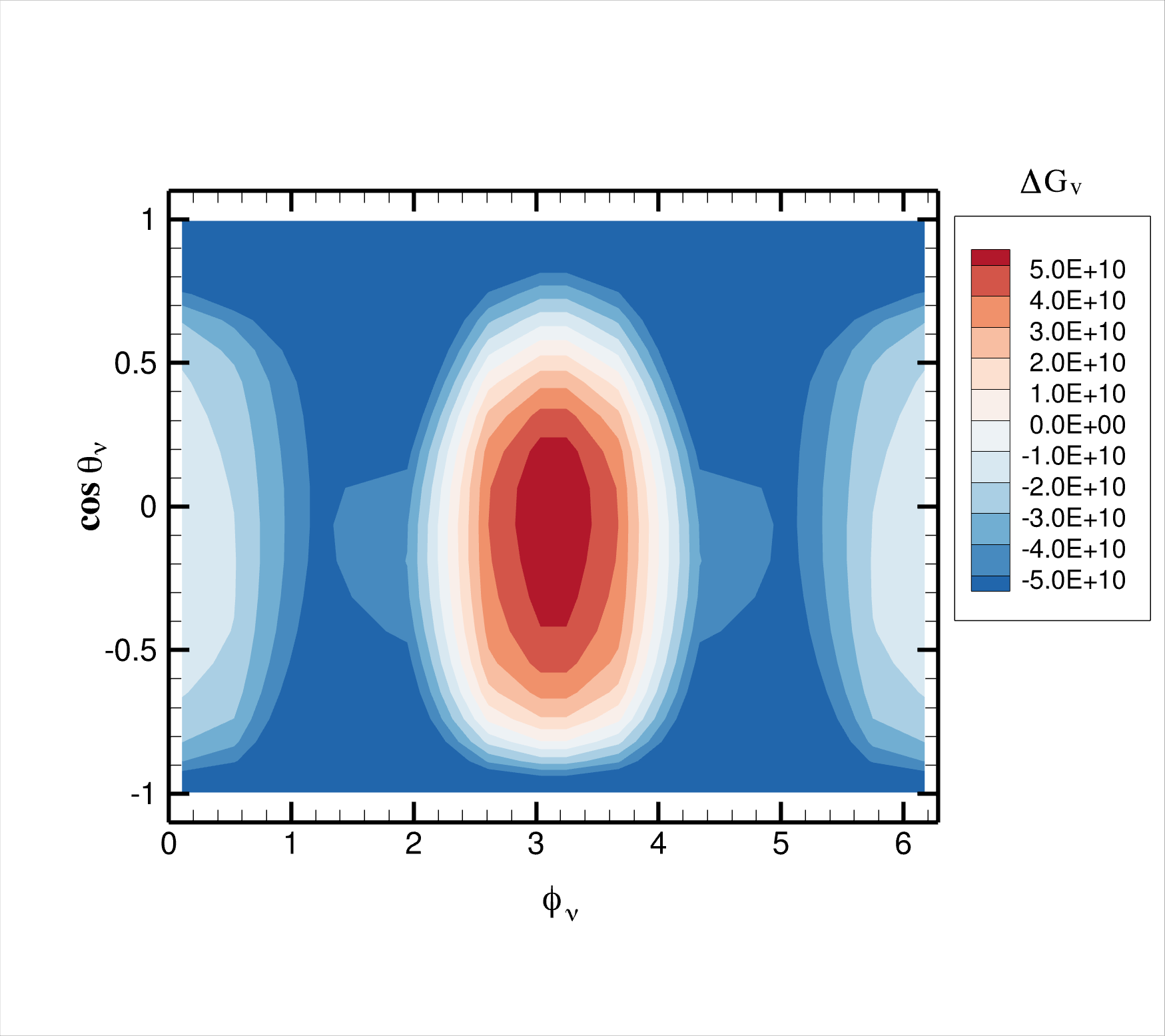}
\end{minipage}
\caption{Color map of ELN angular distributions ($\Delta G_v$ with the unit of $[\mathrm{s}^{-1}]$, see Eq.~(\ref{eq:def_G_v}) for the definition) as functions of $\cos {\theta_{\nu}}$ and $\phi_{\nu}$ for the time snapshot of $t=0\,$s. The spatial position is marked by a cross mark in Figs.~\ref{fig:FFIgrowthdistri}~and~\ref{fig:nu-FFI-0.0s-c1.mark} ($X=19.0\,$km and $Z=8.2\,$km). Red and blue colors denote positive and negative ELN, respectively.}
\label{fig:DifELNAngdist}
\end{figure}

Other important remarks are in order. First, this type of ELN angular crossings does not emerge if the central compact object is a BH due to the absence of strong $\bar{\nu}_e$ emission. Second, an accretion torus with $\mu_{\nu}^{\rm eq}>0$ plays a pivotal role in supplying $\nu_e$-dominant emission, suggesting that it can be a necessary condition for occurrences of ELN angular crossings. As discussed later, this claim holds in other mechanisms. Third, these ELN angular crossings exhibit strong non-axisymmetric structures in neutrino momentum space. As an example, we provide Fig.~\ref{fig:DifELNAngdist} displaying a two-dimensional color map of ELN angular distribution at a spatial position where it is highlighted by a cross mark in Fig~\ref{fig:nu-FFI-0.0s-c1.mark}. ELN becomes positive in angular regions of $\cos{\theta_{\nu}} \sim 0$ and $\phi_{\nu} \sim \pi$, which correspond to the neutrino flight directions of Ray~4~and~7 in Fig.~\ref{fig:CrossingbyNuedif}.

Finally, it is important to emphasize that multi-angle treatments in neutrino transport are essential to detect this type of ELN angular crossings. Two moment approximations, for example, can provide the flux direction but the disk component may be smeared out by intense neutrino emission from HMNS. In addition to this, the incoming and outgoing neutrinos from disks cannot be distinguished only by the first angular moment, which is a major obstacle to detect this type of crossings.

\subsubsection{ELN crossings with a disk shadow}\label{subsubsec:shadowmecha}

\begin{figure}
\begin{minipage}{0.5\textwidth}
\centering
\includegraphics[width=1.0\linewidth]{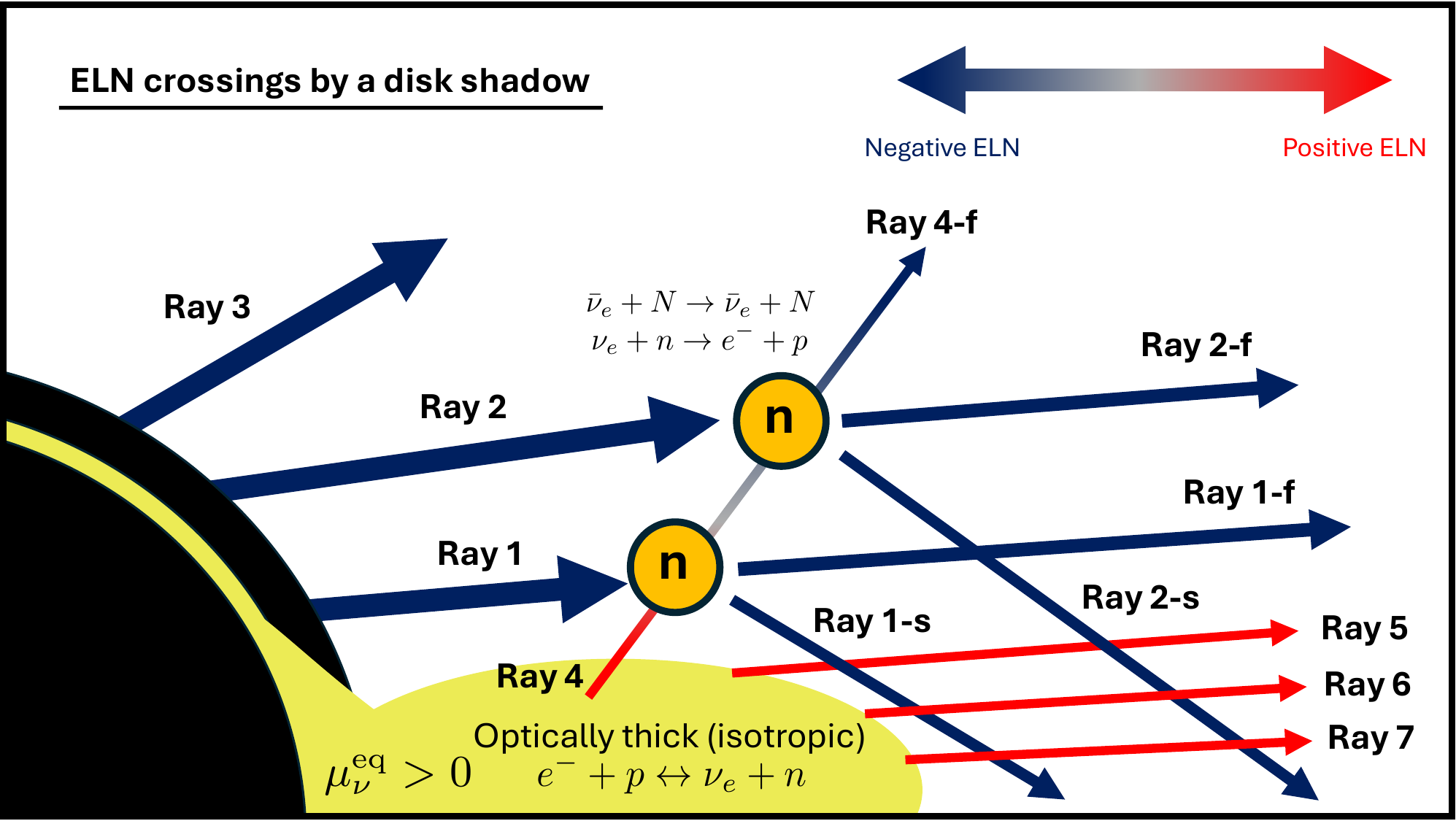}
\end{minipage}
\caption{Same as Fig.~\ref{fig:CrossingbyNuedif} but for a case with a disk shadow. In this figure, the letter "-f" is appended for forward scattered neutrinos (or no scatterings). We note that a larger spatial region is displayed in this figure than Fig.~\ref{fig:CrossingbyNuedif}.}
\label{fig:Crossingbyshadow}
\end{figure}

\begin{figure}
\begin{minipage}{0.45\textwidth}
\centering
\includegraphics[width=1.0\linewidth]{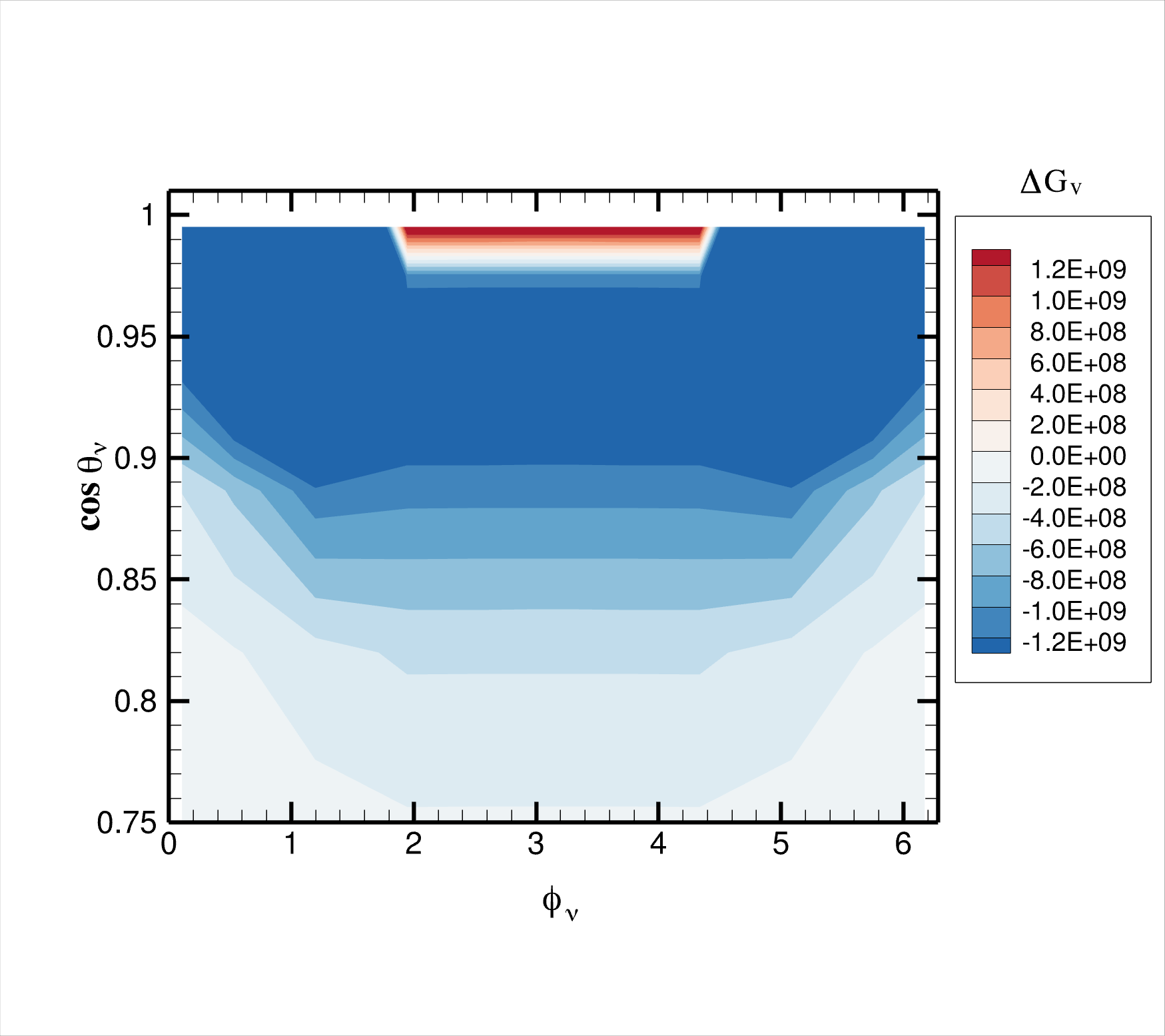}
\end{minipage}
\caption{Same as Fig.~\ref{fig:DifELNAngdist} but for a different spatial position ($X=228.3\,$km and $Z=22.9\,$km). The time is $t=0\,$s. The corresponding spatial position is also highlighted by a red cross mark in the top left panel of Fig.~\ref{fig:FFIgrowthdistri}. In this figure, we zoom in the forward angles ($\cos \theta_{\nu}>0.75$). This result supports our interpretation of ELN angular crossings by a disk shadow mechanism; see the text for more details.}
\label{fig:ShadowELNAngdist}
\end{figure}

At the time snapshot of $t=0\,$s, ELN crossings appear in the equatorial region at a radius greater than $\sim 60\,$km (see the top left panel in Fig.~\ref{fig:FFIgrowthdistri}). As shown below, the inner region of accretion disk with $\mu_{\nu}^{\rm eq} > 0$ is responsible for them.

We illustrate the mechanism with Fig.~\ref{fig:Crossingbyshadow}. As mentioned already, $\bar{\nu}_e$ emission dominates over $\nu_e$ at the surface of HMNS. Although these neutrinos undergo scatterings with nucleons in the vicinity of HMNS (see Ray 1 and 2 in Fig.~\ref{fig:Crossingbyshadow}), ELN sustains negative for the forward scattered (or unscattered) neutrinos (see Ray 1-f and 2-f). In the equatorial region, however, the optically thick disk with $\mu_{\nu}^{\rm eq} > 0$ surrounds the HMNS, which blocks $\bar{\nu}_e$ emission from the HMNS. Instead, ELN becomes positive for neutrinos emitted from the optically thick disk with $\mu_{\nu}^{\rm eq} > 0$ (see Ray 5 to 7 in Fig.~\ref{fig:Crossingbyshadow}). Meanwhile, scattered neutrinos from higher latitudes, in which $\bar{\nu}_e$ dominates over $\nu_e$ (see Ray 1-s and 2-s), propagate into the equatorial region. These two components generate ELN angular crossings.

One may wonder why ELN angular crossings are not observed in higher latitudes (above the disk). This can be illustrated by Ray 4 and Ray 4-f in Fig.~\ref{fig:Crossingbyshadow}. In the vicinity of energy sphere of $\nu_e$, ELN is certainly positive which can generate ELN angular crossings. In fact, ELN angular crossings appear at the edge of optically thick region with $\mu_{\nu}^{\rm eq} > 0$ (see top left panel in Fig.~\ref{fig:FFIgrowthdistri}), whose mechanism is essentially the same as the one with $\nu_e$ diffusion (see Sec.~\ref{subsubsec:nuedif}). As already mentioned above, however, neutron rich environments above the disk attenuate $\nu_e$ due to charged-current reactions ($\nu_e$ absorption by neutrons), and $\bar{\nu}_e$ emitted from HMNS is more likely to scatter off nucleons. Consequently, ELN eventually becomes negative as neutrinos propagate towards higher latitudes (see the color transition from Ray 4 to Ray 4-f in Fig.~\ref{fig:Crossingbyshadow}), causing disappearance of ELN angular crossings.

Figure~\ref{fig:ShadowELNAngdist} portrays ELN angular distribution in the spatial position of $X=228.3\,$km and $Z=22.9\,$km at $t=0\,$s (see also the top left panel of Fig.~\ref{fig:FFIgrowthdistri}). As shown in the plot, ELN is positive at $\cos \theta_{\nu} \sim 1$ (radially forward direction) and $\phi_{\nu} \sim \pi$, while they are negative in other regions. We note that neutrino rays in the forward direction with $\phi_{\nu} \sim \pi$ predominantly originate from the disk with $\mu_{\nu}^{\rm eq} > 0$, whereas those with $\phi_{\nu} \sim 0$ experience scatterings with nucleons above the disk. This supports our proposed mechanism to generate ELN angular crossings.

It should be emphasized that HMNS also plays a crucial role on the mechanism. In fact, if HMNS has already collapsed into a BH, there would be no angular regions with negative ELN at the corresponding region. Let us also point out that the emergence of $\mu_{\nu}^{\rm eq} > 0$ region plays an important role similar to other mechanisms. However, this is merely a necessary condition for appearance of this type of ELN angular crossings. Similar to the argument in Sec.~\ref{subsubsec:nuedif}, the excess of $\nu_e$ compared to $\bar{\nu}_e$ in low energy ($\lesssim 10\,$MeV) regions needs to be sufficiently high to counteract the negative ELN in high energy regions. This condition is not trivial and hinges on the matter distribution around the disk. We also note that the disk needs to be optically thick to block neutrinos from HMNS for the shadow mechanism, indicating that the ELN crossing does not appear in the late phase after the disk becomes transparent. In fact, this type of crossings is no longer observed after $0.2\,$s in our model.

\subsubsection{ELN crossings by $\nu_e$ contamination}\label{subsubsec:ELNnuecontami}

\begin{figure}
\begin{minipage}{0.5\textwidth}
\centering
\includegraphics[width=1.0\linewidth]{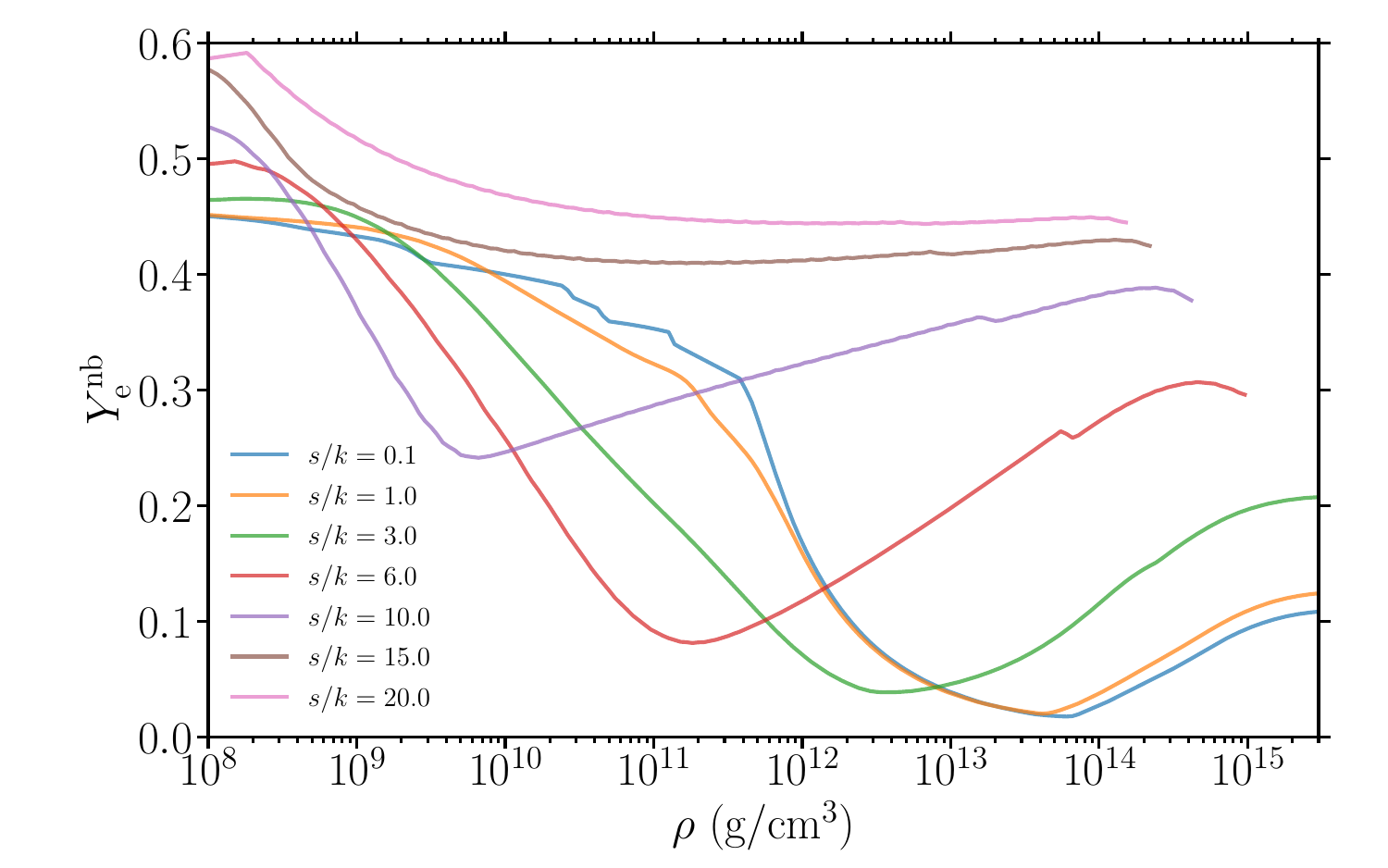}
\end{minipage}
\caption{$Y_e^{\rm nb}$ as a function of $\rho$, where $Y_e^{\rm nb}$ corresponds to electron fractions with $\mu_{\nu}^{\rm eq}=0$. Different lines with different color distinguish entropy per baryon ($s/k$).}
\label{fig:Ye_fixS}
\end{figure}

After $\sim 0.2\,$s, an optically thick accretion disk with $\mu_{\nu}^{\rm eq} > 0$ disappears, while regions with $\mu_{\nu}^{\rm eq} \sim 0$ emerge above the disk. Before delving into ELN angular crossings, we briefly discuss them, since they play an important role in the generation of such crossings.

As shown in Fig.~\ref{fig:nuechemipote}, we observe rather commonly high positive values of $\mu_{\nu}^{\rm eq}$ in the polar region. This is attributed to the $\nu_e$ irradiation. The $\nu_e$ absorption by neutrons increases $Y_e$ (see also Fig.~\ref{fig:Ye}), leading to $\mu_{\nu}^{\rm eq}>0$; see \cite{2017PhRvD..96l3012S} for more details. We also find in Fig.~\ref{fig:nuechemipote} that $\mu_{\nu}^{\rm eq}$ is nearly zero (but positive) with a band-like structure along regions with $\theta \sim \pi/3$ and $r<200\,$km. This band structure expands in the equatorial direction with time, and it eventually covers the entire disk (see the bottom right panel in Fig.~\ref{fig:nuechemipote}). A key quantity to understand this feature is $Y_e^{\rm nb}$ which denotes electron fraction satisfying $\mu_{\nu}^{\rm eq}=0$. In Fig.~\ref{fig:Ye_fixS}, we plot $Y_e^{\rm nb}$ as a function of baryon mass density ($\rho$), with different colors distinguishing entropy per baryon ($s/k$). As shown in this figure, $Y_e^{\rm nb}$ becomes lower with increasing $\rho$ along adiabatic lines for densities below $\sim 10^{10} {\rm g/cm^3}$. This suggests that the adiabatic compression of fluid element leads to deleptonization (mainly through electron capture by free protons). Indeed, fluid elements forming the band-like structure of $\mu_{\nu}^{\rm eq} \gtrsim 0$ in our BNSM model advect towards the HMNS, exhibiting the compression. This is consistent with the above argument\footnote{Strictly speaking, the fluid compression in BNSM models is not adiabatic due to neutrino cooling and fluid viscosity. However, the adiabatic condition is a reasonable approximation in this density and temperature regime, since the neutrino emission is relatively weak and the fluid viscosity compensates for the entropy loss by neutrino cooling.}.

\begin{figure}
\begin{minipage}{0.5\textwidth}
\centering
\includegraphics[width=1.0\linewidth]{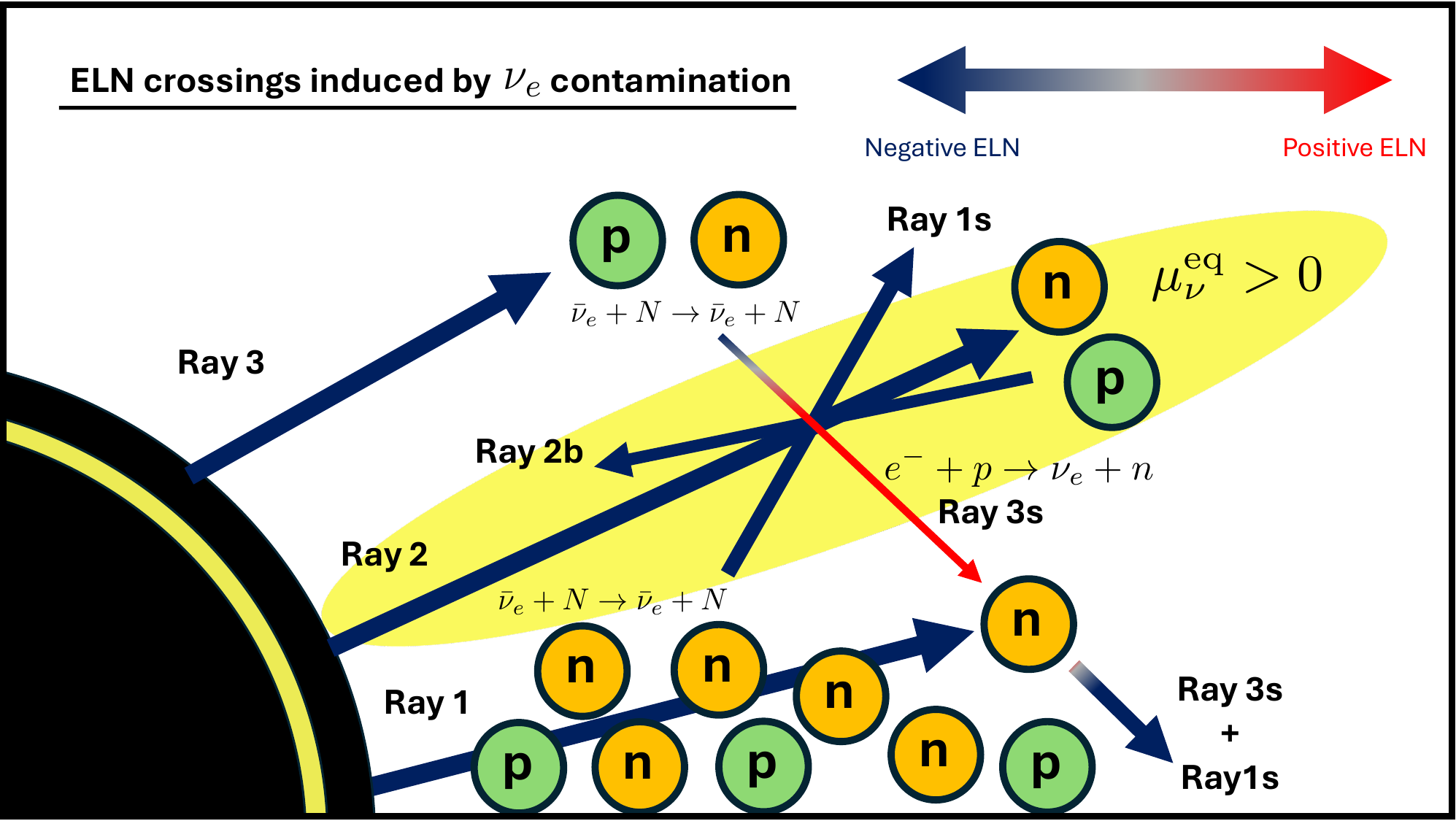}
\end{minipage}
\caption{Same as Fig.~\ref{fig:CrossingbyNuedif} but for ELN angular crossing generation by $\nu_e$ contamination.}
\label{fig:Crossingbycontami}
\end{figure}

\begin{figure}
\begin{minipage}{0.45\textwidth}
\centering
\includegraphics[width=1.0\linewidth]{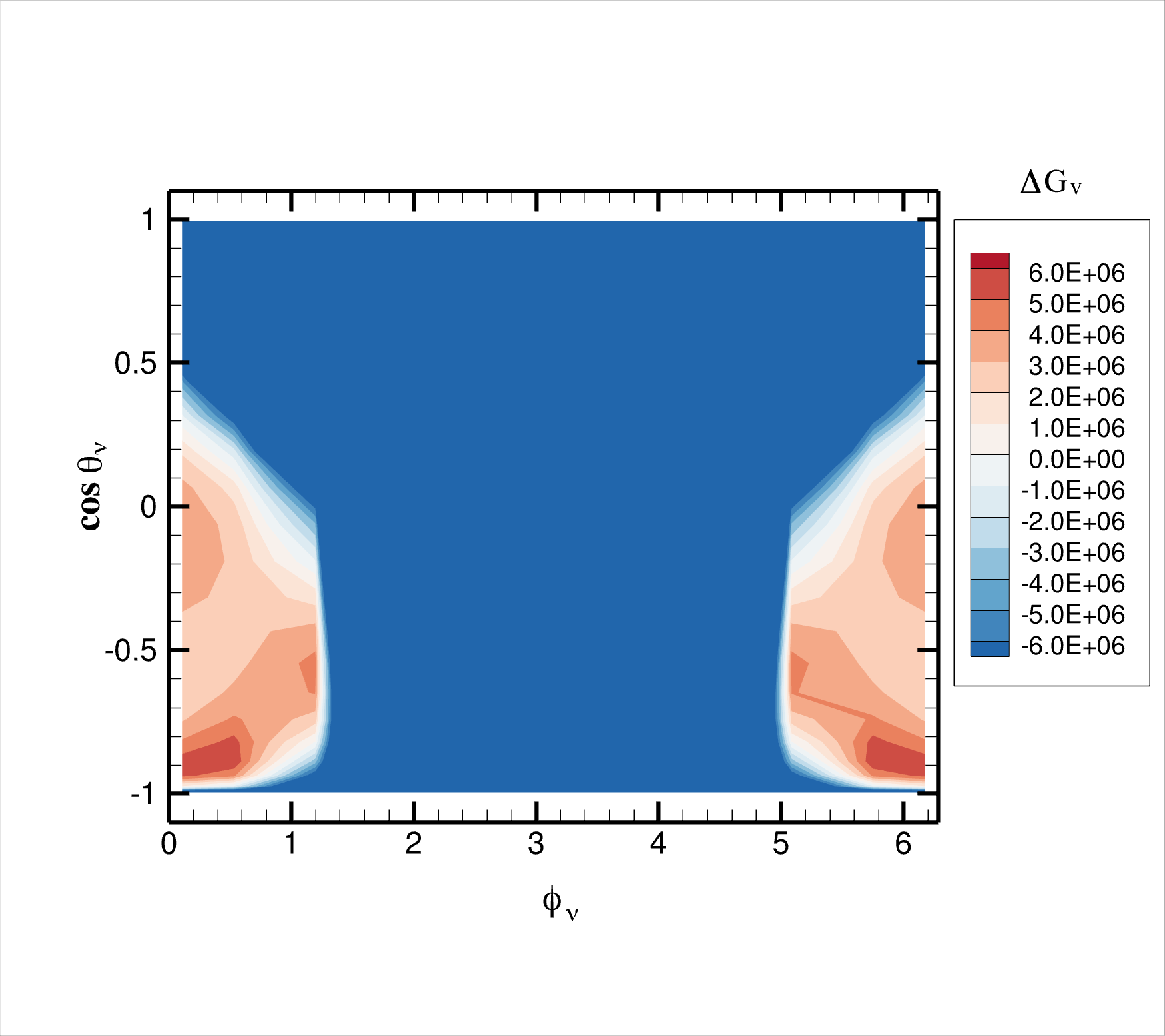}
\end{minipage}
\caption{Same as Fig.~\ref{fig:DifELNAngdist} but for a different spatial position ($X=97.6\,$km and $Z=32.6\,$km). The time is $t=0.6\,$s. In Fig.~\ref{fig:FFIgrowthdistri}, we put a red cross mark at the corresponding position in the panel for $t=0.6\,$s.}
\label{fig:contamiELNAngdist}
\end{figure}

Let us consider ELN angular crossings associated with the band like structure of $\mu_{\nu}^{\rm eq} \gtrsim 0$. Fig.~\ref{fig:Crossingbycontami} presents a schematic illustration of the underlying mechanism. This figure is sketched to represent the case with $t=0.6\,$s (see the bottom left panels in Figs.~\ref{fig:nuechemipote}~and~\ref{fig:FFIgrowthdistri}) but the mechanism is essentially the same for other time snapshots ($t=0.8,$ and $1.3\,$s). As already pointed out in previous subsections, $\bar{\nu}_e$ emission from the surface of HMNS dominates over $\nu_e$, and the former undergoes more frequent scatterings with nucleons than the latter. This makes the ELN angular distributions be negative in the full angles. On the other hand, $\nu_e$ emission dominates over $\bar{\nu}_e$ in the band-like flow with $\mu_{\nu}^{\rm eq} \gtrsim 0$, which potentially changes the sign of ELN. In the radially forward directions, the intensity of $\bar{\nu}_e$ is strong (since they are emitted from the vicinity of HMNS; see Ray~2 in Fig.~\ref{fig:Crossingbycontami}), implying that ELN becomes negative. In the backward directions, $\bar{\nu}_e$ is still more abundant than $\nu_e$ (see Ray~2b). This is attributed to the angular dependence of nucleon scatterings, which favors backward scatterings. Although $\bar{\nu}_e$ intensity is much lower than the forward direction, it still leads to negative ELN. Let us, then, consider the neutrinos coming from the disk region. In the equatorial region, $\bar{\nu}_e$ also dominates. This is due to stronger charged current reaction for $\bar{\nu}_e$ and more frequent $\bar{\nu}_e$ scatterings than those for $\nu_e$. As a result, more $\bar{\nu}_e$ can escape vertically from the equatorial region, that also makes ELN negative in this direction (see Ray~1s). However, neutrinos which propagate from higher latitude to the equatorial plane have a distinct behavior. In polar regions where matter density is very low, neutrino scatterings are less active (see Ray~3 and 3s). As a result, $\nu_e$ emission in the band-like flow with $\mu_{\nu}^{\rm eq} \gtrsim 0$ can exceed the scattered $\bar{\nu}_e$, resulting in a positive ELN. In Fig.~\ref{fig:contamiELNAngdist}, we show that the ELN becomes positive in the angular region of $\phi_{\nu} \sim 0 (2 \pi)$ and $\cos{\theta_{\nu}} \lesssim 0$, which supports our interpretation.

It is worth mentioning that this type of ELN crossings tends to vanish in the equatorial region, which is illustrated as Ray~3s + Ray~1s in Fig.~\ref{fig:Crossingbycontami}. In this region, the neutron abundance is higher than in higher latitudes, leading to many scattered $\bar{\nu}_e$ and local $\bar{\nu}_e$ emission through a charged current process. As a result, ELN becomes negative for Ray~3s + Ray~1s. This accounts for no ELN angular crossings in the equatorial region at $t=0.6\,$s time snapshot. It should be mentioned that the competition between $\bar{\nu}_e$ and $\nu_e$ hinges on the matter profile. The scattered neutrinos decrease with time due to lower neutrino emission from HMNS and low matter density in the disk (which also decrease with time). Consequently, $\bar{\nu}_e$ cannot overcome $\nu_e$ for Ray~3s + Ray~1s in Fig.~\ref{fig:Crossingbycontami}, suggesting that ELN crossings can be observed in the equatorial region at later phases. This accounts for the spatial distributions for ELN angular crossings for later time snapshots of $t=0.8\,$s and $1.3\,$s (see Fig.~\ref{fig:FFIgrowthdistri}).

Let us make another argument to explain no ELN angular crossings at $\gtrsim 20\,$km for the time snapshots of $t=0.2\,$s and $0.4\,$s, despite the fact that there is a band-like structure of $\mu_{\nu}^{\rm eq} \gtrsim 0$. During these relatively early post-merger phases, $\bar{\nu}_e$ emission from HMNS is still strong, and the matter density in the polar region is also higher than in the late phase. As a result, the scattered neutrinos from the polar region (Ray~3s in Fig.~\ref{fig:Crossingbycontami}) is abundant, which increase $\bar{\nu}_e$ intensity. As a result, $\nu_e$ emission in the region of $\mu_{\nu}^{\rm eq} \gtrsim 0$ cannot dominate over $\bar{\nu}_e$, suppressing ELN angular crossings.

Here is a short summary of this subsection regarding detailed analyses of FFI. The appearance of $\mu_{\nu}^{\rm eq} \gtrsim 0$ underlies all mechanism to generate ELN angular crossings. However, it is important to emphasize that this condition is not sufficient on its own to guarantee occurrences of FFI as discussed above, and multi-angle neutrino transport simulations are necessary for the reliable assessment. We also illustrated in this section that intense $\bar{\nu}_e$ emission from HMNS leads to distinct crossing generation mechanisms from those in BH accretion disk systems.


\begin{figure*}
\begin{minipage}{1.0\textwidth}
\centering
\includegraphics[width=0.45\linewidth]{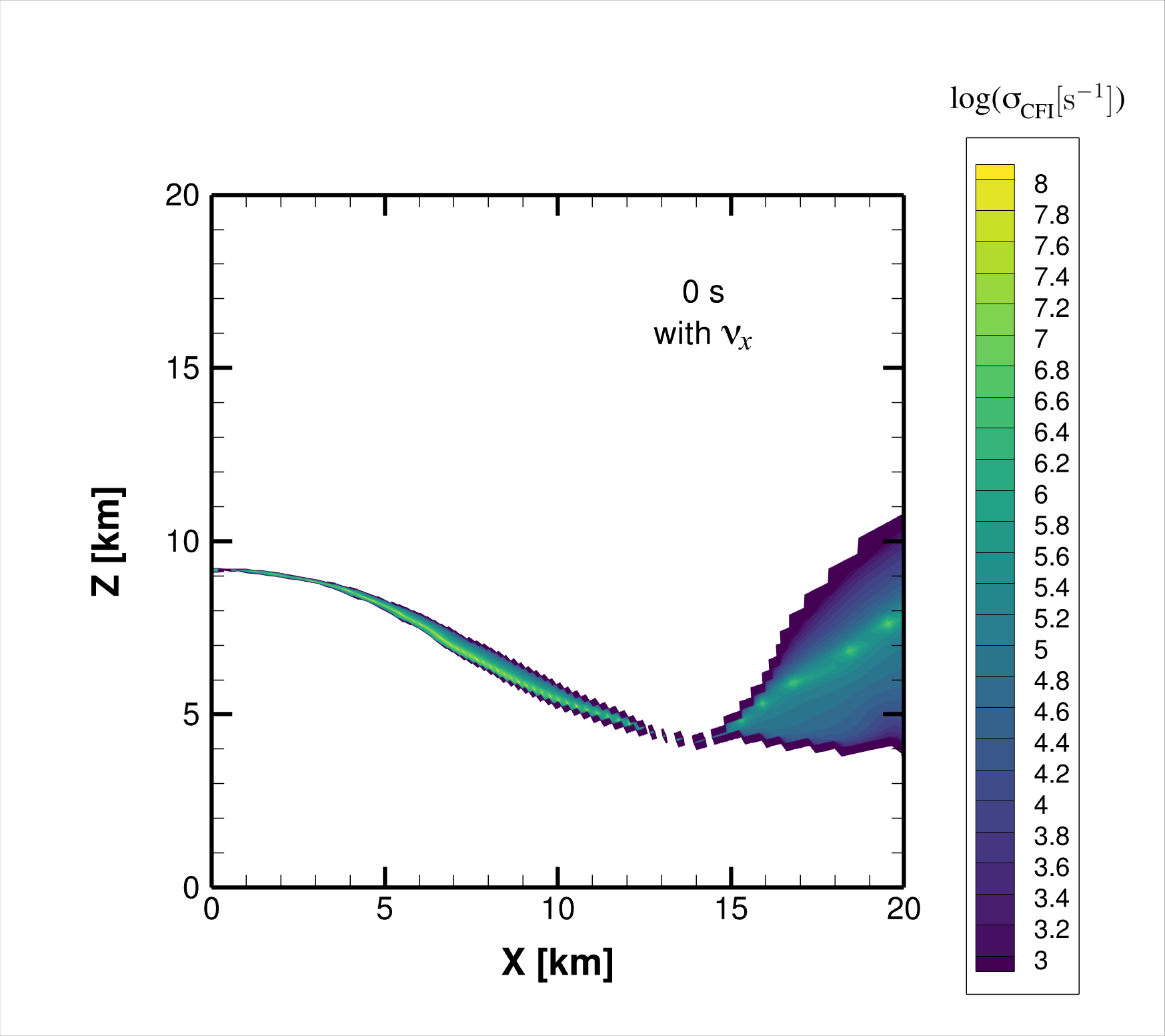}
\includegraphics[width=0.45\linewidth]{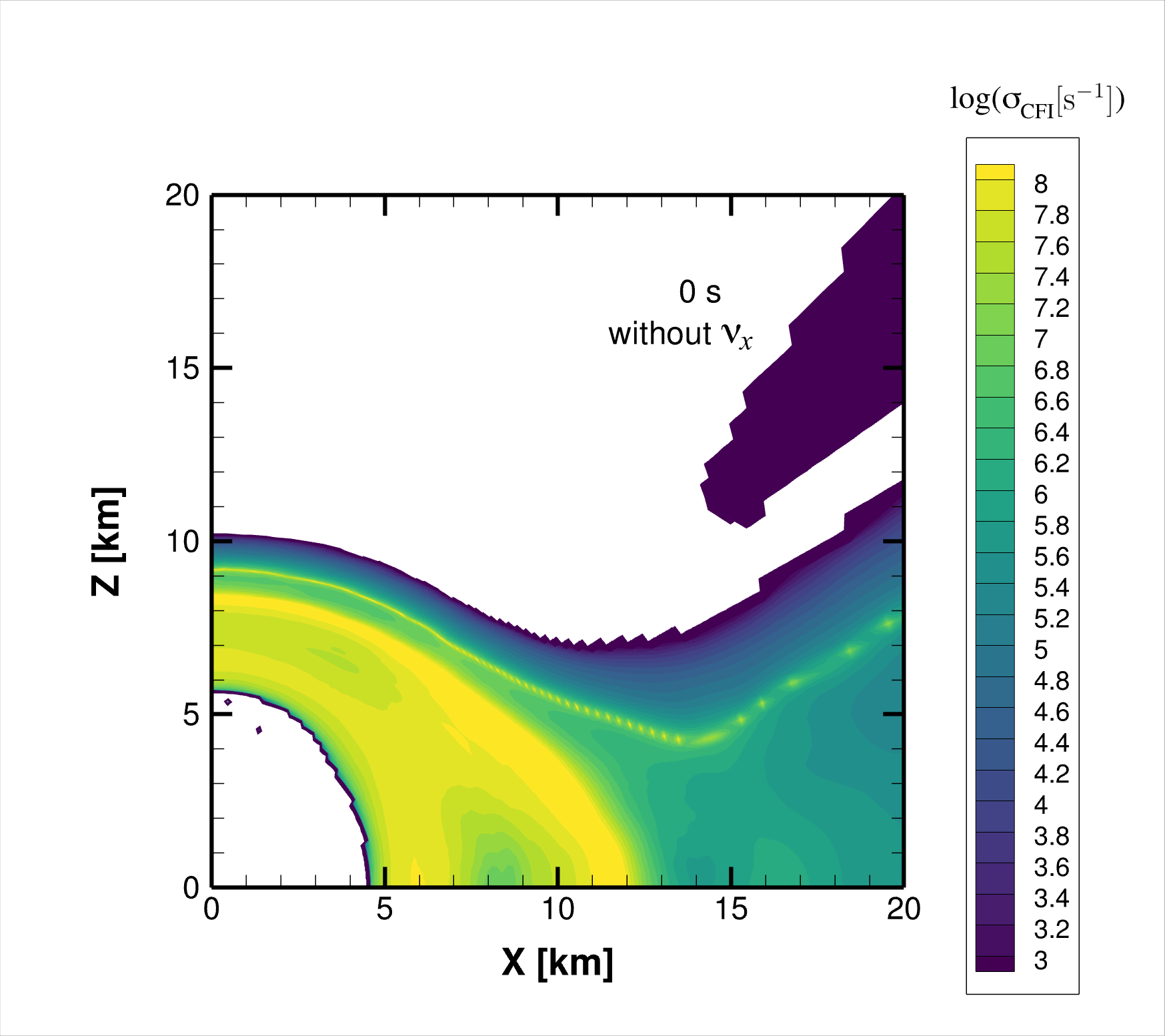}
\end{minipage}
\caption{Comparison of CFI growth rate between with (left) and without (right) $\nu_x$ contributions. We display the plot in the spatial box of $0{\rm km} < X < 20 {\rm km}$ and $0{\rm km} < Z < 20 {\rm km}$. The time corresponds to $t=0\,$s.}
\label{fig:CFI_growth_t0_withwithoutnux}
\end{figure*}

\subsection{CFI analysis}\label{subsec:CFIanalysis}
Here, we turn our attention to CFI. It is important to note that the overall trend regarding occurrences of CFI aligns well with the findings in \cite{2023PhRvD.108h3002X}, which examined CFI in cases with a BH accretion disk. Our result suggests that CFI can occur in BNSM remnants regardless of central compact object, as long as the accretion disk is formed. On the other hand, some new intriguing features arise in the present study. In the following, we highlight them with detailed analyses.

\subsubsection{Resonance-like CFI}\label{subsubsec:resoCFI}

In Sec.~\ref{subsubsec:alphadrivenmecha}, we discussed ELN angular crossings in the optically thick region with $\mu_{\nu}^{\rm eq} \sim 0$. The same region is also subject to resonance-like CFI. This is because the resonance-like CFI occurs in the region with $A^2\ll |G\alpha|$ (see also Sec.~\ref{subsec:flavorinstaana}), while $A$ is nearly zero at the region of $\mu_{\nu}^{\rm eq} \sim 0$ ($n_{\nu_e} \sim n_{\bar{\nu}_e}$). As shown in the top left panel of Fig.~\ref{fig:CFIgrowthdistri} (time snapshot of $t=0\,$s), the high growth rate of CFI can be observed along the band with $\mu_{\nu}^{\rm eq} \sim 0$ (see also Fig.~\ref{fig:nuechemipote}), which marks the resonance-like CFI.

An important remark regarding $\nu_x$ contribution should be made here. In Fig.~\ref{fig:CFI_growth_t0_withwithoutnux}, we provide zoom-in plots for growth rates of CFI at the time snapshot of $t=0\,$s. Left and right panels show the case with and without $\nu_x$ contribution, respectively\footnote{More specifically, we set $\Gamma_{x}=\bar{\Gamma}_{x}=0$ and $n_x=\bar{n}_x=0$ in the case without $\nu_x$ contribution.}. As shown in the plots clearly, the result with no $\nu_x$ contributions shows the overestimation of both growth rate and area of unstable regions of CFI. We note that this extended unstable region emerged in cases with no $\nu_x$ contribution corresponds to non-resonance CFIs; see also the similar argument in \cite{2023PhRvD.108l3024L}. It is also worth mentioning that resonance-like CFI should be suppressed if $\nu_x$ is in an exact thermal equilibrium with matter (see discussions in \cite{2024PhRvD.110d3039L}). The occurrences of CFI indicates that $\nu_x$ starts to deviate from the equilibrium. Our result shows the importance of $\nu_x$ for accurate estimation of CFI.

\subsubsection{Stable configuration of CFI}\label{subsubsec:stableCFI}
Although the growth rate of CFI is much lower than FFI, the present study shows that CFI plays a complementary role on flavor conversions. For instances, the CFI structure is more persistent than FFI for $t \lesssim 1\,$s. In fact, the spatial region of CFI tends to be wider than FFI, and CFI can occur in time snapshots of $t=0.2\,$s and $0.4\,$s but FFI does not occur at outer radii ($\gtrsim 20\,$km) at these time snapshots. This result suggests that there are some phases which CFI must dominate over FFI.

\begin{figure}
\begin{minipage}{0.45\textwidth}
\centering
\includegraphics[width=1.0\linewidth]{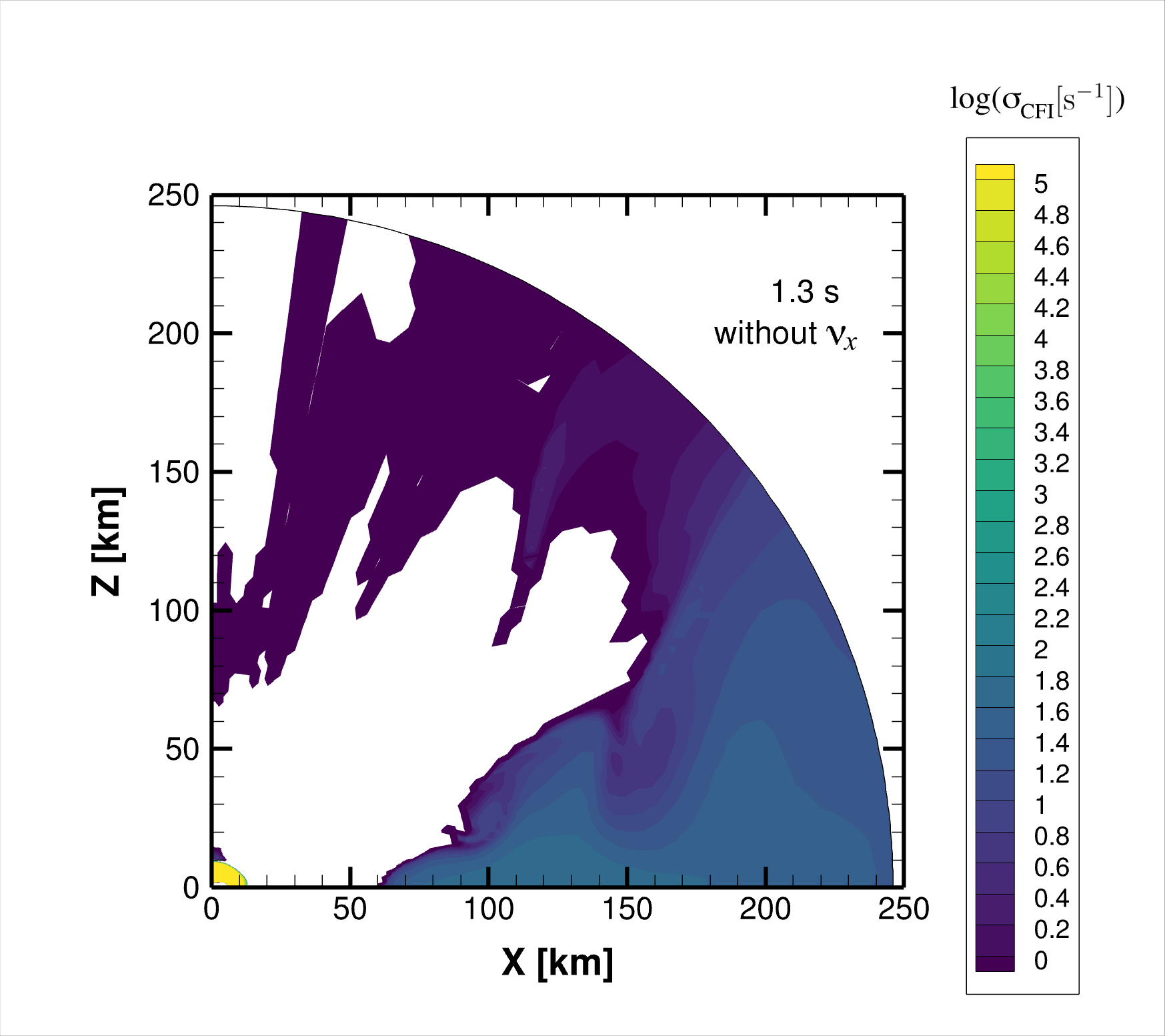}
\end{minipage}
\caption{Same as Fig.~\ref{fig:CFIgrowthdistri} but for the case without $\nu_x$ contribution. The time is $t=1.3\,$s.}
\label{fig:CFIgrowth_t1.3_nomu}
\end{figure}

\begin{figure*}
\begin{minipage}{1.0\textwidth}
\centering
\includegraphics[width=0.45\linewidth]{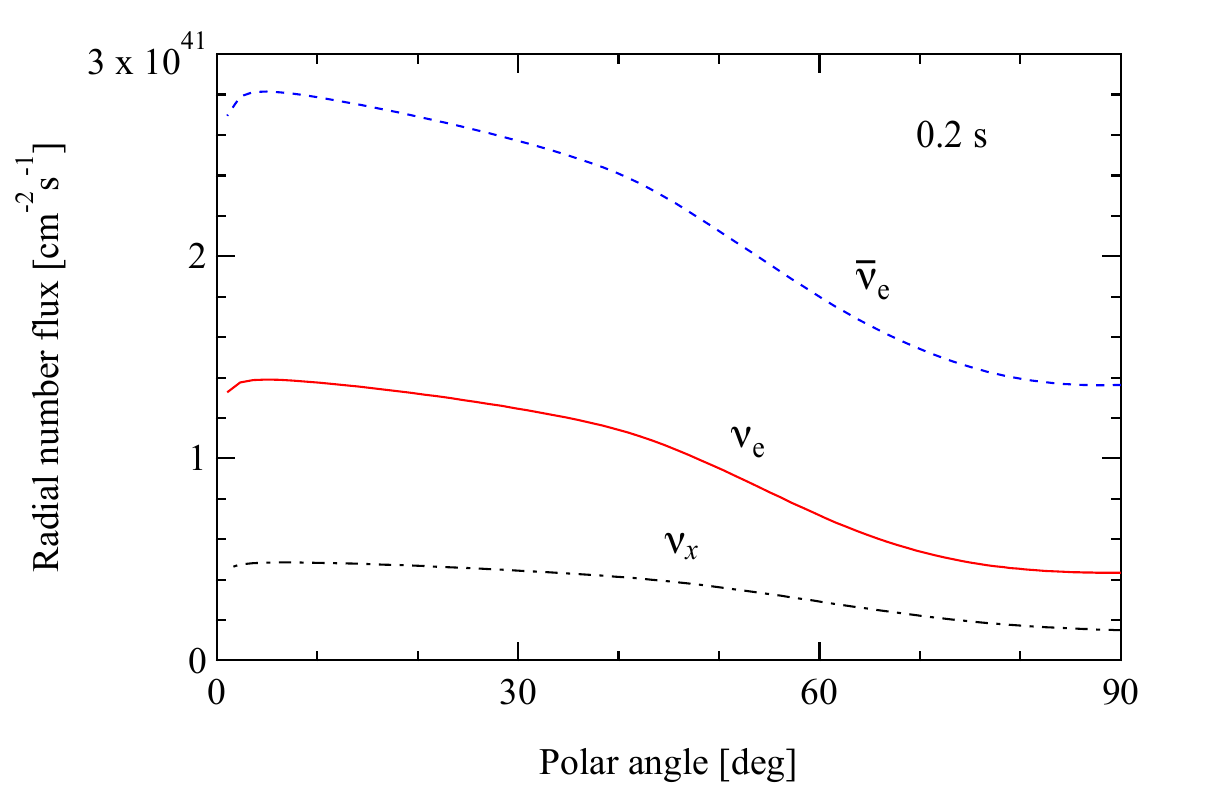}
\includegraphics[width=0.45\linewidth]{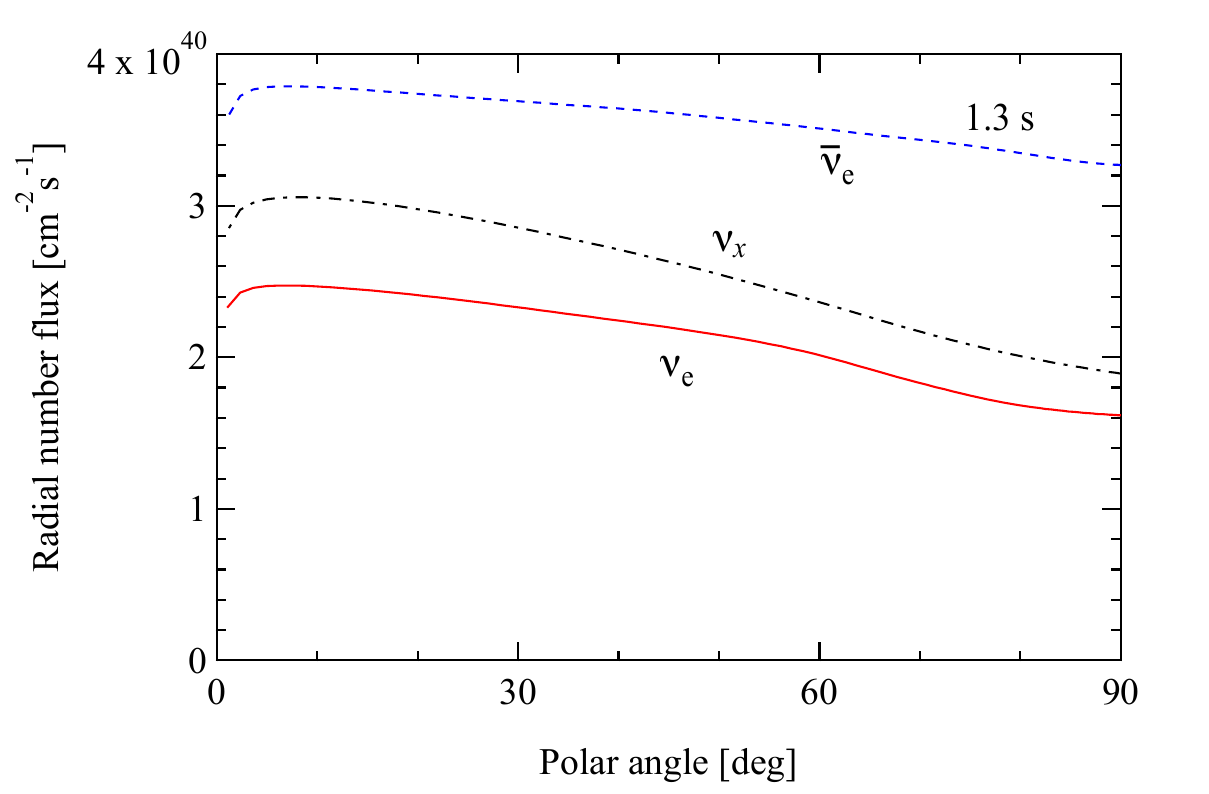}
\end{minipage}
\caption{Neutrino number flux as a function of polar angle, measured at the outer boundary. Left and right panels correspond to time snapshot of $t=0.2\,$s and $t=1.3\,$s, respectively. Line color and type distinguish neutrino species.}
\label{fig:Numflux02_13}
\end{figure*}

Another intriguing feature observed in growth rates of CFI from Fig.~\ref{fig:CFIgrowthdistri} is that CFI tends to be suppressed in regions with $\mu_{\nu}^{\rm eq} \gtrsim 0$. This trend can be explained by a trend of $\alpha$ (see Eq.~(\ref{eq_G_A}) for the definition). In the corresponding region with $\mu_{\nu}^{\rm eq} \gtrsim 0$, the disparity between $\Gamma$ and $\bar{\Gamma}$ becomes smaller than regions with $\mu_{\nu}^{\rm eq} \ll 0$, which suppresses CFI. It is interesting to point out that ELN angular crossings are generated in the vicinity of a band-like structure with $\mu_{\nu}^{\rm eq} \gtrsim 0$ (see Sec.~\ref{subsubsec:ELNnuecontami}), which exhibits an anti-correlation between CFI and this type of ELN angular crossing. This shows that CFI is complementary to FFI in some regions.

Two important remarks should be made here. First, as shown in Fig.~\ref{fig:CFIgrowthdistri}, the growth rate of CFI in optically thin region is lower than $c/r$, where $c$ and $r$ denote the speed of light and radius, respectively. This indicates that the advection time scale is shorter than the growing time scale of CFI, implying that neutrinos pass through the CFI unstable region, before the CFI reaches in non-linear phase. This suggests that CFI in this region does not change the neutrino radiation field and thus has a negligible impact on BNSM dynamics.

Second, although the growth rate of CFI in the optically thick region is much lower than FFI, it potentially has a substantial impact on BNSM dynamics. This is because CFI can reach non-linear phase and the time scale of flavor conversion is much shorter than that for neutrino diffusion. This argument is supported by a previous study by \cite{2023PhRvD.107h3016X}, which demonstrated that neutrino radiation fields are radically changed by CFI in CCSN environments. It should be noted that quantifying the impact of CFI on neutrino radiation field in BNSM environments lies beyond the scope of this paper, so we defer the detailed analysis based on quantum kinetic neutrino transport to future work.

\subsubsection{Disappearance of CFI in the late phase}\label{subsubsec:disapCFI}

As shown in the bottom right panel of Fig.~\ref{fig:CFIgrowthdistri}, CFI is not observed outside of HMNS at the time snapshot of $t=1.3\,$s. To investigate this trend further, we make the same plot as Fig.~\ref{fig:CFIgrowthdistri} but without $\nu_x$ in Fig.~\ref{fig:CFIgrowth_t1.3_nomu}. Below, we discuss the origin of the CFI suppression by using this figure.

Let us at first point out that CFI (without $\nu_x$ contribution) displayed in Fig.~\ref{fig:CFIgrowth_t1.3_nomu} has a correlation to $Y_e$ distribution; see the bottom right panel in Fig.~\ref{fig:Ye}. As these plots illustrate, CFI is suppressed in regions with $Y_e \sim 0.5$. Because number densities of neutrons and protons are nearly equal in these regions, there are small differences in charged-current reactions between $\nu_e$ and $\bar{\nu}_e$, which suppresses CFI. On the other hand, CFI can occur if there are no $\nu_x$ in low $Y_e$ environments. Below, we provide more quantitative discussions regarding roles of $\nu_x$ on CFI.

In Fig.~\ref{fig:Numflux02_13}, we show the species-dependent neutrino number luminosity measured at the outer boundary as a function of a polar angle. For comparison, we display them for two different time snapshots: $t=0.2\,$s (left) and $1.3\,$s (right). As can be seen in this figure, the number luminosity of $\nu_x$ exceeds $\nu_e$ for $t=1.3\,$s, while $\bar{\nu}_e$ is the highest number flux. This suggests that the sign of $g$ and $\bar{g}$ (see Eq.~(\ref{eq:gdef})) is negative and positive, respectively. This leads to $|G| < |A|$ (see Eq.~(\ref{eq_G_A})), which makes the system stable with respect to CFI (see Eqs.~(\ref{eq:CFIgrowthrate_isopre})~and~(\ref{eq:CFIgrowthrate_isobreak})). Our finding suggests that the relative number densities of different neutrino species is a good diagnostic to assess occurrences of CFI.

One caveat is that the above analysis is based on our approximate estimation of the growth rate of CFI (Eqs.~(\ref{eq:CFIgrowthrate_isopre})~and~(\ref{eq:CFIgrowthrate_isobreak})). However, the isotropic condition for neutrino angular distributions is obviously broken in spatial regions where we discuss here. This suggests that anisotropic CFI modes could be unstable in these regions. We also note that nucleon scatterings, which are neglected in this study, may also contribute to CFI. It should be mentioned, however, that, even if such a CFI occurs, their growth rates should be relatively low. This is because weak reactions in the outer radii at $t=1.3\,$s is subtle. Therefore we conclude that CFI at outer radii does not significantly impact neutrino radiation fields in the late phase of BNSM remnant.

\section{Summary and conclusion}\label{sec:summary}

BNSMs offer unique astrophysical environments responsible for strong GW and neutrino emission, r-process nucleosynthesis, and short GRB. Although great progress has been made in numerical simulations, collective neutrino oscillations, which could be prevalent in the remnant system, have posed a great challenge in BNSM modeling. Addressing these issues require modeling quantum kinetic neutrino transport, but these numerical simulations are, unfortunately, infeasible due to exorbitant computational cost. On the other hand, linear stability analysis of neutrino flavor instabilities are tractable problems and it is valuable to understand when, where, and how collective neutrino oscillations occur. The similar study has already been attempted in CCSN models (see, e.g., \cite{2019ApJ...886..139N,2021PhRvD.103f3033A,2021PhRvD.104h3025N,2024PhRvD.109b3012A}) and BNSM remnant systems with a BH accretion disk \cite{2024ApJ...974..110M}. 

However, we still have limited understanding of mechanisms for occurrences of flavor instabilities in BNSM remnant systems, which require detailed analyses with multi-energy, multi-angle, multi-species neutrino transport based on sophisticated multi-dimensional BNSM models. One of the compelling questions is whether the type of central compact remnant (BH or HMNS) can qualitatively affect occurrences of flavor instabilities. In the light of this, we conduct an in-depth stability analysis for FFI and CFI, corresponding to the most notable flavor instabilities potentially happening in BNSM environments, by using a model of long-term numerical relativity simulations with a long-lived HMNS remnant. To increase the reliability of our analysis, we run Boltzmann neutrino transport simulations on top of matter distributions in a post-processing manner. This approach is especially important for searching for ELN angular crossings, which correspond to indicators to the onset of FFI.

Our key findings are summarized as follows.
\begin{enumerate}
\item We confirm the emergence of FFI and CFI in the remnant of BNSM (see Sec.~\ref{subsec:overall}), which is consistent with previous studies. However, detailed features in ELN angular crossings cannot be explained by a simple geometric argument proposed by \cite{2017PhRvD..95j3007W} and ELN crossing analyses based on more detailed simulations in \cite{2024ApJ...974..110M}. Regarding CFI, we also find some new insights regarding their detailed properties.
\item In our models, four different mechanisms operate in generating ELN angular crossings (see Sec.~\ref{subsec:mechaELNgene}). In each case, occurrences of ELN angular crossings are associated with regions with $\mu_{\nu}^{\rm eq}>0$. This is attributed to the fact that the overall radiation field is dominated by $\bar{\nu}_e$ due to the presence of HMNS. In such environments, $\nu_e$ supply by charged-current reactions in regions with $\mu_{\nu}^{\rm eq}>0$ can make ELN positive in along certain neutrino trajectories, which is indicative of appearance of ELN angular crossings. However, it is important to emphasize that ELN crossings do not occur unless the emission of $\nu_e$ is strong enough to exceed $\bar{\nu}_e$, indicating that the appearance of $\mu_{\nu}^{\rm eq}>0$ is merely a necessary condition for FFI (see Sec.~\ref{subsec:mechaELNgene} for more details). We also note that three schematic pictures (Figs.~\ref{fig:CrossingbyNuedif},~\ref{fig:Crossingbyshadow}, and~\ref{fig:Crossingbycontami}) help readers to understand complex generation mechanisms graphically.
\item The overall property of CFI is in line with \cite{2023PhRvD.108h3002X}. This indicates that CFI is less sensitive to the difference of the central remnant (BH or HMNS) than FFI. We also find that, although the time scale of CFI observed in our models is longer than FFI, CFI persistently exists in accretion disks at $t \lesssim 1\,$s, suggesting that impacts of CFI could be more significant than FFI. The sustained CFI appearance in accretion disks is attributed to the fact that the low $Y_e$ environment leads to a large disparity of neutrino matter reactions between $\nu_e$ and $\bar{\nu}_e$, that offers preferable conditions for occurrences of CFI.
\item We demonstrate that neglecting $\nu_x$ results in overestimating both the area of unstable regions and growth rates of CFI, exhibiting the importance of accurate modeling of not only $\nu_e$ and $\bar{\nu}_e$ radiation fields but also $\nu_x$ one for CFI. 
\item We also find that unstable region of CFI disappears even inside the disk at very late phase ($t \gtrsim 1\,$s). This is partially due to the leptonization (or protonization) in the inner disk. For outer radii, $\nu_x$ is more abundant than $\nu_e$, which leads to suppress CFI (see Sec.~\ref{subsubsec:disapCFI} for more details).
\end{enumerate}


While the present study supports a claim of FFI and CFI occurring in BNSM remnants, gauging impacts of neutrino flavor conversions associated with these flavor instabilities is a different issue. Further progress can be made by going beyond local- and stability analyses. A promising path forward lies in conducting numerical simulations on global quantum kinetic neutrino transport, which can be technically handled by appropriate approximations/prescriptions, e.g., an attenuation method \cite{2022PhRvL.129z1101N,2023PhRvD.107h3016X,2023PhRvD.108j3014N,2023PhRvL.130u1401N,2023PhRvD.108l3003N,2024PhRvD.109l3008X}. These simulations will provide a reference data for developing more accurate subgrid models of flavor conversions. 

Useful surrogate formulae of QKE for subgrid models have already been proposed in the literature (see, e.g., BGK subgrid model in \cite{2024PhRvD.109h3013N}). We also note that time-dependent neutrino-radiation hydrodynamic simulations with incorporating effects of flavor conversions have already been attempted \cite{2021PhRvL.126y1101L,2022PhRvD.105h3024J,2022PhRvD.106j3003F}, while recent two works in \cite{2025arXiv250311758Q,2025arXiv250323727L} adopted more physically accurate prescriptions of flavor instabilities than previous ones, exhibiting the rapid progress in this research area. Regardless of the difference of detailed prescriptions of flavor conversions, these works commonly demonstrated that flavor conversions could have an influence on BNSM dynamics. For more quantitative arguments, further efforts should be made to improve the physical fidelity of flavor conversions in BNSM simulations.

Last but not least, it should be mentioned that quantitative discussions regarding impacts of FFI and CFI require six-species neutrino (three flavors plus their anti-partners) analysis. More specifically, heavy leptonic neutrinos can no longer be handled collectively, when flavor conversions start to develop. In fact, if on-shell muons appear in BNSM remnants, flavor instabilities would be more complex, since flavor instabilities can occur in all three coherent sectors (see, e.g., \cite{2024PhRvD.110d3039L,2025arXiv250318145L}). At the moment, we leave the problem open but will tackle the important issue regarding roles of on-shell muons on flavor conversions in future work.

\begin{acknowledgments}
We are supported by Grant-in-Aid for Scientific Research (23K03468, 23H04900, 24K00632, 25H01273). For providing high performance computing resources, Computing Research Center, KEK, JLDG on SINET of NII, Research Center for Nuclear Physics, Osaka University, Yukawa Institute of Theoretical Physics, Kyoto University, and Information Technology Center, University of Tokyo are acknowledged. This work is also supported by MEXT as "Program for Promoting Researches on the Supercomputer Fugaku" (Structure and Evolution of the Universe Unraveled by Fusion of Simulation and AI, JPMXP1020230406, Project ID: hp230204, hp240219,hp250226), the HPCI System Research Project (Project ID: hp230056, hp230270, hp240041, hp240264, hp240079, hp250166, hp250006). Numerical computations were performed at Sakura, Cobra, and Raven of the Max Planck Computing and Data Facility.
\end{acknowledgments}
\bibliography{bibfile}

\appendix

\begin{figure*}[ht]
\begin{minipage}{1.0\textwidth}
\centering
\includegraphics[width=0.32\linewidth]{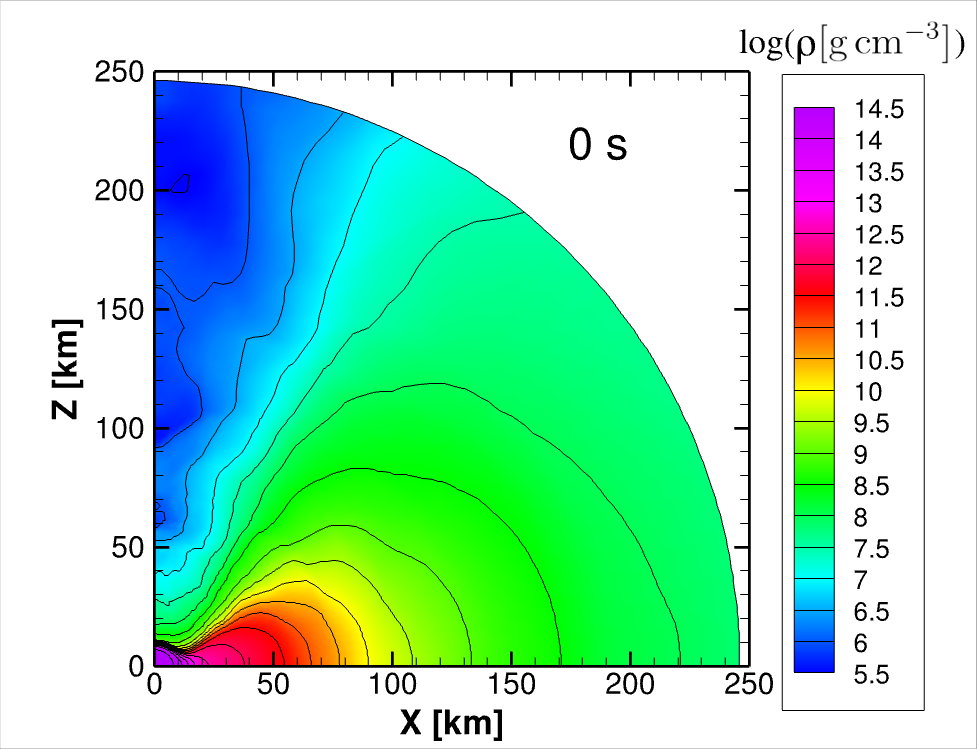}
\includegraphics[width=0.32\linewidth]{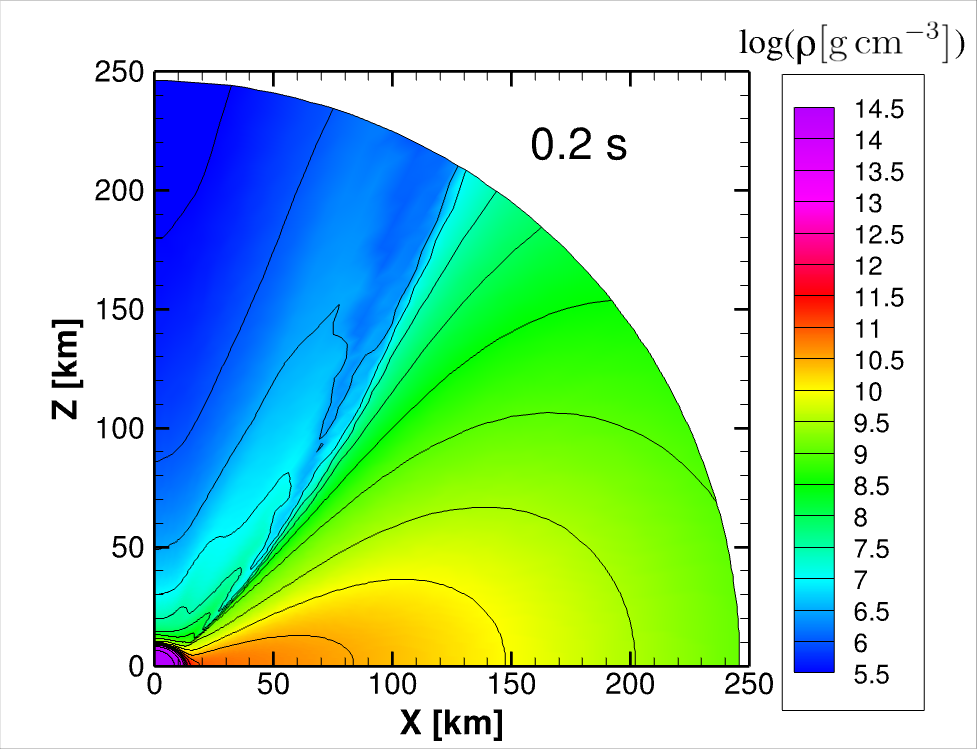}
\includegraphics[width=0.32\linewidth]{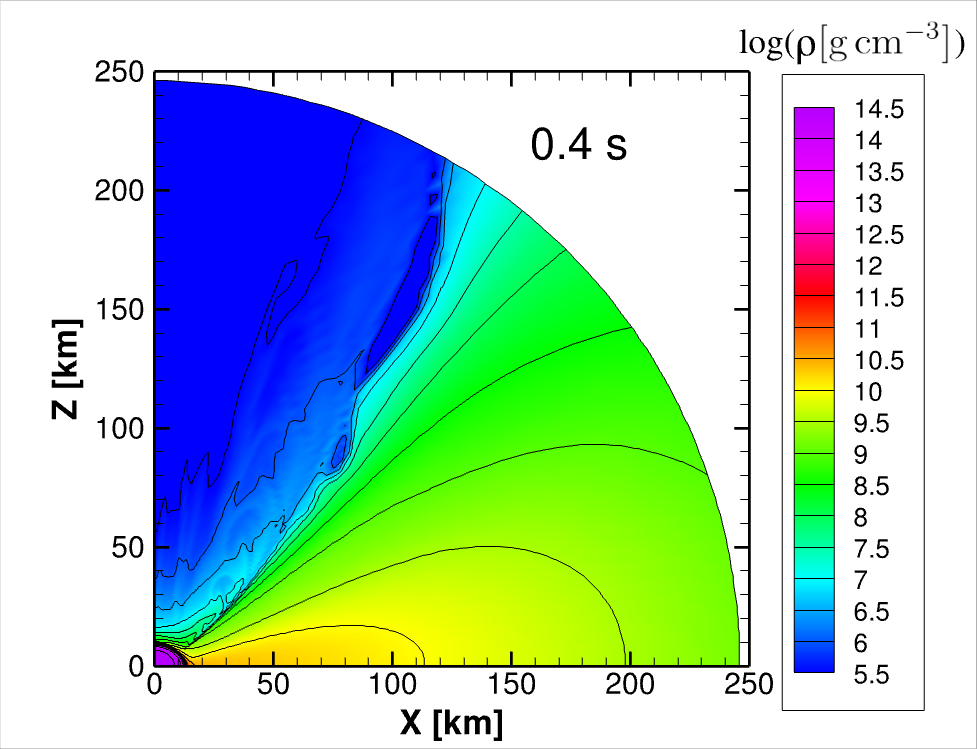}
\includegraphics[width=0.32\linewidth]{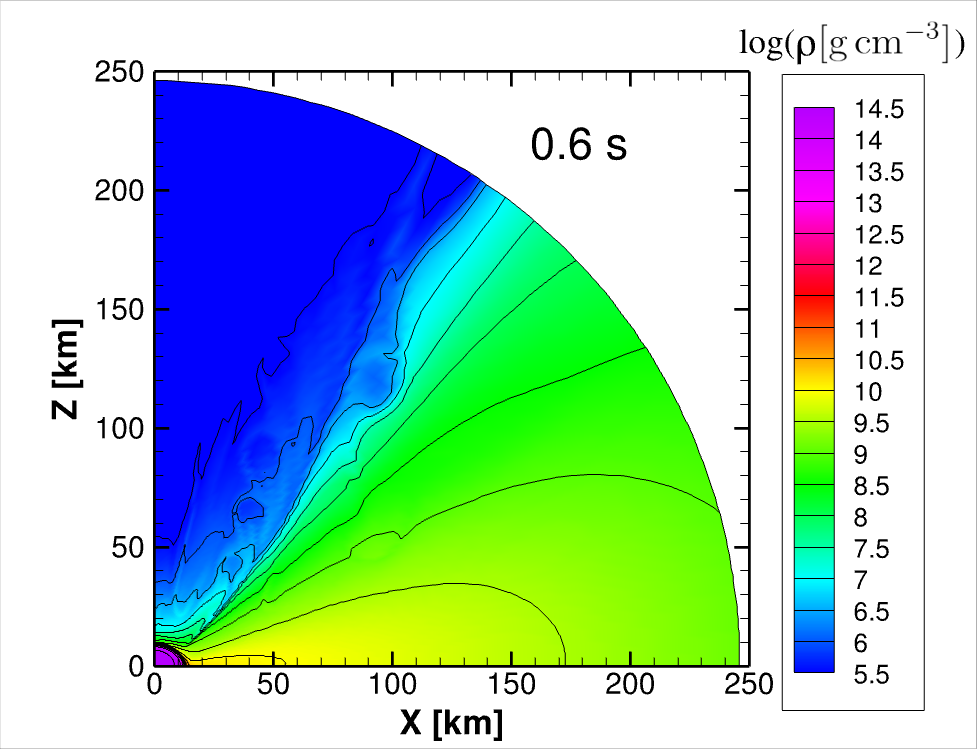}
\includegraphics[width=0.32\linewidth]{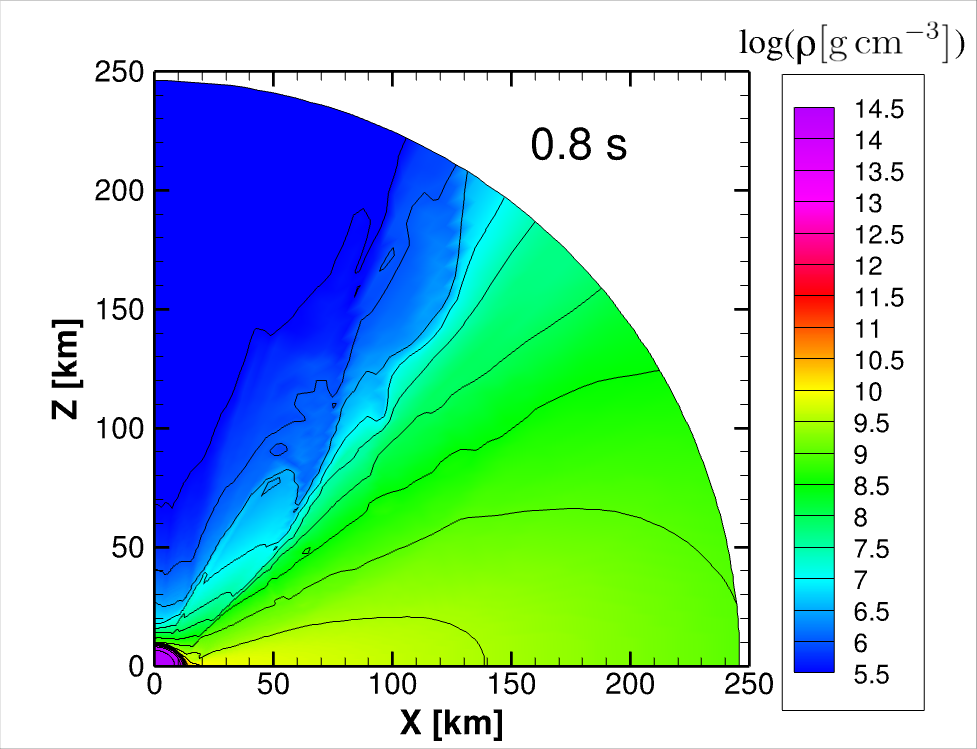}
\includegraphics[width=0.32\linewidth]{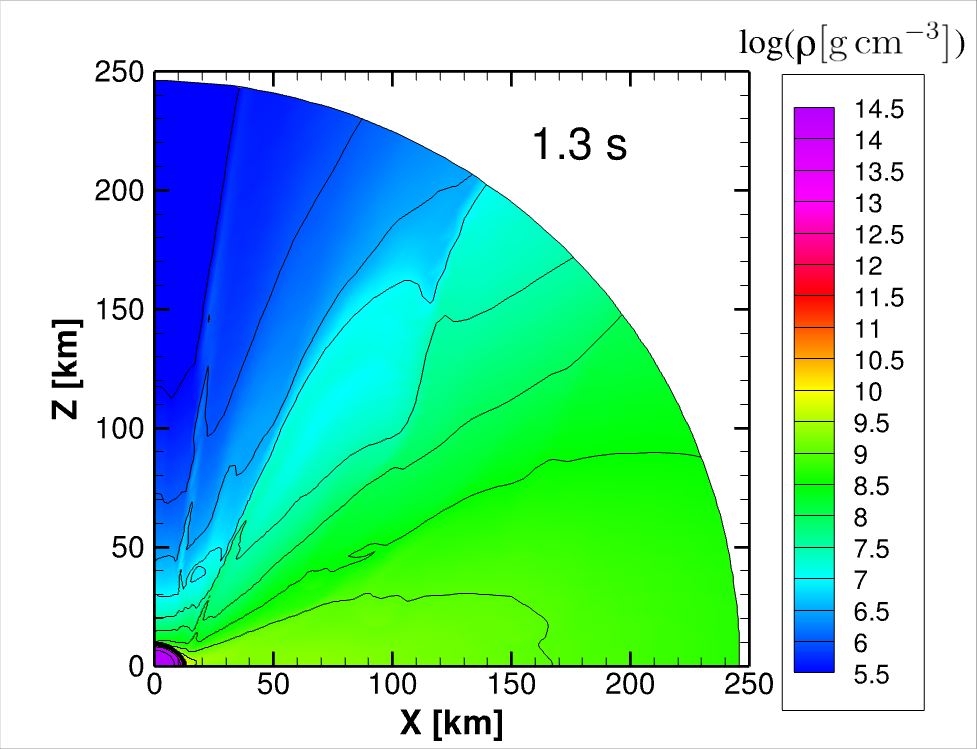}
\end{minipage}
\caption{Color maps of rest-mass density of fluid ($\rho$).}
\label{fig:matterdensity}
\end{figure*}

\begin{figure*}[ht]
\begin{minipage}{1.0\textwidth}
\centering
\includegraphics[width=0.32\linewidth]{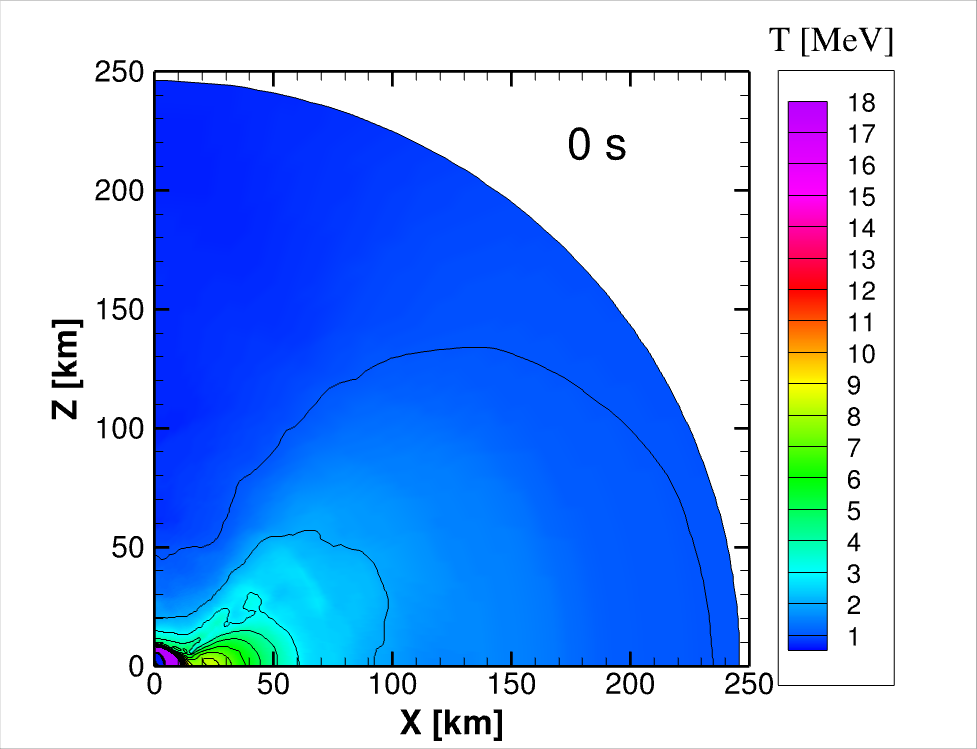}
\includegraphics[width=0.32\linewidth]{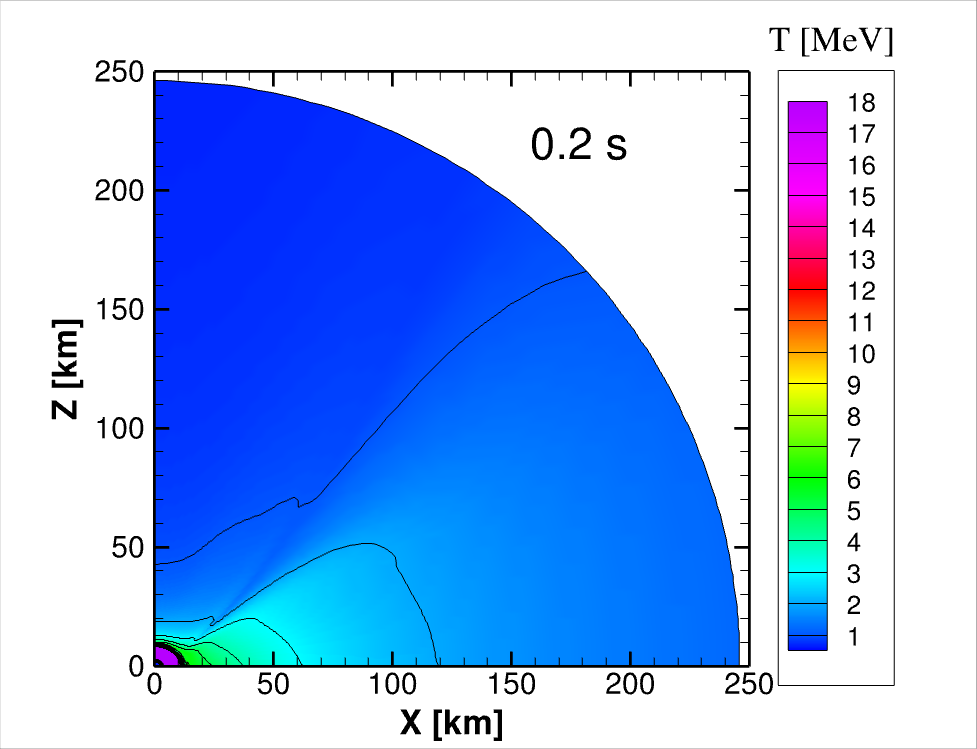}
\includegraphics[width=0.32\linewidth]{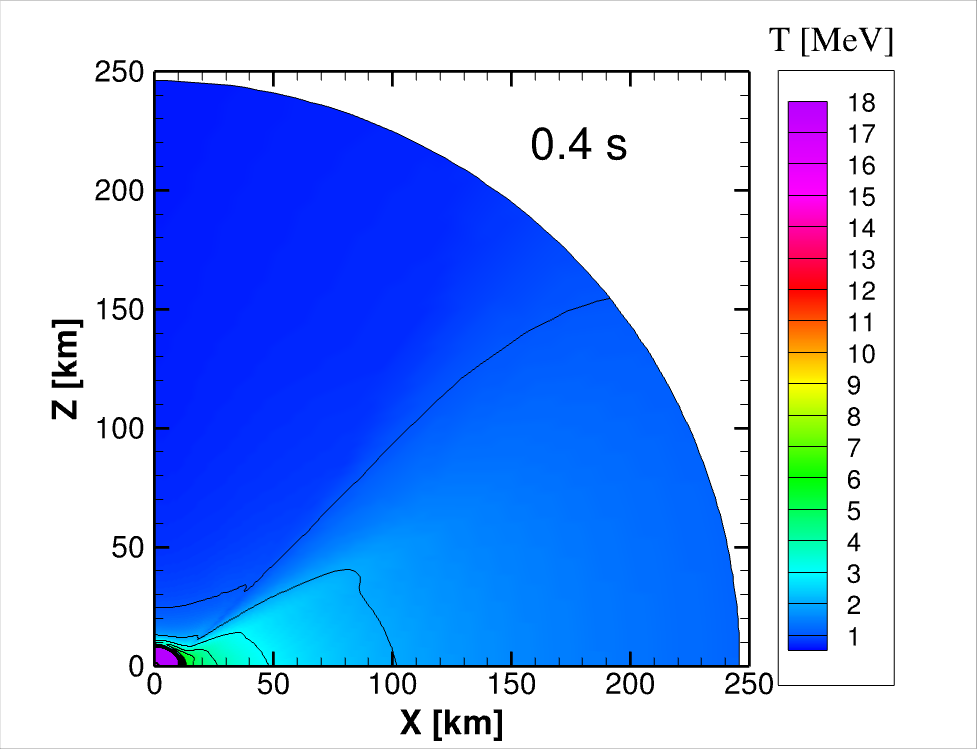}
\includegraphics[width=0.32\linewidth]{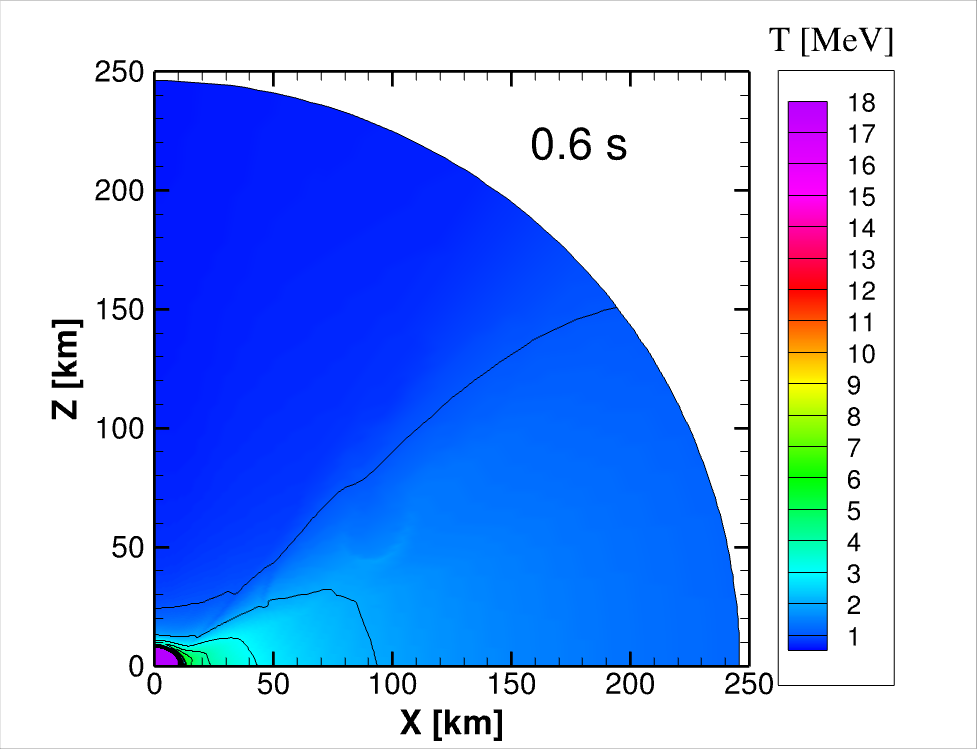}
\includegraphics[width=0.32\linewidth]{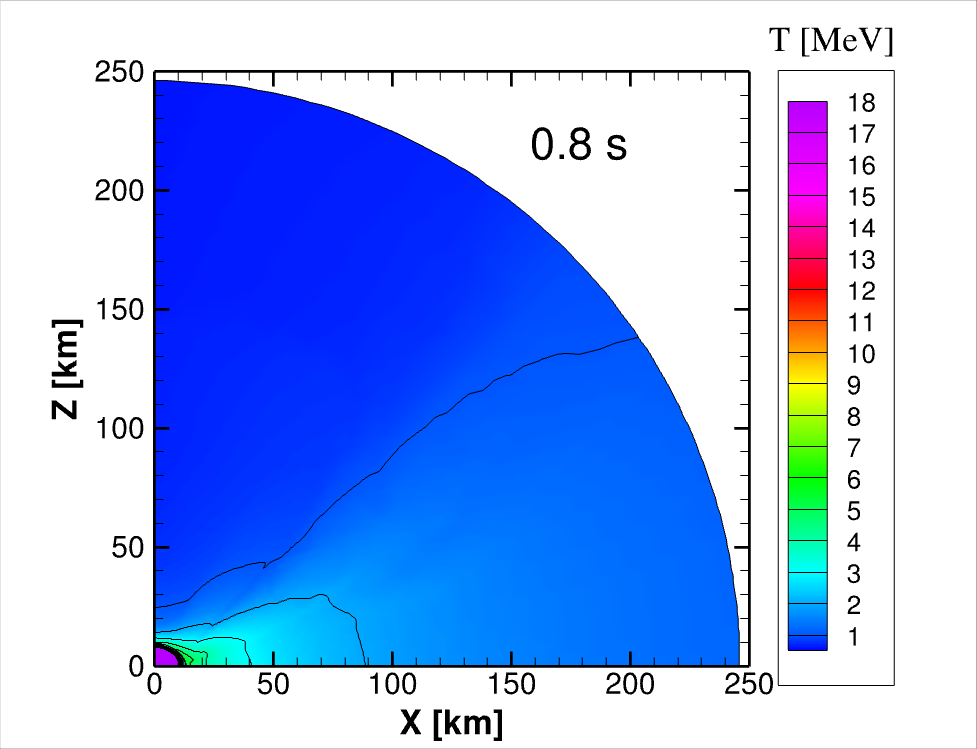}
\includegraphics[width=0.32\linewidth]{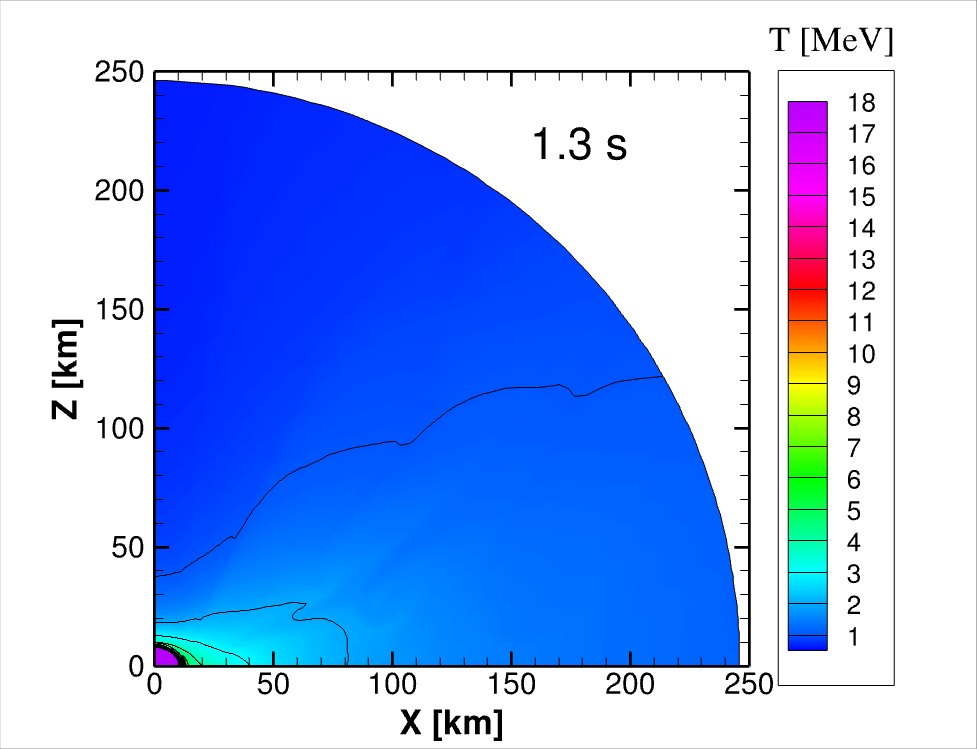}
\end{minipage}
\caption{Same as Fig.~\ref{fig:matterdensity} but for matter temperature (${\rm T}$).}
\label{fig:tempera}
\end{figure*}

\begin{figure}[ht]
\centering
\begin{minipage}{0.5\textwidth}
\includegraphics[width=0.9\linewidth]{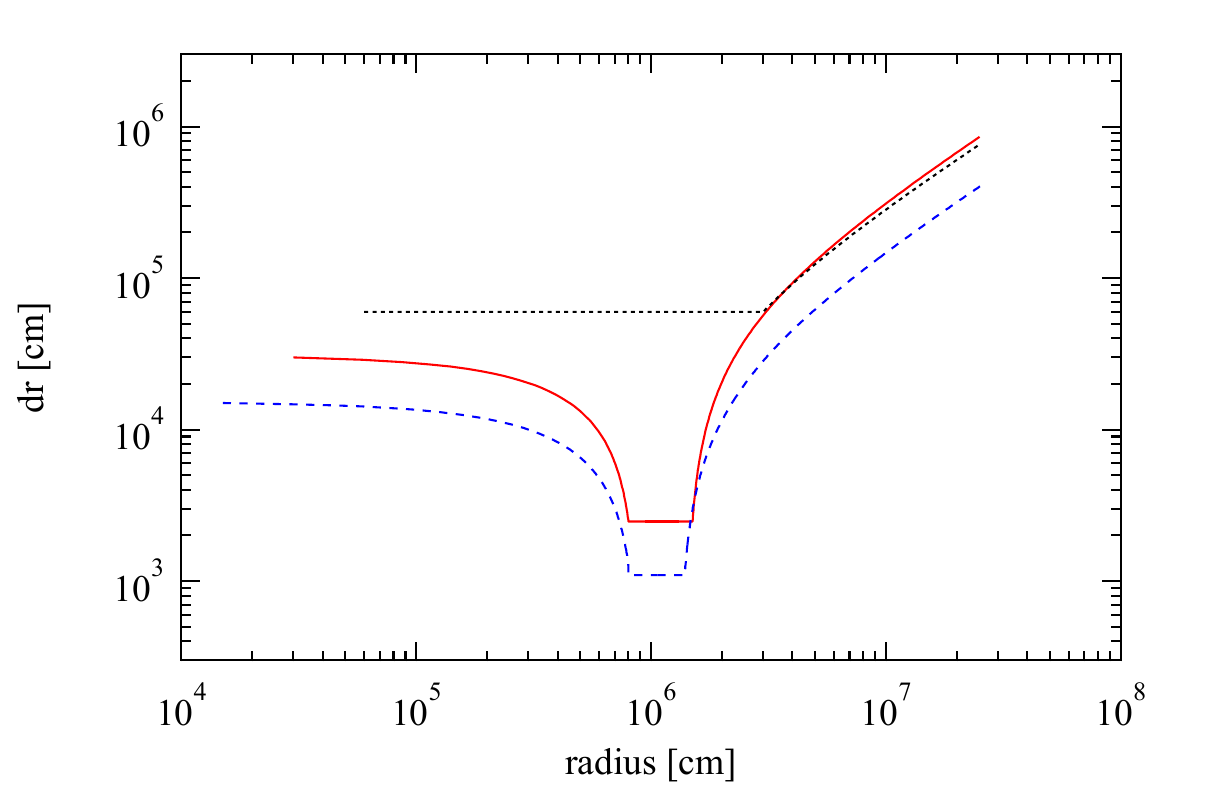}
\end{minipage}
\caption{Radial grid structure employed in Boltzmann simulations. The dotted line displays the radial grid spacing in the previous study (128 grid points) \cite{2021ApJ...907...92S}. The solid and dashed lines correspond to the radial grid width in our reference model (512 grid points) and high resolution one (1024 grid points), respectively.}
\label{fig:radial-mesh_comparison}
\end{figure}

\begin{figure}[ht]
\centering
\begin{minipage}{0.5\textwidth}
\includegraphics[width=0.9\linewidth]{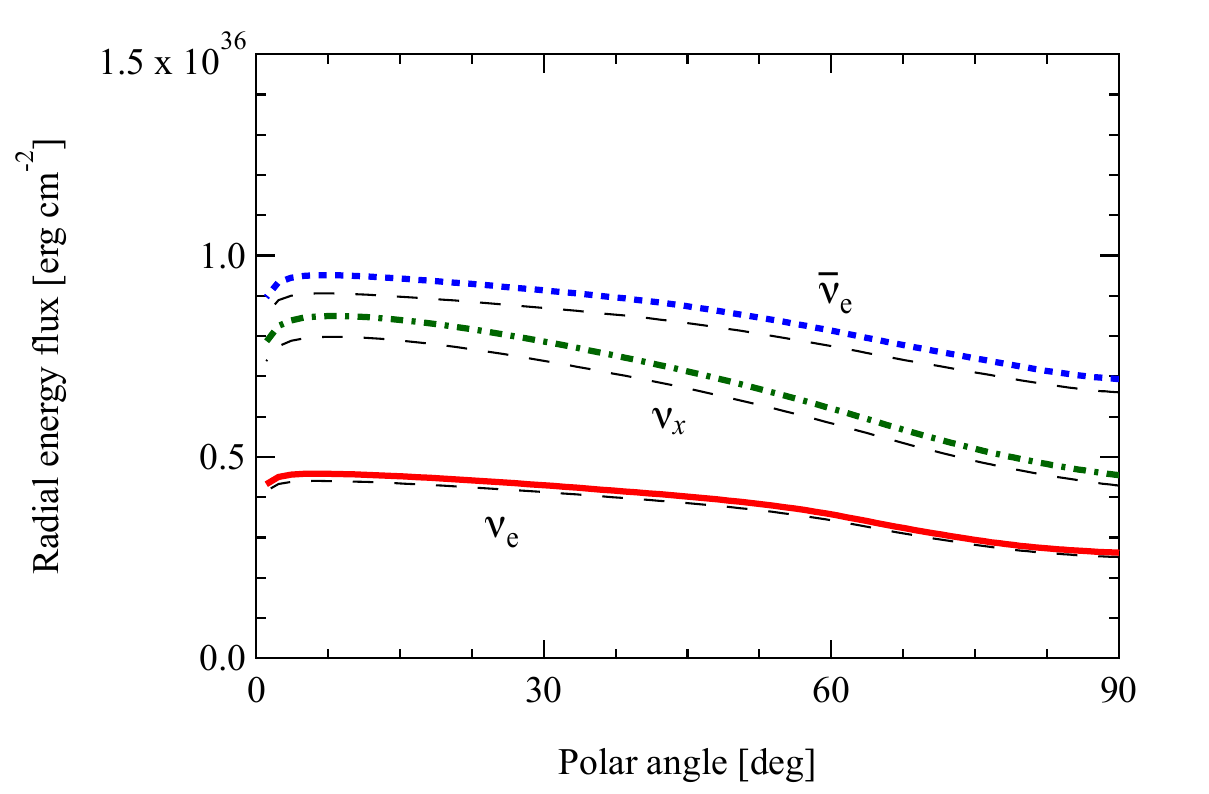}
\end{minipage}
\caption{The radial component of energy flux of neutrinos, measured at the outer boundary, as a function of polar angle. The time is $t=1.3\,$s. Neutrino species are distinguished by line types and color: $\nu_e$ (red solid), $\bar{\nu}_e$ (blue dotted), and $\nu_{x}$ (green dash-dotted). Results of high resolution simulation are shown by thin long-dashed lines at each panel.}
\label{fig:polar-mesh_comparison}
\end{figure}

\section{Fluid profiles} \label{appendix:matterdist}

We provide matter profiles used in this study, which are obtained by numerical relativity simulations in \cite{2020ApJ...901..122F} (DD2-135M model). We note that our previous work in \cite{2021ApJ...907...92S} focused on the phase with $0 < t < 135\,$ms, while the present study extends over a longer timescale ($>1\,$s). Figure~\ref{fig:matterdensity} portrays spatial distributions of baryon mass density at different time snapshots. We also provide distributions of matter temperature in Fig.~\ref{fig:tempera}. In analysis of flavor instabilities, these figures can be used as a reference displaying the overall trend of time evolution of the accretion disk. See also Figs.~\ref{fig:nuechemipote}~and~\ref{fig:Ye} for chemical potential of $\nu_e$ and $Y_e$, respectively.

\section{Grid structure and resolution study in Boltzmann neutrino transport simulations} \label{appendix:resodepe}

As mentioned in Sec.~\ref{subsec:Boltzmann}, high spatial resolutions are necessary in modeling of neutrino transport in the vicinity of HMNS, where matter profiles are substantially changed over the short distance. We note that HMNS gradually shrinks with time due to neutrino cooling, and consequently the density gradient around the envelope becomes sharper. This indicates that the required spatial resolution needs to be higher in the later phase of BNSM remnant. According to our resolution study, insufficient spatial resolutions results in the overestimation of neutrino flux, and the least required resolution around the envelop of HMNS at $T=1.3\,$s (the final time snapshot in our simulation) is $\sim 20\,$m to accurately model the neutrino transport. 

On the other hand, such a fine resolution is not required inside HMNS and in the outer envelope, where matter gradient is more moderate. We, hence, adopt a grid structure tailored to the problem. Figure~\ref{fig:radial-mesh_comparison} displays radial grid width as a function of radius for three different cases. The black dotted curve represents the case used in our previous study \cite{2021ApJ...907...92S}, while the red solid curve corresponds to the one adopted in the present study. As depicted, the new grid structure has a $\sim 20\,$m resolution at $r \sim 10\,$km where the density gradient is the steepest. On the other hand, the radial grid width at the center is roughly ten times larger than the finest resolution, and it increases logarithmically with radius at $r \gtrsim 15\,$km. This gridding strategy efficiently reduces the number of grid points while maintaining high spatial resolutions in the outer region of the HMNS.

To validate the grid structure and our reference spatial resolution, we perform higher resolution simulations at a time snapshot of $t=1.3\,$s. The radial grid structure for the high resolution simulation can be seen in a blue dashed curve in Fig.~\ref{fig:radial-mesh_comparison}. In Fig.~\ref{fig:polar-mesh_comparison}, we compare the radial component of energy flux measured at the outer boundary between the reference- and high resolution models. We confirm that the error is within a few percents, demonstrating a good performance of our gridding technique. We also note that, regarding the resolution study in neutrino angular directions, a detailed check has already been made in our previous study. We refer interested readers to Appendix~B in \cite{2021ApJ...907...92S} for more details.

\end{document}